\documentclass[12pt] {article}
\begin{document}
\setcounter{page}{1}
\def\theequation{\arabic{section}.\arabic{equation}}
\def\theequation{\thesection.\arabic{equation}}
\setcounter{section}{0}

\title{Dynamics of low--energy nuclear forces for electromagnetic and
weak reactions with the deuteron in the Nambu--Jona--Lasinio model of
light nuclei}

\author{A. N. Ivanov~\thanks{E--mail: ivanov@kph.tuwien.ac.at, Tel.:
+43--1--58801--14261, Fax: +43--1--58801--14299}~${\textstyle
^\ddagger}$, H. Oberhummer~\thanks{E--mail: ohu@kph.tuwien.ac.at,
Tel.: +43--1--58801--14251, Fax: +43--1--58801--14299} ,
N. I. Troitskaya~\thanks{Permanent Address: State Technical
University, Department of Nuclear Physics, 195251 St. Petersburg,
Russian Federation} , M. Faber~\thanks{E--mail:
faber@kph.tuwien.ac.at, Tel.: +43--1--58801--14261, Fax:
+43--1--58801--14299}}

\date{\today}

\maketitle

\begin{center}
{\it Institut f\"ur Kernphysik, Technische Universit\"at Wien,\\
Wiedner Hauptstr. 8-10, A-1040 Vienna, Austria}
\end{center}

\begin{center}
\begin{abstract}
A dynamics of low--energy nuclear forces is investigated for
low--energy electromagnetic and weak nuclear reactions with the
deuteron in the Nambu--Jona--Lasinio model of light nuclei by example
of the neutron--proton radiative capture (M1--capture) n + p $\to$ D +
$\gamma$, the photomagnetic disintegration of the deuteron $\gamma$ +
D $\to$ n + p and weak reactions of astrophysical interest. These are
the solar proton burning p + p $\to$ D + e$^+$ + $\nu_{\rm e}$, the
pep--process p + e$^-$ + p $\to$ D + $\nu_{\rm e}$ and the neutrino
and anti--neutrino disintegration of the deuteron caused by charged
$\nu_{\rm e}$ + D $\to$ e$^-$ + p + p, $\bar{\nu}_{\rm e}$ + D $\to$
e$^+$ + n + n and neutral $\nu_{\rm e}(\bar{\nu}_{\rm e})$ + D $\to$
$\nu_{\rm e}(\bar{\nu}_{\rm e})$ + n + p weak currents.
\end{abstract}
\end{center}

\begin{center}
\end{center}

\newpage

\section{Introduction}
\setcounter{equation}{0}

\hspace{0.2in} Recently [1] we have developed the Nambu--Jona--Lasinio
model of light nuclei [2], or differently the nuclear
Nambu--Jona--Lasinio (NNJL) model, invented for the description of
low--energy nuclear forces at the quantum field theoretic level. We
have shown that the NNJL model is fully motivated by QCD [1]. The
deuteron appears in {\it the nuclear phase of QCD} as a
neutron--proton collective excitation, a Cooper np--pair, caused by a
phenomenological local four--nucleon interaction. Strong low--energy
interactions of the deuteron coupled to itself and other particles are
described in terms of one--nucleon loop exchanges. The one--nucleon
loop exchanges allow to transfer nuclear flavours from an initial to a
final nuclear state by a minimal way and to take into account
contributions of nucleon--loop anomalies determined completely by
one--nucleon loop diagrams. The dominance of contributions of
nucleon--loop anomalies is justified in the large $N_C$ approach to
the description of non--perturbative QCD with $SU(N_C)$ gauge group at
$N_C\to\infty$, where $N_C$ is the number of quark colours.

In this paper we apply the NNJL model to the description of
low--energy nuclear forces for electromagnetic and weak reactions with
the deuteron by example of the evaluation of the cross sections for
the neutron--proton radiative capture n + p $\to$ D + $\gamma$ for
thermal neutrons caused by the ${^1}{\rm S}_0 \to {^3}{\rm S}_1$
transition (the M1--transition), the photomagnetic disintegration
of the deuteron $\gamma$ + D $\to$ n + p and weak reactions of
astrophysical interest: 1) the solar proton burning p + p $\to$ D +
e$^+$ + $\nu_{\rm e}$, 2) the pep--process p + e$^-$ + p $\to$ D +
$\nu_{\rm e}$ and 3) the reactions of neutrino and anti--neutrino
disintegration of the deuteron caused by charged $\nu_{\rm e}$ + D
$\to$ e$^-$ + p + p, $\bar{\nu}_{\rm e}$ + D $\to$ e$^+$ + n + n and
neutral $\nu_{\rm e}(\bar{\nu}_{\rm e})$ + D $\to$ $\nu_{\rm
e}(\bar{\nu}_{\rm e})$ + n + p weak currents.

\noindent{\bf Low--energy electromagnetic nuclear reactions with the
deuteron}. It is well--known that the reaction of the neutron--proton
radiative capture n + p $\to$ D + $\gamma$ for thermal neutrons plays
an important role for the primordial nucleosynthesis in the Big Bang
model [3]. Indeed, the deuterons produced via the neutron--proton
radiative capture n + p $\to$ D + $\gamma$ burn to ${^4}{\rm He}$
through the reactions D + p $\to$ ${^3}{\rm He}$ + $\gamma$ and
${^3}{\rm He}$ + n $\to$ ${^4}{\rm He}$ + $\gamma$. Therefore, the
correct description of the neutron--proton radiative capture n + p
$\to$ D + $\gamma$ is essential for the theoretical prediction of the
amount of the matter in Universe.

The reaction of the photomagnetic disintegration of the deuteron
$\gamma$ + D $\to$ n + p is related to the neutron--proton radiative
capture n + p $\to$ D + $\gamma$ via time--reversal invariance of
strong and electromagnetic forces. The investigation of this reaction
is conceived to get an additional check of our result for the
M1--capture related to the analysis of the energy dependence of the
cross section calculated in the NNJL model at energies far from
threshold.

The cross section $\sigma({\rm np \to D \gamma})(T_{\rm n})$ for the
neutron--proton radiative capture has been measured for thermal
neutrons at the laboratory kinetic energy $T_{\rm n} = 0.0252\,{\rm
eV}$ that corresponds to the laboratory velocity $v_{\rm n}/c =
7.34\times 10^{-6}$ (the absolute value is $v_{\rm n} = 2.2\times
10^5\,{\rm cm/s}$) [4]:
\begin{eqnarray}\label{label1.1}
\sigma({\rm np \to D \gamma})_{\exp}(T_{\rm n}) = (334.2\pm 0.5)\,{\rm mb}.
\end{eqnarray}
For the first time the cross section for the neutron--proton radiative
capture has been calculated by Austern [5] in the Potential model
approach (PMA):
\begin{eqnarray}\label{label1.2}
\sigma({\rm np \to D \gamma})(T_{\rm n}) = (303\pm 4)\,{\rm mb}.
\end{eqnarray}
The observed discrepancy about 10$\%$ has been then explained by Riska
and Brown [6] in terms of the contributions of the exchange currents
and the $\Delta(1232)$ resonance.

Recently the evaluation of the cross section for the neutron--proton
radiative capture n + p $\to$ D + $\gamma$ has been carried out by
using Chiral perturbation theory in the framework of the Effective
Field Theory (EFT) approach [7--10] formulated by Weinberg within
Effective Chiral Lagrangian description of nuclear forces [11] (see
also Refs.\,[12]). The theoretical results obtained in Refs.\,[7--10]
for the cross section for the neutron--proton radiative capture are
rather contradictory. Indeed, in Ref.\,[7] the experimental value of
the cross section for the M1--capture has been reproduced without free
parameters, the definition of which demands a fit of experimental
data, with an accuracy better than 1$\%$, $\sigma({\rm np \to D
\gamma})(T_{\rm n}) = (334\pm 3)\,{\rm mb}$. In turn, in the more
recent publications [8--10] the predictions are not so much
optimistic. Indeed, in Ref.\,[8] there has been obtained the value
$\sigma({\rm np \to D \gamma})(T_{\rm n}) = 297.2\,{\rm mb}$ which is
about 11$\%$ less compared with the experimental one
Eq.(\ref{label1.1}). As has been stated in Ref.\,[8] for the
evaluation of the correct value of the cross section for the
M1--capture within the EFT one needs to add a free parameter undefined
in the approach. This parameter should be fixed from the fit of the
experimental value Eq.(\ref{label1.1}). This program has been realized
in Ref.\,[9]. Then, in Refs.\,[10] the neutron--proton radiative
capture has been calculated for the center of mass energies of the np
pair up to $T_{\rm np} \le 1\,{\rm MeV}$, $T_{\rm n} = 2\,T_{\rm np}$,
by including the contribution of the E1--transition in addition to the
M1.  This has added new free parameters with respect to that
introduced in Refs.\,[8,9], where only the M1--transition has been
taken into account.

In the NNJL model we calculate the cross section for the M1--capture
both in the tree--meson approximation and by including the
contributions of chiral one--meson loop corrections and the
$\Delta(1232)$ resonance. For the evaluation of chiral one--meson loop
corrections we apply Chiral perturbation theory at the quark level
(CHPT)$_q$ [13] developed within the extended Nambu--Jona--Lasinio
(ENJL) model with a linear realization of chiral $U(3)\times U(3)$
symmetry.

In our consideration the $\Delta(1232)$ resonance is the Rarita--Schwinger
field [14] $\Delta^a_{\mu}(x)$, the isotopical index $a$ runs over $a =
1,2,3$, having the following free Lagrangian [15,16]:
\begin{eqnarray}\label{label1.3}
\hspace{-0.5in} {\cal L}^{\Delta}_{\rm kin}(x) =
\bar{\Delta}^a_{\mu}(x) [-(i\gamma^{\alpha}\partial_{\alpha} -
M_{\Delta}) \,g^{\mu\nu} +
\frac{1}{4}\gamma^{\mu}\gamma^{\beta}(i\gamma^{\alpha}\partial_{\alpha}
- M_{\Delta}) \gamma_{\beta}\gamma^{\nu}] \Delta^a_{\nu}(x),
\end{eqnarray}
where $M_{\Delta} = 1232\,{\rm MeV}$ is the mass of the $\Delta$
resonance field $\Delta^a_{\mu}(x)$. In terms of the eigenstates of
the electric charge operator the fields $\Delta^a_{\mu}(x)$ are given
by [15,16]
\begin{eqnarray}\label{label1.4}
\begin{array}{llcl}
&&\Delta^1_{\mu}(x) = \frac{1}{\sqrt{2}}\Biggr(\begin{array}{c}
\Delta^{++}_{\mu}(x) - \Delta^0_{\mu}(x)/\sqrt{3} \\
\Delta^+_{\mu}(x)/\sqrt{3} - \Delta^-_{\mu}(x)
\end{array}\Biggl)\,,\,
\Delta^2_{\mu}(x) = \frac{i}{\sqrt{2}}\Biggr(\begin{array}{c}
\Delta^{++}_{\mu}(x) + \Delta^0_{\mu}(x)/\sqrt{3} \\
\Delta^+_{\mu}(x)/\sqrt{3} + \Delta^-_{\mu}(x)
\end{array}\Biggl)\,,\\
&&\Delta^3_{\mu}(x) = -\sqrt{\frac{2}{3}}\Biggr(\begin{array}{c}
\Delta^+_{\mu}(x) \\ \Delta^0_{\mu}(x) \end{array}\Biggl).
\end{array}
\end{eqnarray}
The fields $\Delta^a_{\mu}(x)$ obey the subsidiary constraints:
$\partial^{\mu}\Delta^a_{\mu}(x) = \gamma^{\mu}\Delta^a_{\mu}(x) = 0$
[14--16]. The Green function of the free $\Delta$--field is determined
by
\begin{eqnarray}\label{label1.5}
\hspace{-0.5in}\langle 0|{\rm T}(\Delta_{\mu}(x_1)
\bar{\Delta}_{\nu}(x_2))|0\rangle
= - i S_{\mu\nu}(x_1 - x_2).
\end{eqnarray}
In the momentum representation $S_{\mu\nu}(x)$ reads [15--17]:
\begin{eqnarray}\label{label1.6}
\hspace{-0.5in}S_{\mu\nu}(p) = \frac{1}{M_{\Delta} - \hat{p}}\Bigg( -
g_{\mu\nu} + \frac{1}{3}\gamma_{\mu}\gamma_{\nu} +
\frac{1}{3}\frac{\gamma_{\mu}p_{\nu} -
\gamma_{\nu}p_{\mu}}{M_{\Delta}} +
\frac{2}{3}\frac{p_{\mu}p_{\nu}}{M^2_{\Delta}}\Bigg).
\end{eqnarray}
The most general form of the ${\rm \pi N \Delta}$-- interaction
compatible with the requirements of chiral symmetry reads [18]:
\begin{eqnarray}\label{label1.7}
\hspace{-0.5in}&&{\cal L}_{\rm \pi N \Delta}(x) = \frac{g_{\rm \pi
N\Delta}}{2M_{\rm
N}}\bar{\Delta}^a_{\omega}(x)\Theta^{\omega\varphi}N(x)
\partial_{\varphi}\pi^a(x)
+ {\rm h.c.} = \nonumber\\
\hspace{-0.5in}&&= \frac{g_{\rm \pi N\Delta}}{\sqrt{6}M_{\rm
N}}\Bigg[\frac{1}{\sqrt{2}}\bar{\Delta}^+_{\omega}(x)\Theta^{\omega\varphi}
n(x) \partial_{\varphi}\pi^+(x) -
\frac{1}{\sqrt{2}}\bar{\Delta}^0_{\omega}(x)\Theta^{\omega\varphi}
p(x) \partial_{\varphi}\pi^-(x)\nonumber\\
\hspace{-0.5in}&&- \bar{\Delta}^+_{\omega}(x)\Theta^{\omega\varphi}
p(x) \partial_{\varphi}\pi^0(x) -
\bar{\Delta}^0_{\omega}(x)\Theta^{\omega\varphi} p(x)
\partial_{\varphi}\pi^0(x) + \ldots \Bigg],
\end{eqnarray}
where $M_{\rm N} = M_{\rm n} = M_{\rm p} = 940\,{\rm MeV}$ is the
nucleon mass.  The nucleon field $N(x)$ is the isotopical doublet with
the components $N(x) = (p(x), n(x))$, and $\pi^a(x)$ is the pion field
with the components $\pi^1(x) = (\pi^-(x) + \pi^+(x))/\sqrt{2}$,
$\pi^2(x) = (\pi^-(x) - \pi^+(x))/i\sqrt{2}$ and $\pi^3(x) =
\pi^0(x)$. The tensor $\Theta^{\omega\varphi}$ is given in Ref.\,[15]:
$\Theta^{\omega\varphi} = g^{\omega\varphi} - (Z +
1/2)\gamma^{\omega}\gamma^{\varphi}$, where the parameter $Z$ is
arbitrary. There is no consensus on the exact value of $Z$. From
theoretical point of view $Z=1/2$ is preferred [15].  Phenomenological
studies give only the bound $|Z| \le 1/2$ [18]. The value of
the coupling constant $g_{\rm \pi N\Delta}$ relative to the coupling
constant $g_{\rm \pi NN}$ is $g_{\rm \pi N\Delta} = 2\,g_{\rm \pi
NN}$ [19].

Assuming that the transition $\Delta \to {\rm N} + \gamma$ is
primarily a magnetic one the effective Lagrangian describing the
$\Delta \to {\rm N} + \gamma$ decays can be determined as [19--21]:
\begin{eqnarray}\label{label1.8}
\hspace{-0.5in}&&{\cal L}_{\rm \gamma N \Delta}(x) = i e \frac{g_{\rm
\gamma N\Delta}}{2M_{\rm N}}\bar{N}(x) \gamma_{\alpha}\gamma^5
\Delta^3_{\beta}(x) F^{\beta\alpha}(x) + {\rm h.c.} = \nonumber\\
\hspace{-0.5in}&&= - \frac{ie}{\sqrt{6}}\frac{g_{\rm \gamma
N\Delta}}{M_{\rm N}}[\bar{p}(x)\gamma_{\alpha}\gamma^5
\Delta^+_{\beta}(x) + \bar{n}(x)\gamma_{\alpha}\gamma^5
\Delta^0_{\beta}(x)]\,F^{\beta\alpha}(x) + {\rm h.c.},
\end{eqnarray}
where $F^{\alpha\beta}(x) = \partial^{\alpha}A^{\beta}(x) -
\partial^{\beta}A^{\alpha}(x)$ is the electromagnetic strength field
tensor and $A^{\alpha}(x)$ is a vector potential of the
electromagnetic field. The value of the coupling constant $g_{\rm
\gamma N\Delta}$ relative to the coupling constant $g_{\rm \pi NN}$ is
$g_{\rm \gamma N\Delta} = 0.14\,g_{\rm \pi NN}$ caused by the $SU(6)$
symmetry of strong low--energy interactions [19].

The NNJL model realizes the Lagrange approach to the description of
low--energy nuclear forces [1]. For the evaluation of the effective
Lagrangian of the transition n + p $\to$ D + $\gamma$ it is necessary,
first, to determine the effective Lagrangian of the strong low--energy
transition n + p $\to$ n + p, or more generally N + N $\to$ N + N,
where N = (p, n) is a nucleon field. Since the NNJL model describes
low--energy interactions of the deuteron in terms of one--nucleon loop
exchanges the effective Lagrangian of the transition N + N $\to$ N + N
plays an important role.  Due to the transition n + p $\to$ n + p the
np pair on--mass shell in the initial state transfers itself into the
np pair off--mass shell couples to the deuteron and the photon through
 one--nucleon loop exchanges. Then, the one--nucleon loop diagrams
are calculated at leading order in the $1/M_{\rm N}$ expansion that
corresponds to the large $N_C$ expansion due to the proportionality
$M_{\rm N} \sim N_C$ valid in the multi--colour QCD with $SU(N_C)$
gauge group at $N_C \to \infty$ [22].

Such a procedure of the evaluation of effective Lagrangians in the
NNJL model resembles that has being used in the ENJL model [13,23--25]
for the derivation of effective chiral Lagrangians up to the formal
replacement $q(\bar{q}) \to N(\bar{N})$, where $q(\bar{q})$ is a quark
(anti--quark) field. In the ENJL model the dominance of the leading
order contributions in the $1/M_q$ expansion, where $M_q$ is a
constituent quark mass, has been explained by a dynamics of quark
confinement, whereas in the NNJL model the dominance of the leading
contributions in the $1/M_{\rm N}$ expansion is justified by the large
$N_C$ approach to non--perturbative QCD.

Since relative momenta of the np pair in the reaction of the
M1--capture are smaller compared with the mass of the pion $M_{\pi} =
135\,{\rm MeV}$, for the derivation of the effective Lagrangian of the
strong low--energy transition n + p $\to$ n + p we can follow the
ideology of the EFT [7--12] and integrate out pion degrees of freedom
as well as other heavier degrees of freedom. The result of the
integration can be represented in the form of two contributions, where
the first is the explicit one--pion exchange, whereas the second is a
phenomenological one describing a collective contribution of both
many--pion exchanges and heavier meson degrees of freedom such as
$\sigma(660)$, $\rho(770)$, $\omega(782)$ and so on.

The effective Lagrangian ${\cal L}^{\rm np \to np}_{\rm one-pion}(x)$
of the strong low--energy transition n + p $\to$ n + p in the
one--pion exchange approximation is defined by
\begin{eqnarray}\label{label1.9}
{\cal L}^{\rm np \to np}_{\rm one-pion}(x)= \frac{g^2_{\rm \pi
NN}}{M^2_{\pi}}\,\{[\bar{n}(x)\gamma^5 n(x)][\bar{p}(x)\gamma^5
p(x)]\,-\,2\,[\bar{p}(x)\gamma^5 n(x)][\bar{n}(x)\gamma^5 p(x)]\},
\end{eqnarray}
where $g_{\rm \pi NN} = 13.4$ is the ${\rm \pi NN}$ coupling constant,
$n(x)$ and $p(x)$ are the operators of the interpolating fields of the
neutron and the proton. The effective Lagrangian Eq.(\ref{label1.9})
is local, as we have neglected the squared momentum transfer $- q^2$
with respect to the squared pion mass, $- q^2 \ll M^2_{\pi}$. Since in
the reaction n + p $\to$ D + $\gamma$ the np pair couples to the
deuteron and the photon in the ${^1}{\rm S}_0$ state, we should
rearrange the operators of the neutron and the proton interpolating
fields in the effective interaction Eq.(\ref{label1.9}) by such a way
to introduce the products of the np operators creating the np states
with definite total spins.  This rearrangement can be carried out by
means of the Fierz transformation that gives
\begin{eqnarray}\label{label1.10}
\gamma^5\otimes \gamma^5 &=& \frac{1}{4}\,C\otimes C +
\frac{1}{4}\,\gamma^5C\otimes C\gamma^5 +
\frac{1}{4}\,\gamma^{\mu}C\otimes C\gamma_{\mu} +
\frac{1}{4}\,\gamma^{\mu} \gamma^5C\otimes C\gamma_{\mu} \gamma^5\nonumber\\
&&+ \frac{1}{8}\,\sigma^{\mu\nu}C\otimes C\sigma_{\mu\nu},
\end{eqnarray}
where $\sigma^{\mu\nu} =(\gamma^{\mu}\gamma^{\nu} -
 \gamma^{\nu}\gamma^{\mu})/2$. By virtue of Eq.(\ref{label1.10}) we
 recast the effective Lagrangian Eq.(\ref{label1.9}) into the form
\begin{eqnarray}\label{label1.11}
{\cal L}^{\rm np \to np}_{\rm one-pion}(x)&=&  \frac{g^2_{\rm \pi
NN}}{4M^2_{\pi}}\,\{[\bar{p}(x)\gamma^5 n^c(x)][\bar{n^c}(x) \gamma^5
p(x)] + [\bar{p}(x)\gamma^{\mu}\gamma^5 n^c(x)][\bar{n^c}(x)
\gamma_{\mu}\gamma^5 p(x)]\nonumber\\
&&\quad\quad + [\bar{p}(x) n^c(x)][\bar{n^c}(x)
p(x)] + 3\,[\bar{p}(x)\gamma^{\mu}n^c(x)][\bar{n^c}(x)
\gamma_{\mu}p(x)]\nonumber\\
&&\quad\quad + \frac{3}{2}\,[\bar{p}(x)\sigma^{\mu\nu}
n^c(x)][\bar{n^c}(x)\sigma_{\mu\nu}  p(x)]\}.
\end{eqnarray}
Here $\bar{n^c}(x) = n^T(x)C$ and $n^c(x) = C\bar{n}^T(x)$, where $C$
is a charge conjugation matrix and $T$ is a transposition.  The first
two terms in the effective Lagrangian Eq.(\ref{label1.11}) describe
the strong low--energy ${^1}{\rm S}_0 \to {^1}{\rm S}_0$ transition of
the np pair in the ${^1}{\rm S}_0$ state. Thereby, the effective
Lagrangian providing the ${^1}{\rm S}_0 \to {^1}{\rm S}_0$ transition
of the np pair and caused by the one--pion exchange we would use in
the form
\begin{eqnarray}\label{label1.12}
\hspace{-0.3in}&&{\cal L}^{\rm np \to np}_{\rm one-pion}(x) =\nonumber\\
\hspace{-0.3in}&&= \frac{g^2_{\rm \pi
NN}}{4M^2_{\pi}}\,\{[\bar{p}(x)\gamma^5 n^c(x)][\bar{n^c}(x) \gamma^5
p(x)] + [\bar{p}(x)\gamma^{\mu}\gamma^5 n^c(x)][\bar{n^c}(x)
\gamma_{\mu}\gamma^5 p(x)]\}.
\end{eqnarray}
The phenomenological part of the effective Lagrangian responsible for
the strong low--energy ${^1}{\rm S}_0 \to {^1}{\rm S}_0$ transition of
the np pair in the ${^1}{\rm S}_0$ state we would choose by following
the EFT ideology [7--12] as well and write
\begin{eqnarray}\label{label1.13}
\hspace{-0.3in}&&{\cal L}^{\rm np \to np}_{\rm ph.}(x) =\nonumber\\
\hspace{-0.3in}&&= -\frac{2\pi a_{\rm
np}}{M_{\rm N}}\,\{[\bar{p}(x)\gamma^5 n^c(x)][\bar{n^c}(x) \gamma^5
p(x)] + [\bar{p}(x)\gamma^{\mu}\gamma^5 n^c(x)][\bar{n^c}(x)
\gamma_{\mu}\gamma^5 p(x)]\},
\end{eqnarray}
where $a_{\rm np} =( - 23.75\pm 0.01)\,{\rm fm}$ is the S--wave
scattering length of the elastic np scattering in the ${^1}{\rm
S}_0$ state [17].

The appearance of the S--wave scattering length for the definition of
the phenomenological coupling constant of the strong low--energy
transition n + p $\to$ n + p, or differently low--energy elastic np
scattering is rather natural in the EFT. Indeed, in the EFT pions are
treated perturbatively and at leading order in the pion--exchange
approximation for the description of the S--wave scattering length of
low--energy elastic np scattering in the ${^1}{\rm S}_0$ state
Weinberg has introduced a phenomenological four--nucleon interaction
with a coupling constant [11]
\begin{eqnarray}\label{label1.14}
C_0 = -\frac{4\pi a_{\rm np}}{M_{\rm N}}.
\end{eqnarray}
As has been stated in the EFT this constant is a collective
contribution coming from the integration over all meson degrees of
freedom heavier than the pionic ones. Since in our approach we
separate the pionic degrees of freedom from that with masses heavier
than the pion as well, the integration over these heavy meson degrees
of freedom should have the same form as it is postulated in the
EFT. We have only halved the coupling constant $C_0$ defined by
Eq.(\ref{label1.14}) in order to distribute symmetrically
phenomenological contributions between couplings $[\bar{p}(x)\gamma^5
n^c(x)][\bar{n^c}(x) \gamma^5 p(x)]$ and
$[\bar{p}(x)\gamma^{\mu}\gamma^5 n^c(x)][\bar{n^c}(x)
\gamma_{\mu}\gamma^5 p(x)]$. 

The total effective Lagrangian describing the strong low--energy
transition n + p $\to$ n + p of the np pair in the ${^1}{\rm S}_0$
state is then determined by
\begin{eqnarray}\label{label1.15}
\hspace{-0.3in}&&{\cal L}^{\rm np \to np}_{\rm eff}(x) ={\cal L}^{\rm
np \to np}_{\rm one-pion}(x) + {\cal L}^{\rm np \to np}_{\rm ph.}(x) =
\nonumber\\
\hspace{-0.3in}&&= C_{\rm
NN}\,\{[\bar{p}(x)\gamma^5 n^c(x)][\bar{n^c}(x) \gamma^5 p(x)] +
[\bar{p}(x)\gamma^{\mu}\gamma^5 n^c(x)][\bar{n^c}(x)
\gamma_{\mu}\gamma^5 p(x)]\},
\end{eqnarray}
where the effective coupling constant $C_{\rm NN}$ of the strong
low--energy transition n + p $\to$ n + p is equal to
\begin{eqnarray}\label{label1.16}
C_{\rm NN} = \frac{g^2_{\rm \pi NN}}{4M^2_{\pi}} - \frac{2\pi a_{\rm
np}}{M_{\rm N}} = 3.27\times 10^{-3}\,{\rm MeV}^{-2}.
\end{eqnarray}
Note that the contribution of the phenomenological part to
the effective coupling constant $C_{\rm NN}$ makes up less than
33$\%$. 

Since nuclear forces are isotopically invariant [26], the effective
Lagrangian of the strong low--energy N + N $\to$ N + N transition of
the NN pair in the ${^1}{\rm S}_0$ state can be defined as follows
\begin{eqnarray}\label{label1.17}
\hspace{-0.3in}&&{\cal L}^{\rm NN \to NN}_{\rm eff}(x) = C_{\rm
NN}\,
\nonumber\\
\hspace{-0.3in}&&\{[\bar{p}(x)\gamma^5 n^c(x)][\bar{n^c}(x) \gamma^5
p(x)] + [\bar{p}(x)\gamma^{\mu}\gamma^5 n^c(x)][\bar{n^c}(x)
\gamma_{\mu}\gamma^5 p(x)]\nonumber\\
\hspace{-0.3in}&&+ \frac{1}{2}[\bar{n}(x)\gamma^5 n^c(x)][\bar{n^c}(x) 
\gamma^5
n(x)] + [\bar{n}(x)\gamma^{\mu}\gamma^5 n^c(x)][\bar{n^c}(x)
\gamma_{\mu}\gamma^5 n(x)]\nonumber\\
\hspace{-0.3in}&&+ \frac{1}{2}[\bar{p}(x)\gamma^5 p^c(x)][\bar{p^c}(x)
\gamma^5 p(x)] + [\bar{p}(x)\gamma^{\mu}\gamma^5 p^c(x)][\bar{p^c}(x)
\gamma_{\mu}\gamma^5 p(x)]\}.
\end{eqnarray}
We would like to emphasize the effective Lagrangian
Eq.(\ref{label1.16}) describing strong low--energy N + N $\to$ N + N
transitions is obtained in complete agreement with the EFT ideology.

In the low--energy limit the effective local four--nucleon interaction
Eq.\,(\ref{label1.17}) vanishes due to the reduction
\begin{eqnarray}\label{label1.18}
[\bar{N}(x)\gamma_{\mu}\gamma^5 N^c
(x)][\bar{N^c}(x)\gamma^{\mu}\gamma^5 N(x)] \to - [\bar{N}(x) \gamma^5
N^c(x)][\bar{N^c}(x) \gamma^5 N(x)],
\end{eqnarray}
where $N(x)$ is an operator of the neutron or the proton interpolating
field. Such a vanishing of the one--pion exchange contribution to the
NN potential is well--known in the EFT approach [11,12] and the PMA
[26]. In power counting [11,12] the interaction induced by the
one--pion exchange is of order $O(p^2)$, where $p$ is a relative
momentum of the NN system. The former is due to the Dirac matrix
$\gamma^5$ which leads to the interaction between small components of
Dirac bispinors of nucleon wave functions.

In the one--nucleon loop exchange approach the contributions of the
interactions $[\bar{N}(x)\gamma_{\mu}\gamma^5 N^c
(x)][\bar{N^c}(x)\gamma^{\mu}\gamma^5 N(x)]$ and $ [\bar{N}(x)
\gamma^5 N^c(x)][\bar{N^c}(x) \gamma^5 N(x)]$ to the amplitudes of
nuclear reactions are different and do not cancel each other in the
low--energy limit due to the dominance of nucleon--loop anomalies
[1]. This provides the interaction between large components of Dirac
bispinors of nucleon wave functions that distinguishes the NNJL model
from the EFT.

\noindent{\bf Low--energy weak nuclear reactions with the
deuteron}. The weak nuclear reaction p + p $\to$ D + e$^+$ + $\nu_{\rm
e}$, the solar proton burning, plays an important role in Astrophysics
[3,27]. It gives start for the p--p chain of nucleosynthesis in the
Sun and main--sequence stars [3,27]. In the Standard Solar Model
(SSM) [28] the total (or bolometric) luminosity of the Sun $L_{\odot}
= (3.846\pm 0.008)\times 10^{26}\,{\rm W}$ is normalized to the
astrophysical factor $S_{\rm pp}(0)$ for the solar proton burning. The
recommended value $S_{\rm pp}(0) = 4.00\times 10^{-25}\,{\rm MeV b}$
[29] has been found by averaging over the results obtained in the
Potential model approach (PMA) [30,31] and the Effective Field Theory
(EFT) approach [32,33]. As has been shown recently in Ref.[34] {\it
the inverse and forward helioseismic approach} confirm the recommended
value of $S_{\rm pp}(0)$ within experimental errors on the
helioseismic data and solar neutrino fluxes.

In this paper we apply the NNJL model to the description of
low--energy nuclear forces for weak nuclear reactions with the
deuteron of astrophysical interest: 1) the solar proton burning p + p
$\to$ D + e$^+$ + $\nu_{\rm e}$, 2) the pep--process p + e$^-$ + p
$\to$ D + $\nu_{\rm e}$ and 3) the reactions of neutrino and
anti--neutrino disintegration of the deuteron caused by charged
$\nu_{\rm e}$ + D $\to$ e$^-$ + p + p, $\bar{\nu}_{\rm e}$ + D $\to$
e$^+$ + n + n and neutral $\nu_{\rm e}(\bar{\nu}_{\rm e})$ + D $\to$
$\nu_{\rm e}(\bar{\nu}_{\rm e})$ + n + p weak currents. The reactions
$\nu_{\rm e}$ + D $\to$ e$^-$ + p + p and $\nu_{\rm e}$ + D $\to$
$\nu_{\rm e}$ + n + p caused by charged and neutral weak currents,
respectively, and induced by solar neutrinos are planned to be
measured for solar neutrino experiments at Sudbury Neutrino
Observatory (SNO) [35]. In turn, the cross sections for the reactions
$\bar{\nu}_{\rm e}$ + D $\to$ e$^+$ + n + n and $\bar{\nu}_{\rm e}$ +
D $\to$ $\bar{\nu}_{\rm e}$ + n + p caused by charged and neutral weak
currents, respectively, and induced by reactor anti--neutrinos have
been recently measured by the Reines's experimental group [36]. In the
sense of charge independence of weak interaction strength the
observation of the reaction $\bar{\nu}_{\rm e}$ + D $\to$ e$^+$ + n +
n is equivalent to the observation of the reaction of the solar proton
burning p + p $\to$ D + e$^+$ + $\nu_{\rm e}$ in the terrestrial
laboratories.

The paper is organized as follows. In Section\,2 we evaluate the
amplitude of the M1--capture n + p $\to$ D + $\gamma$ in the
tree--meson approximation. The contribution of low--energy elastic np
scattering to the amplitude of the process n + p $\to$ D + $\gamma$ is
obtained in agreement with low--energy nuclear phenomenology. In
Section\,3 we evaluate contributions of chiral one--meson loop
corrections to the amplitude of the M1--capture in Chiral perturbation
theory at the quark level (CHPT)$_q$ developed within the ENJL model
with a linear realization of chiral $U(3)\times U(3)$ symmetry. In
Section\,4 we include the contribution of the $\Delta(1232)$ resonance
and analyse the total cross section for the neutron--proton radiative
capture for thermal neutrons and compare it with experimental data. In
Section\,5 we treat the reaction of the photomagnetic disintegration
of the deuteron $\gamma$ + D $\to$ n + p related to the
neutron--proton radiative capture n + p $\to$ D + $\gamma$ via
time--reversal invariance and analyse the energy dependence of the
cross section at energies far from threshold. In Sect.\,6 we evaluate
the amplitude of the solar proton burning. We show that the
contribution of low--energy elastic pp scattering in the ${^1}{\rm
S}_0$ state with the Coulomb repulsion is described in agreement with
low--energy nuclear phenomenology in terms of the S--wave scattering
length and the effective range. In Sect.\,7 we evaluate the
astrophysical factor for the solar proton burning and obtain the value
$S_{\rm pp}(0) = 4.08\times 10^{-25}\,{\rm MeV\, b}$ agreeing good
with the recommended one $S_{\rm pp}(0) = 4.00\times 10^{-25}\,{\rm
MeV\, b}$.  In Sect.\,8 we evaluate the cross section for the neutrino
disintegration of the deuteron $\nu_{\rm e}$ + D $\to$ e$^-$ + p + p
with respect to $S_{\rm pp}(0)$. In Sect.\,9 we adduce the evaluation
of the astrophysical factor $S_{\rm pep}(0)$ for the pep--process
relative to $S_{\rm pp}(0)$. In Sects.\,10 and 11 we evaluate the
cross sections for the anti--neutrino disintegration of the deuteron
$\bar{\nu}_{\rm e}$ + D $\to$ e$^+$ + n + n and $\bar{\nu}_{\rm e}$ +
D $\to$ $\bar{\nu}_{\rm e}$ + n + p and average them over the
anti--neutrino energy spectrum. The average values of the cross
sections agree well with experimental data [36]. The cross sections
for the weak nuclear reactions of astrophysical interest are
calculated at zero contribution of the nucleon tensor current
[1]. This makes the description of low--energy nuclear forces within
the NNJL model compatible with the predictions of the SSM [27,28]. In
more detail this is discussed the Conclusion and Appendix.  In
the Conclusion we discuss the obtained results.  In Appendix we evaluate the
effective Lagrangian ${\cal L}^{\rm pp \to De^+\nu_{\rm e}}_{\rm
eff}(x)$ of the low--energy weak transition p + p $\to$ D + e$^+$ +
$\nu_{\rm e}$. The contribution of the nucleon tensor current [1] to
the effective Lagrangians of low--energy weak nuclear transitions of
astrophysical interest like p + p $\to$ D + e$^+$ + $\nu_{\rm e}$ and
so on is evaluated and discussed.

\section{The M1--capture in the tree--meson approximation}
\setcounter{equation}{0}

\hspace{0.2in} Since the NNJL model realizes a Lagrange approach to
the description of low--energy nuclear forces [1], first thing what we
have to do is to evaluate the effective Lagrangian ${\cal L}^{\rm np
\to D\gamma}_{\rm eff}(x)$ of the transition n + p $\to$ D +
$\gamma$. In the tree--meson approximation the effective Lagrangian
${\cal L}^{\rm np \to D\gamma}_{\rm eff}(x)$ is defined by
one--nucleon loop diagrams depicted in Figs.\,1 and 2. The evaluation
of these diagrams at leading order in the large $N_C$ expansion yields
\begin{eqnarray}\label{label2.1}
{\cal L}^{\rm np \to D\gamma}_{\rm eff}(x) &=&(\mu_{\rm p} -
\mu_{\rm n})\,\frac{e}{2M_{\rm N}}\,\frac{g_{\rm V}}{4\pi^2}\,C_{\rm
NN}\,D^{\dagger}_{\mu\nu}(x){^*\!F^{\mu\nu}(x)}[\bar{p^c}(x)\gamma^5
n(x)]\nonumber\\
&+& i\,(\mu_{\rm p} - \mu_{\rm n})\,\frac{e}{2M_{\rm N}}\,\frac{g_{\rm
V}}{4\pi^2}\,C_{\rm NN}\,M_{\rm
N}\,D^{\dagger}_{\mu}(x){^*\!F^{\mu\nu}(x)}[\bar{p^c}(x)\gamma_{\nu}\gamma^5
n(x)],
\end{eqnarray}
where ${^*\!F^{\mu\nu}(x)}=
\frac{1}{2}\,\varepsilon^{\mu\nu\alpha\beta}F_{\alpha\beta}(x)$,
$\mu_{\rm p} = 2.793$ and $\mu_{\rm n} = -\,1.913$ are the magnetic
dipole moments of the proton and the neutron measured in nuclear
magnetons, $D_{\mu}(x)$ is the operator of the interpolating field of
the deuteron and $D_{\mu\nu}(x) = \partial_{\mu} D_{\nu}(x) -
\partial_{\nu} D_{\mu}(x)$, $g_{\rm V}$ is a phenomenological coupling
constant of the NNJL model related to the electric quadrupole moment
of the deuteron $Q_{\rm D} = 0.286\,{\rm fm}^2$ [1]: $g^2_{\rm V} =
2\pi^2 Q_{\rm D}M^2_{\rm N}$.

The matrix element of the transition n + p $\to$ D + $\gamma$ we
define by a usual way 
\begin{eqnarray}\label{label2.2}
&&\int d^4x\langle D(k_{\rm D})\gamma(k)|{\cal L}^{\rm np\to
D\gamma}_{\rm eff}(x)|n(p_1)p(p_2)\rangle = \nonumber\\
&&= (2\pi)^4\delta^{(4)}( k_{\rm D} + k - p_1 - p_2)\,\frac{{\cal M}({\rm
n} + {\rm p}\to {\rm D} + \gamma)}{\displaystyle \sqrt{2E_1V\,2E_2V\,
2E_{\rm D}V\,2\omega V}},
\end{eqnarray}
where $E_i\,(i =1,2,{\rm D})$ and $\omega$ are the energies of the
neutron, the proton, the deuteron and the photon, $V$ is the
normalization volume. 

The wave functions of the initial $|n(p_1) p(p_2)\rangle$ and final
$\langle D(k_{\rm D})\gamma(k)|$ state we take in the usual form
\begin{eqnarray}\label{label2.3}
|n(p_1) p(p_2)\rangle &=& a^{\dagger}_{\rm
 n}(\vec{p}_1,\sigma_1)\,a^{\dagger}_{\rm
 p}(\vec{p}_2,\sigma_2)|0\rangle,\nonumber\\
\langle D(k_{\rm D})\gamma(k)| &=& \langle 0|a_{\rm D}(\vec{k}_{\rm D},\lambda_{\rm D})\,a(\vec{k},\lambda),
\end{eqnarray}
where $a^{\dagger}_{\rm n}(\vec{p}_1,\sigma_1)$ and $a^{\dagger}_{\rm
p}(\vec{p}_2,\sigma_2)$ are the operators of creation of the
neutron and the proton, and $a_{\rm D}(\vec{k}_{\rm D},\lambda_{\rm
D})$ and $a(\vec{k},\lambda)$ are the operators of annihilation of
the deuteron and the photon.

The matrix element of the transition n + p $\to$ D + $\gamma$ reads
\begin{eqnarray}\label{label2.4}
{\cal M}({\rm n + p\to D + \gamma})&=&(\mu_{\rm p} - \mu_{\rm n})
\,\frac{e}{2 M_{\rm N}}\,\frac{g_{\rm V}}{4\pi^2}\,C_{\rm
NN}\,\varepsilon^{\alpha\beta\mu\nu} k_{\alpha}\,
e^*_{\beta}(k,\lambda)\, e^*_{\mu}(k_{\rm D},\lambda_{\rm
D})\nonumber\\ &&\times [\bar{u^c}(p_2)(2\, k_{\rm D \nu} - M_{\rm N}\gamma_{\nu})\gamma^5
u(p_1)],
\end{eqnarray}
where $e^*_{\beta}(k,\lambda)$ and $e^*_{\mu}(k_{\rm D},\lambda_{\rm
D})$ are the 4--vectors of the polarization of the photon and the
deuteron, then $\bar{u^c}(p_2)$ and $u(p_1)$ are the bispinorial wave
functions of the proton and the neutron, respectively, normalized by
$\bar{u^c}(p_2)u^c(p_2) = -\,2\,M_{\rm N}$ and $\bar{u}(p_1)u(p_1) =
2\,M_{\rm N}$.

In the low--energy limit when
\begin{eqnarray}\label{label2.5}
[\bar{u^c}(p_2)\gamma_{\nu}\gamma^5 u(p_1)] \to - g_{\nu
0}\,[\bar{u^c}(p_2)\gamma^5 u(p_1)]
\end{eqnarray}
and $k_{\rm D \nu} \to g_{\nu 0}\,2 M_{\rm N}$ the matrix element
Eq.(\ref{label2.4}) acquires the form
\begin{eqnarray}\label{label2.6}
{\cal M}({\rm n + p\to D + \gamma})&=& e\,(\mu_{\rm p} - \mu_{\rm n})
\,\frac{5 g_{\rm V}}{8\pi^2}\,C_{\rm NN}\,(\vec{k}\times
\vec{e}^{\,*}(\vec{k},\lambda))\cdot \vec{e}^{\,*}(\vec{k}_{\rm
D},\lambda_{\rm D})\nonumber\\
&&\times [\bar{u^c}(p_2)\gamma^5 u(p_1)].
\end{eqnarray}
The evaluation of the matrix element Eq.(\ref{label2.6}), the
effective vertex of the n + p $\to$ D + $\gamma$ transition, we have
carried out with the wave functions of the neutron and the proton in
the form of the plane waves. However, a physical ${^1}{\rm S}_0$ state
of the np pair is defined by low--energy nuclear forces. For the
description of the contribution of low--energy nuclear forces to the
physical ${^1}{\rm S}_0$ state of the np pair coupled to the deuteron
and the photon we suggest to sum an infinite series of one--nucleon
bubbles with vertices defined by the effective Lagrangian ${\cal
L}^{\rm np \to np}_{\rm eff}(x)$ Eq.(\ref{label1.15}). The result of
the summation can be represented in the following form
\begin{eqnarray}\label{label2.7}
&&{\cal M}({\rm n + p\to D + \gamma}) = e\,(\mu_{\rm p} - \mu_{\rm n})
\,\frac{5 g_{\rm V}}{8\pi^2}\,C_{\rm NN}\,(\vec{k}\times
\vec{e}^{\,*}(\vec{k},\lambda))\cdot \vec{e}^{\,*}(\vec{k}_{\rm D},\lambda_{\rm D})
\,[\bar{u^c}(p_2)\gamma^5 u(p_1)]\nonumber\\ 
&&\times
\frac{1}{\displaystyle 1 + \frac{C_{\rm NN}}{16\pi^2}\int
\frac{d^4q}{\pi^2i}\,{\rm tr}\Bigg\{\gamma^5 \frac{1}{M_{\rm N} -
\hat{q} - \hat{P} - \hat{Q}}\gamma^5 \frac{1}{M_{\rm N} - \hat{q} -
\hat{Q}}\Bigg\}},
\end{eqnarray}
where $P = p_1 + p_2 = (2\sqrt{p^2 + M^2_{\rm N}}, \vec{0}\,)$ is the
4--momentum of the np pair in the center of mass frame. Then, $Q =a\,P
+ b\,K = a\,(p_1 + p_2) + b\,(p_1 - p_2)$ is an arbitrary shift of
virtual momentum with arbitrary parameters $a$ and $b$, and in the
center of mass frame $K = p_1 - p_2 = (0,2\,\vec{p}\,)$. The explicit
dependence of the momentum integral on $Q$ can be evaluated by means
of the Gertsein--Jackiw procedure [37] (see also Ref.\,[1]). It is
given by
\begin{eqnarray}\label{label2.8}
\hspace{-0.5in} &&\int \frac{d^4q}{\pi^2i}\,{\rm tr}\Bigg\{\gamma^5
\frac{1}{M_{\rm N} - \hat{q} - \hat{P} - \hat{Q}}\gamma^5
\frac{1}{M_{\rm N} - \hat{q} - \hat{Q}}\Bigg\} =\nonumber\\
\hspace{-0.5in} &&=\int \frac{d^4q}{\pi^2i}\,{\rm tr}\Bigg\{\gamma^5
\frac{1}{M_{\rm N} - \hat{q} - \hat{P}}\gamma^5 \frac{1}{M_{\rm N} -
\hat{q}}\Bigg\} - 2\, a\,(a + 1)\,P^2 - 2\,b^2\,K^2.
\end{eqnarray}
For the evaluation of the momentum integral over $q$ we would keep
only the leading order contributions in the $1/M_{\rm N}$ expansion
caused by the large $N_C$ expansion [1]. This yields
\begin{eqnarray}\label{label2.9}
\hspace{-0.5in} &&\int \frac{d^4q}{\pi^2i}\,{\rm tr}\Bigg\{\gamma^5
\frac{1}{M_{\rm N} - \hat{q} - \hat{P} - \hat{Q}}\gamma^5
\frac{1}{M_{\rm N} - \hat{q} - \hat{Q}}\Bigg\} =\nonumber\\
\hspace{-0.5in} &&=- 8\, a\,(a + 1)\,M^2_{\rm N} + 8\,(b^2 - a\,(a +
1))\,p^2 - i\,8\pi\,M_{\rm N}\,p.
\end{eqnarray}
The expression Eq.(\ref{label2.7}) we reduce to the form
\begin{eqnarray}\label{label2.10}
{\cal M}({\rm n + p \to D + \gamma})&=& e\,(\mu_{\rm p} - \mu_{\rm n})
\,\frac{5 g_{\rm V}}{8\pi^2}\,C_{\rm NN}\,(\vec{k}\times
\vec{e}^{\,*}(\vec{k},\lambda))\cdot 
\vec{e}^{\,*}(\vec{k}_{\rm D},\lambda_{\rm D})
\,[\bar{u^c}(p_2)\gamma^5 u(p_1)]\nonumber\\ 
&&\times\, \frac{{\cal Z}}{\displaystyle 1 -
\frac{1}{2} r_{\rm np} a_{\rm np} p^2 + i a_{\rm np} p}.
\end{eqnarray}
Here we have denoted 
\begin{eqnarray}\label{label2.11}
a_{\rm np} &=& - \frac{C_{\rm NN}M_{\rm N}}{2\pi}\,{\cal Z}\quad,\quad
r_{\rm np} = (b^2 - a\,(a + 1))\,\frac{2}{\pi}\,\frac{1}{M_{\rm N}},
\nonumber\\ \frac{1}{{\cal Z}}&=& 1 - \frac{a(a+1)}{2\pi^2}\,C_{\rm 
NN}\,M^2_{\rm N},
\end{eqnarray}
where $r_{\rm np} = 2.75\pm 0.05\,{\rm fm}$ is the effective range of
low--energy elastic np scattering [17].

Renormalizing the wave functions of nucleons $\sqrt{{\cal Z}}\,u(p_1)
\to u(p_1)$ and $\sqrt{{\cal Z}}\,u(p_2) \to u(p_2)$ we arrive at the
expression
\begin{eqnarray}\label{label2.12}
{\cal M}({\rm n + p \to D + \gamma})&=& e\,(\mu_{\rm p} - \mu_{\rm n})
\,\frac{5 g_{\rm V}}{8\pi^2}\,C_{\rm NN}\,(\vec{k}\times
\vec{e}^{\,*}(\vec{k},\lambda))\cdot 
\vec{e}^{\,*}(\vec{k}_{\rm D},\lambda_{\rm D})
\,[\bar{u^c}(p_2)\gamma^5 u(p_1)]\nonumber\\
&&\times\frac{1}{\displaystyle 1 -
\frac{1}{2} r_{\rm np} a_{\rm np} p^2 + i\,a_{\rm np} p},
\end{eqnarray}
where the factor $1/(1 - \frac{1}{2} r_{\rm np} a_{\rm np} p^2
+i\,a_{\rm np} p)$ describes the contribution of low--energy nuclear
forces to the physical ${^1}{\rm S}_0$ state of the np pair coupled to
the deuteron and the photon. It has the form of the amplitude of
low--energy elastic np scattering in the ${^1}{\rm S}_0$ state in
complete agreement with low--energy nuclear phenomenology [26]. By using
the relation expressing the phase shift $\delta_{\rm np}(p)$ of
low--energy elastic np scattering in terms of the S--wave scattering
length $a_{\rm np}$ and the effective range $r_{\rm np}$ [26]
\begin{eqnarray}\label{label2.13}
{\rm ctg}\,\delta_{\rm np}(p) = -\,\frac{1}{a_{\rm np}} + \frac{1}{2}\,r_{\rm
np}\,p^2
\end{eqnarray}
we can recast the factor $1/(1 - \frac{1}{2} r_{\rm np} a_{\rm np} p^2
+i\,a_{\rm np} p)$ into the form
\begin{eqnarray}\label{label2.14}
\frac{1}{\displaystyle 1 - \frac{1}{2} r_{\rm np} a_{\rm np} p^2 +
i\,a_{\rm np} p} = e^{\textstyle i\delta_{\rm np}(p)}\frac{\sin
\delta_{\rm np}(p)}{-a_{\rm np}\,p}.
\end{eqnarray}
In terms of the phase shift $\delta_{\rm np}(p)$ the expression
Eq.(\ref{label2.12}) reads
\begin{eqnarray}\label{label2.15}
{\cal M}({\rm n + p \to D + \gamma})&=& e\,(\mu_{\rm p} - \mu_{\rm n})
\,\frac{5 g_{\rm V}}{8\pi^2}\,C_{\rm NN}\,(\vec{k}\times
\vec{e}^{\,*}(\vec{k},\lambda))\cdot \vec{e}^{\,*}(\vec{k}_{\rm D},\lambda_{\rm D})\nonumber\\
&&\times\,[\bar{u^c}(p_2)\gamma^5 u(p_1)]\,e^{\textstyle i\delta_{\rm
np}(p)}\frac{\sin \delta_{\rm np}(p)}{-a_{\rm np}\,p}.
\end{eqnarray}
In the NNJL model low--energy nuclear forces between the neutron and
the proton in the physical deuteron state are described by the
one--nucleon loop exchanges in terms of the phenomenological coupling
constant $g_{\rm V}$ which is defined by the electric quadrupole
moment of the deuteron $Q_{\rm D}$, $g^2_{\rm V} = 2\pi^2 Q_{\rm
D}M^2_{\rm N}$ [1].  The electric quadrupole moment of the deuteron is
caused by nuclear tensor forces [26]. Therefore, the relation
$g^2_{\rm V} = 2\pi^2 Q_{\rm D}M^2_{\rm N}$ confirms at the quantum
field theoretic level the fact pointed out by Blatt and Weisskopf that
{\it the existence of a bound triplet state of the neutron--proton
system would be entirely due to the tensor force} [38]. Thus, in NNJL
model through the phenomenological coupling constant $g_{\rm V}$
tensor forces govern the existence of the deuteron as a bound
neutron--proton triplet spin state and strength of low--energy
interactions of the deuteron with nucleons and other particles in
terms of the one--nucleon loop exchanges. The evaluation of
one--nucleon loop diagrams at leading order in the $1/M_{\rm N}$
expansion, or in the large $N_C$ expansion, reduces a momentum
dependence of nucleon diagrams to the trivial form accounting for only
the Lorentz covariant properties of the interaction. In this approach
the deuteron looks like a point--like particle. Such a representation
is enough for the evaluation of effective Lagrangians of different
low--energy nuclear transitions with the deuteron in an initial or a
final state describing effective vertices of low--energy nuclear
transitions defined at their thresholds. However, for the evaluation
of amplitudes of low--energy nuclear reactions for energies far from
threshold one needs to take into account a spatial smearing of the
physical deuteron caused by a finite radius $r_{\rm D} =
1/\sqrt{\varepsilon_{\rm D}M_{\rm N}}= 4.319\,{\rm MeV}$ [17]
determined by the binding energy of the deuteron $\varepsilon_{\rm D}
= 2.225\,{\rm MeV}$. The spatial smearing of the physical deuteron we
introduce phenomenologically in the form
\begin{eqnarray}\label{label2.16}
F_{\rm D}(p) = \frac{1}{1 + r^2_{\rm D}p^2},
\end{eqnarray}
that is nothing more than the momentum representation of the
approximate ${^3}{\rm S}_1$ wave state of the deuteron
[26].

Substituting Eq.(\ref{label2.16}) in Eq.(\ref{label2.15}) we obtain
the amplitude of the M1--capture calculated in the NNJL model:
\begin{eqnarray}\label{label2.17}
{\cal M}({\rm n + p \to D + \gamma})&=& e\,(\mu_{\rm p} - \mu_{\rm n})
\,\frac{5 g_{\rm V}}{8\pi^2}\,C_{\rm NN}\,(\vec{k}\times
\vec{e}^{\,*}(\vec{k},\lambda))\cdot \vec{e}^{\,*}(\vec{k}_{\rm
D},\lambda_{\rm D})\nonumber\\ &&\times\,[\bar{u^c}(p_2)\gamma^5
u(p_1)]\,e^{\textstyle i\delta_{\rm np}(p)}\frac{\sin \delta_{\rm
np}(p)}{-a_{\rm np}\,p}\,\frac{1}{1 + r^2_{\rm D}p^2}.
\end{eqnarray}
For thermal neutrons the kinetic energy of the relative movement of
the np pair is of order $T_{\rm np} \sim 10^{-8}\,{\rm MeV}$ [9]. This
yields the relative momentum of the np pair to be smaller compared
with the binding energy of the deuteron, $p \sim 3\times 10^{-3}\,{\rm
MeV} \ll \varepsilon_{\rm D} = 2.225\,{\rm MeV}$. Therefore, for
thermal neutrons without loss of generality we can calculate the
amplitude of the M1--capture setting $p=0$:
\begin{eqnarray}\label{label2.18}
{\cal M}({\rm n + p \to D + \gamma}) &=& e\,(\mu_{\rm p} - \mu_{\rm n})
\,\frac{5 g_{\rm V}}{8\pi^2}\,C_{\rm NN}\,(\vec{k}\times
\vec{e}^{\,*}(\vec{k},\lambda))\cdot \vec{e}^{\,*}(\vec{k}_{\rm
D},\lambda_{\rm D})\nonumber\\
&&\times\,[\bar{u^c}(p_2)\gamma^5 u(p_1)].
\end{eqnarray}
Thus, we have obtained that the amplitude of the neutron--proton
radiative capture n + p $\to$ D + $\gamma$ for thermal neutrons
coincides with the matrix element of the effective Lagrangian ${\cal
L}^{\rm np \to D\gamma}_{\rm eff}(x)$ given by Eq.(\ref{label2.6}).

The cross section for the M1--capture for thermal neutrons calculated
in the tree--meson approximation is then defined by
\begin{eqnarray}\label{label2.19}
\sigma({\rm np \to D\gamma})(T_{\rm n}) = \frac{1}{v_{\rm n}}\,(\mu_{\rm p}-\mu_{\rm
n})^2\,\frac{25}{64}\,\frac{\alpha}{\pi^2}\,Q_{\rm D}\,C^2_{\rm 
NN}\,M_{\rm N}\,\varepsilon^3_{\rm D}\Bigg(1 + \frac{1}{2}\,\frac{T_{\rm n}}{\varepsilon_{\rm D}}\Bigg)^3 = 276\,{\rm m b}.
\end{eqnarray}
The theoretical value $\sigma({\rm np \to D\gamma})(T_{\rm n}) =
276\,{\rm m b}$ is about 17$\%$ smaller compared with the experimental
one $\sigma({\rm np \to D\gamma})_{\exp}(T_{\rm n})=(334.2\pm
0.5)\,{\rm m b}$. Thus, in the tree--meson approximation the NNJL
model predicts the cross section for the M1--capture for thermal
neutrons somewhat worse than the EFT, $\sigma({\rm np \to
D\gamma})(T_{\rm n}) = 297.2\,{\rm m b}$ [8]. In order to improve the
agreement with the experimental data we have to include chiral
one--meson loop corrections [7--9] and the contribution of the
$\Delta(1232)$ resonance [6,7].

\section{Chiral one--meson loop corrections to the amplitude 
of the M1--capture} \setcounter{equation}{0}

\hspace{0.2in} For the evaluation of chiral meson--loop corrections in
the NNJL we use (CHPT)$_q$ developed in Refs.[13] within the ENJL
model with a linear realization of chiral $U(3)\times U(3)$
symmetry. Below we consider the contributions of chiral one--meson
loop corrections induced by the virtual meson transitions $\pi \to a_1
\gamma$, $a_1 \to \pi\,\gamma$, $\pi \to (\omega, \rho) \gamma$,
$(\omega, \rho) \to \pi \gamma$, $\sigma \to (\omega, \rho)\gamma$ and
$(\omega, \rho) \to \sigma \gamma$, where $\sigma$ is a scalar partner
of pions under chiral $SU(2)\times SU(2)$ transformations in
(CHPT)$_q$ with a linear realization of chiral $U(3)\times U(3)$
symmetry [13].

The effective Lagrangians $\delta {\cal L}^{\rm pp\gamma}_{\rm
eff}(x)$ and $\delta {\cal L}^{\rm nn\gamma}_{\rm eff}(x)$, caused by
the virtual meson transitions $\pi \to a_1\,\gamma$, $a_1 \to
\pi\,\gamma$, $\pi \to (\omega, \rho) \gamma$, $(\omega, \rho) \to \pi
\gamma$, $\sigma \to (\omega, \rho)\gamma$ and $(\omega, \rho) \to
\sigma \gamma$, we evaluate at leading order in the large $N_C$
expansion [1]. The results of the evaluation contain divergent
contributions. In order to remove these divergences we apply the
renormalization procedure developed in (CHPT)$_q$ for the evaluation
of chiral meson--loop corrections (see {\it Ivanov} in
Refs. [13]). Since the renormalized expressions should vanish in the
chiral limit $M_{\pi} \to 0$ [13], only the virtual meson transitions
with intermediate $\pi$--meson give non--trivial contributions. The
contributions of the virtual meson transitions with the intermediate
$\sigma$--meson are found finite in the chiral limit and subtracted
according to the renormalization procedure [13]. Such a cancellation
of the $\sigma$--meson contributions in the one--meson loop
approximation agrees with Chiral perturbation theory using a
non--linear realization of chiral symmetry, where $\sigma$--meson--like
exchanges can appear only in two--meson loop approximation.  Then, the
sum of the contributions of the virtual meson transitions $\pi^- \to
\rho^- \gamma$, $\pi^0 \to \rho^0 \gamma$ and $\pi^0 \to \omega
\gamma$ to the effective coupling ${\rm nn\gamma}$ is equal to
zero. As a result the effective Lagrangians $\delta {\cal L}^{\rm
pp\gamma}_{\rm eff}(x)$ and $\delta {\cal L}^{\rm nn\gamma}_{\rm
eff}(x)$ are given by
\begin{eqnarray}\label{label3.1}
\delta {\cal L}^{\rm pp\gamma}_{\rm eff}(x)&=& \frac{ie}{4M_{\rm
N}}\Bigg[g_{\rm A}g_{\rm \pi
NN}\frac{\alpha_{\rho}}{16\pi^3}\,\frac{M_{\rm
N}}{F_{\pi}}M^2_{\pi}\,J_{\rm \pi a_1 N } + g_{\rm \pi
NN}\,\frac{N_C\alpha_{\rho}}{16\pi^3}\frac{M_{\rm N}}{F_{\pi}}
\,M^2_{\pi}\,J_{\rm \pi VN}\Bigg],\nonumber\\
&&\times\,
[\bar{p}(x)\sigma_{\mu\nu}p(x)]\,F^{\mu\nu}(x),\nonumber\\ \delta
{\cal L}^{\rm nn\gamma}_{\rm eff}(x)&=&\frac{ie}{4M_{\rm
N}}\Bigg[-\,g_{\rm A}g_{\rm \pi
NN}\frac{\alpha_{\rho}}{16\pi^3}\,\frac{M_{\rm
N}}{F_{\pi}}M^2_{\pi}\,J_{\rm \pi a_1 N }\Bigg]\, 
[\bar{n}(x)\sigma_{\mu\nu}n(x)]\,F^{\mu\nu}(x),
\end{eqnarray}
where $\alpha_{\rho} = g^2_{\rho}/4\pi = 2.91$ is the effective
coupling constant of the $\rho \to \pi \pi$ decay, $F_{\pi} =
92.4\,{\rm MeV}$ is the leptonic coupling constant of pions, and
$g_{\rm A} =1.267$ [39].  Then, $J_{\rm \pi a_1 N}$ and$J_{\rm \pi
VN}$ are the momentum integrals determined by 
\begin{eqnarray}\label{label3.2}
J_{\rm \pi a_1  N}&=&\int \frac{d^4p}{\pi^2}\,\frac{1}{(M^2_{\pi} +
p^2)(M^2_{a_1} + p^2)(M^2_{\rm N} + p^2)} =
0.017\,M^{-2}_{\pi},\nonumber\\ J_{\rm \pi VN} &=&\int
\frac{d^4p}{\pi^2}\,\frac{1}{(M^2_{\pi} + p^2)(M^2_{\rm V} +
p^2)(M^2_{\rm N} + p^2)}= 0.024\,M^{-2}_{\pi},
\end{eqnarray}
where $p$ is Euclidean 4--momentum, $M_{\rm V} = M_{\rho} = M_{\omega}
= 770\,{\rm MeV}$ [39] and $M_{a_1} =\sqrt{2}\,M_{\rho}$ [13].

At $N_C = 3$ the cross section for the M1--capture accounting for the
contribution of the effective interaction Eq.(\ref{label3.1}) amounts
to
\begin{eqnarray}\label{label3.3}
\hspace{-0.5in}&&\sigma({\rm np \to D\gamma})(T_{\rm n}) =
\frac{1}{v}\,(\mu_{\rm p}-\mu_{\rm
n})^2\,\frac{25}{64}\,\frac{\alpha}{\pi^2}\,\,Q_{\rm D}\,C^2_{\rm
NN}\,M_{\rm N}\,\varepsilon^3_{\rm D}\,\nonumber\\
\hspace{-0.5in}&&\times\,\Bigg[1 + \frac{g^2_{\rm \pi NN}}{\mu_{\rm p}
- \mu_{\rm
n}}\,\frac{M^2_{\pi}}{8\pi^2}\,\frac{\alpha_{\rho}}{\pi}\Bigg(J_{\rm
\pi a_1 N} + \frac{3}{2 g_{\rm A}}\,J_{\rm \pi VN}\Bigg)\Bigg]^2 =
287.2\,{\rm m b},
\end{eqnarray}
where we have used the relation $g_{\rm \pi NN} \simeq g_{\rm
A}\,M_{\rm N}/F_{\pi}$. The theoretical value of the cross section for
the neutron--proton radiative capture given by Eq.(\ref{label3.3})
differs from the experimental one by about 14$\%$. This discrepancy we
should describe by taking into account the contribution of the
$\Delta(1232)$ resonance.

\section{The $\Delta(1232)$ resonance}
\setcounter{equation}{0}

\hspace{0.2in} For the evaluation of the contribution of the
$\Delta(1232)$ resonance to the amplitude of the M1--capture in the
NNJL model we have to obtain the effective Lagrangian ${\cal L}^{\rm
np \to \Delta N}_{\rm eff}(x)$ describing the strong low--energy n + p
$\to$ $\Delta$ + N transition. For this aim we should use the
procedure having been applied to the evaluation of the effective
Lagrangian ${\cal L}^{\rm np \to np}_{\rm eff}(x)$ given by
Eq.(\ref{label1.15}). In Refs.[6,7] the evaluation of the contribution
of the $\Delta(1232)$ resonance in terms of exchange currents has been
carried out in the one--pion exchange approximation.  Thereby,
following Refs.\,[6,7] we suppose to evaluate the effective Lagrangian
${\cal L}^{\rm np \to \Delta N}_{\rm eff}(x)$ of the strong
low--energy n + p $\to$ $\Delta$ + N transition in the one--pion
exchange approximation. This gives
\begin{eqnarray}\label{label4.1}
\hspace{-0.5in}&&{\cal L}^{\rm np \to \Delta N}_{\rm eff}(x) = -
\frac{i}{\sqrt{6}}\frac{g_{\rm \pi N\Delta}}{M_{\rm N}}\frac{g_{\rm
\pi NN}}{4 M^2_{\pi}}\int d^4z\,\frac{\partial}{\partial
z_{\varphi}}\delta^{(4)}(z-x)\,\{[\bar{\Delta}^+_{\omega}(z)\,
{\Theta^{\omega}}_{\varphi}\, n^c(x)]\,\nonumber\\
\hspace{-0.5in}&& \times\,[\bar{n^c}(z)\gamma^5 p(x) +
\bar{n^c}(x)\gamma^5 p(z)]
-[\bar{\Delta}^0_{\omega}(z)\,{\Theta^{\omega}}_{\varphi}\,
p^c(x)]\,[\bar{n^c}(z)\gamma^5 p(x) +
\bar{n^c}(x)\gamma^5 p(z)]\nonumber\\
\hspace{-0.5in}&& + 1 \otimes \gamma^5 \to -\gamma_{\nu} \otimes
\gamma^{\nu}\gamma^5\},
\end{eqnarray}
where we have kept only the terms contributing to the transition of
the np pair in the ${^1}{\rm S}_0$ state into the $\Delta N$ state.
Using then the phenomenological Lagrangian
\begin{eqnarray}\label{label4.2}
{\cal L}_{\rm npD}(x) = -\,i\,g_{\rm V}[\bar{p^c}(x)\gamma^{\mu}n(x) -
\bar{n^c}(x)\gamma^{\mu}p(x)] D^{\dagger}_{\mu}(x)
\end{eqnarray}
the effective Lagrangian describing the contribution of the
$\Delta(1232)$ resonance to the low--energy transition n + p
$\to$ D + $\gamma$ is defined by
\begin{eqnarray}\label{label4.3}
\hspace{-0.5in}&&\int d^4x\,{\cal L}^{\rm np \to \Delta N \to
D\gamma}_{\rm eff}(x) = - \int d^4x_1 d^4x_2 d^4x_3\,\langle{\rm T}({\cal L}^{\rm np
\to \Delta N}_{\rm eff}(x_1){\cal L}_{\rm npD}(x_2){\cal L}_{\rm
\gamma N\Delta}(x_3))\rangle = \nonumber\\
\hspace{-0.5in}&&= - \frac{i}{6}\frac{eg_{\rm V}}{M^2_{\rm
N}}\frac{g_{\rm \pi N\Delta}}{g_{\rm \pi NN}}\frac{g_{\rm \gamma
N\Delta}}{g_{\rm \pi NN}}\frac{g^3_{\rm \pi NN}}{4M^2_{\pi}}\int d^4x_1 d^4x_2 d^4x_3\int d^4z\,\frac{\partial}{\partial
z_{\varphi}}\delta^{(4)}(z-x_1)\,\nonumber\\
\hspace{-0.5in}&&\times\,{\rm T}([\bar{p^c}(x_1)\gamma^5 n(z) +
\bar{p^c}(z)\gamma^5
n(x_1)]\,D^{\dagger}_{\mu}(x_2)
F^{\alpha\beta}(x_3))\nonumber\\
\hspace{-0.5in}&&\times\Big\{\langle0|{\rm
T}([\bar{\Delta}^+_{\omega}(z)\,
{\Theta^{\omega}}_{\varphi}\,n^c(x_1)][\bar{p^c}(x_2)\gamma^{\mu}n(x_2) -
\bar{n^c}(x_2)\gamma^{\mu}p(x_2)][\bar{p}(x_3)\gamma_{\beta}\gamma^5
\Delta^+_{\alpha}(x_3)])|0\rangle\nonumber\\
\hspace{-0.5in}&& - \langle0|{\rm
T}([\bar{\Delta}^0_{\omega}(z)\,{\Theta^{\omega}}_{\varphi}\,
p^c(x)][\bar{p^c}(x_2)\gamma^{\mu}n(x_2) -
\bar{n^c}(x_2)\gamma^{\mu}p(x_2)]\nonumber\\
\hspace{-0.5in}&&\times\,[\bar{n}(x_3)\gamma_{\beta}\gamma^5
\Delta^0_{\alpha}(x_3)])|0\rangle + (\gamma^5 \otimes 1 \to -
\gamma_{\nu}\gamma^5 \otimes \gamma^{\nu})\Big\} =\nonumber\\
\hspace{-0.5in}&&= \frac{i}{3}\frac{eg_{\rm V}}{M_{\rm N}}\frac{g_{\rm
\pi N\Delta}}{g_{\rm \pi NN}}\frac{g_{\rm \gamma N\Delta}}{g_{\rm \pi
NN}}\frac{g^3_{\rm \pi NN}}{4M^2_{\pi}}\int d^4x_1
d^4x_2 d^4x_3\int
d^4z\,\frac{\partial}{\partial z_{\varphi}}\delta^{(4)}(z-x_1)\nonumber\\
\hspace{-0.5in}&&\times\,{\rm T}([\bar{p^c}(x_1)\gamma^5 n(z) + 
\bar{p^c}(z)\gamma^5 n(x_1)]\,D^{\dagger}_{\mu}(x_2)
F^{\alpha\beta}(x_3))\nonumber\\
\hspace{-0.5in}&&\times\,\Big\{\langle0|{\rm
T}([\bar{\Delta}^+_{\omega}(z)\,
{\Theta^{\omega}}_{\varphi}\, n^c(x_1)][\bar{n^c}(x_2)\gamma^{\mu}p(x_2)]
[\bar{p}(x_3)\gamma_{\beta}\gamma^5
\Delta^+_{\alpha}(x_3)])|0\rangle\nonumber\\
\hspace{-0.5in}&& + \langle0|{\rm
T}([\bar{\Delta}^0_{\omega}(z)\,{\Theta^{\omega}}_{\varphi}\, p^c(x_1)]
[\bar{p^c}(x_2)\gamma^{\mu}n(x_2)] [\bar{n}(x_3)\gamma_{\beta}\gamma^5
\Delta^0_{\alpha}(x_3)])|0\rangle \nonumber\\
\hspace{-0.5in}&& + (\gamma^5 \otimes 1 \to - \gamma_{\nu}\gamma^5
\otimes \gamma^{\nu})\Big\} =\nonumber\\
\hspace{-0.5in}&&= \frac{2}{3}\frac{ieg_{\rm V}}{M^2_{\rm
N}}\frac{g_{\rm \pi N\Delta}}{g_{\rm \pi NN}}\frac{g_{\rm \gamma
N\Delta}}{g_{\rm \pi NN}}\frac{g^3_{\rm \pi NN}}{4M^2_{\pi}}\int
d^4x_1 d^4x_2 d^4x_3\int
d^4z\,\frac{\partial}{\partial z_{\varphi}}\delta^{(4)}(z-x_1)\nonumber\\
\hspace{-0.5in}&&\times\,\Big\{{\rm T}([\bar{p^c}(x_1)\gamma^5 n(z) +
\bar{p^c}(z)\gamma^5
n(x_1)]\,D^{\dagger}_{\mu}(x_2)
F^{\alpha\beta}(x_3))\nonumber\\
\hspace{-0.5in}&&\times\,\frac{1}{i}{\rm tr}\{S_{\alpha\omega}(x_3 - z)\,{\Theta^{\omega}}_{\varphi}\, S^c_{\rm F}(x_1 - x_2) \gamma^{\mu}
S_{\rm F}(x_2 - x_3) \gamma_{\beta}\gamma^5\}\nonumber\\
\hspace{-0.5in}&&-{\rm T}([\bar{p^c}(x_1)\gamma_{\nu}\gamma^5 n(z) +
\bar{p^c}(z)\gamma_{\nu}\gamma^5
n(x_1)]\,D^{\dagger}_{\mu}(x_2)
F^{\alpha\beta}(x_3))\nonumber\\
\hspace{-0.5in}&&\times\,\frac{1}{i}{\rm tr}\{S_{\alpha\omega}(x_3 - z) \, {\Theta^{\omega}}_{\varphi}\,\gamma^{\nu} S^c_{\rm F}(x_1 - x_2) \gamma^{\mu} S_{\rm F}(x_2 - x_3) \gamma_{\beta}\gamma^5\}\Big\}.
\end{eqnarray}
Thus, the effective Lagrangian ${\cal L}^{\rm np \to \Delta N \to
D\gamma}_{\rm eff}(x)$ is equal to
\begin{eqnarray}\label{label4.4}
\hspace{-0.5in}&&\int d^4x\,{\cal L}^{\rm np \to \Delta N \to
D\gamma}_{\rm eff}(x) = \frac{2}{3}\frac{ieg_{\rm V}}{M^2_{\rm N}}\frac{g_{\rm
\pi N\Delta}}{g_{\rm \pi NN}}\frac{g_{\rm \gamma N\Delta}}{g_{\rm \pi
NN}}\frac{g^3_{\rm \pi NN}}{4M^2_{\pi}}\int d^4x_1 d^4x_2 d^4x_3\int d^4z\,\frac{\partial}{\partial z_{\varphi}}\delta^{(4)}(z-x_1)\nonumber\\
\hspace{-0.5in}&&\times\,\Big\{{\rm T}([\bar{p^c}(x_1)\gamma^5 n(z) +
\bar{p^c}(z)\gamma^5 n(x_1)]\, D^{\dagger}_{\mu}(x_2) F^{\alpha\beta}(x_3))\nonumber\\
\hspace{-0.7in}&&\times\,\frac{1}{i}{\rm tr}\{S_{\alpha\omega}(x_3-z)\, {\Theta^{\omega}}_{\varphi}\, S^c_{\rm F}(x_1-x_2) \gamma^{\mu} S_{\rm F}(x_2 - x_3) \gamma_{\beta}\gamma^5\}\nonumber\\
\hspace{-0.5in}&&-{\rm T}([\bar{p^c}(x_1)\gamma_{\nu}\gamma^5 n(z) +
\bar{p^c}(z)\gamma_{\nu}\gamma^5
n(x_1)]\, D^{\dagger}_{\mu}(x_2) F^{\alpha\beta}(x_3))\nonumber\\
\hspace{-0.5in}&&\times\,\frac{1}{i}{\rm
tr}\{S_{\alpha\omega}(x_3-z)\,
{\Theta^{\omega}}_{\varphi}\,\gamma^{\nu} S^c_{\rm F}(x_1-x_2)
\gamma^{\mu} S_{\rm F}(x_2 - x_3) \gamma_{\beta}\gamma^5\}\Big\}.
\end{eqnarray}
In the momentum representation of the baryon Green functions the
effective Lagrangian Eq.(\ref{label4.4}) reads
\begin{eqnarray}\label{label4.5}
\hspace{-0.5in}&&\int d^4x\,{\cal L}^{\rm np \to \Delta N \to
D\gamma}_{\rm eff}(x) = \frac{2}{3}\frac{ieg_{\rm V}}{M^2_{\rm N}}\frac{g_{\rm
\pi N\Delta}}{g_{\rm \pi NN}}\frac{g_{\rm \gamma N\Delta}}{g_{\rm \pi
NN}}\frac{g^3_{\rm \pi NN}}{4M^2_{\pi}}\int d^4x_1\int
d^4z\,\frac{\partial}{\partial z_{\varphi}}\delta^{(4)}(z-x_1)
\nonumber\\
\hspace{-0.5in}&&\times\,\int\frac{d^4x_2d^4k_2}{(2\pi)^4}
\frac{d^4x_3d^4k_3}{(2\pi)^4}\,e^{\textstyle -ik_2\cdot (x_2 - x_1)}\,e^{\textstyle -ik_3\cdot
(x_3 - z)}\nonumber\\
\hspace{-0.5in}&&\times\Big\{{\rm T}([\bar{p^c}(x_1)\gamma^5 n(z) + \bar{p^c}(z)\gamma^5 n(x_1)]\, D^{\dagger}_{\mu}(x_2) F_{\alpha\beta}(x_3))\nonumber\\
\hspace{-0.5in}&&\times\,\int\frac{d^4k_1}{\pi^2i}\,e^{\textstyle
ik_1\cdot (x_1 - z)}\,{\rm tr}\{S^{\alpha\omega}(k_1 + k_3)\,
\Theta_{\omega\varphi}\,\frac{1}{M_{\rm N} - \hat{k}_1 + \hat{k}_2} \gamma^{\mu}
\frac{1}{M_{\rm N} - \hat{k}_1}
\gamma^{\beta}\gamma^5\}\nonumber\\
\hspace{-0.5in}&&-{\rm T}([\bar{p^c}(x_1)\gamma_{\nu}\gamma^5 n(z) +
\bar{p^c}(z)\gamma_{\nu}\gamma^5
n(x_1)] \, D^{\dagger}_{\mu}(x_2) F^{\alpha\beta}(x_3))\nonumber\\
\hspace{-0.5in}&&\times\,\int\frac{d^4k_1}{\pi^2i}\,e^{\textstyle
ik_1\cdot (x_1 - z)}\,{\rm
tr}\{S^{\alpha\omega}(k_1 + k_3)\,
\Theta_{\omega\varphi}\,\frac{1}{M_{\rm N} - \hat{k}_1 + \hat{k}_2}
\gamma^{\mu} \frac{1}{M_{\rm N} -
\hat{k}_1}\gamma^{\beta}\gamma^5\}\Big\}.
\end{eqnarray}
The effective Lagrangian Eq.(\ref{label4.5}) defines the contribution
of the $\Delta(1232)$ resonance to the low--energy transition n + p
$\to$ D + $\gamma$.

The matrix element of the neutron--proton radiative capture caused by
the contribution of the $\Delta(1232)$ resonance exchange is equal to
\begin{eqnarray}\label{label4.6}
\hspace{-0.5in}&& {\cal M}({\rm n + p \to \Delta N \to D + \gamma})
=\nonumber\\
\hspace{-0.5in}&& = -\frac{ie}{2M^2_{\rm N}}\frac{g_{\rm
V}}{6\pi^2}\frac{g_{\rm \pi N\Delta}}{g_{\rm \pi NN}}\frac{g_{\rm
\gamma N\Delta}}{g_{\rm \pi NN}}\frac{g^3_{\rm \pi
NN}}{4M^2_{\pi}}[\bar{u^c}(p_2)\gamma^5 u(p_1)]\,(k_{\alpha}
e^*_{\beta}(k) - k_{\beta} e^*_{\alpha}(k))\,e^*_{\mu}(k_{\rm
D})\nonumber\\
\hspace{-0.5in}&&\times\,{\cal J}^{\mu\beta\alpha}_5(k_{\rm
D},k)\nonumber\\
\hspace{-0.5in}&&+\frac{ie}{2M^2_{\rm N}}\frac{g_{\rm
V}}{6\pi^2}\frac{g_{\rm \pi N\Delta}}{g_{\rm \pi NN}}\frac{g_{\rm
\gamma N\Delta}}{g_{\rm \pi NN}}\frac{g^3_{\rm \pi
NN}}{4M^2_{\pi}}[\bar{u^c}(p_2)\gamma_{\nu}\gamma^5
u(p_1)]\,(k_{\alpha} e^*_{\beta}(k) - k_{\beta}
e^*_{\alpha}(k))\,e^*_{\mu}(k_{\rm D})\nonumber\\
\hspace{-0.5in}&&\times\,{\cal J}^{\nu\mu\beta\alpha}_5(k_{\rm D},k),
\end{eqnarray}
where the structure functions ${\cal J}^{\mu\beta\alpha}_5(k_{\rm
D},k)$ and ${\cal J}^{\nu\mu\beta\alpha}_5(k_{\rm D},k)$ are defined
by the momentum integrals
\begin{eqnarray}\label{label4.7}
\hspace{-0.5in}&& {\cal J}^{\mu\beta\alpha}_5(k_{\rm D},k) =\nonumber\\
\hspace{-0.5in}&&=\int\frac{d^4k_1}{\pi^2i}\,{\rm tr}\{(k_1 + k_{\rm
D})^{\varphi}S^{\alpha\omega}(k_1 + k)\,
\Theta_{\omega\varphi}\,\frac{1}{M_{\rm N} - \hat{k}_1+ \hat{k}_{\rm
D}} \gamma^{\mu} \frac{1}{M_{\rm N} - \hat{k}_1 }
\gamma^{\beta}\gamma^5\},\nonumber\\
\hspace{-0.5in}&&{\cal J}^{\nu\mu\beta\alpha}_5(k_{\rm D},k) =\nonumber\\
\hspace{-0.5in}&&=\int\frac{d^4k_1}{\pi^2i}\,{\rm tr}\{(k_1 + k_{\rm
D})^{\varphi}S^{\alpha\omega}(k_1 + k)\,
\Theta_{\omega\varphi}\,\gamma^{\nu}\frac{1}{M_{\rm N} - \hat{k}_1 +
\hat{k}_{\rm D}} \gamma^{\mu} \frac{1}{M_{\rm N} - \hat{k}_1}
\gamma^{\beta}\gamma^5\}.
\end{eqnarray}
At leading order in the large $N_C$ expansion the structure functions Eq.(\ref{label4.7}) read
\begin{eqnarray}\label{label4.8}
{\cal J}^{\mu\beta\alpha}_5(k_{\rm D},k)&=& \frac{4}{3}\,\Bigg(Z
-\frac{1}{2}\Bigg)\,i\,M_{\rm N}\,
\varepsilon^{\mu\beta\alpha\lambda}\,k_{\rm D\lambda},\nonumber\\
{\cal J}^{\nu\mu\beta\alpha}_5(k_{\rm D},k)&=& \frac{2}{3}\,\Bigg(Z
-\frac{1}{2}\Bigg)\,i\,M^2_{\rm N}\, \varepsilon^{\mu\beta\alpha\nu}.
\end{eqnarray}
We have neglected the mass difference between the masses of the
$\Delta(1232)$ resonance and the nucleon.  The matrix element of the
low--energy transition n + p $\to$ D + $\gamma$ caused by the
$\Delta(1232)$ resonance contribution is equal to
\begin{eqnarray}\label{label4.9}
\hspace{-0.5in}&&{\cal M}_{\Delta}({\rm n + p \to D + \gamma}) =
\frac{e}{2M_{\rm N}}\frac{g_{\rm V}}{4\pi^2}\Bigg[\Bigg(\frac{1}{2} -
Z\Bigg)\,\frac{8}{9}\,\frac{g_{\rm \pi N\Delta}}{g_{\rm \pi
NN}}\,\frac{g_{\rm \gamma N\Delta}}{g_{\rm \pi NN}}\,\frac{g^3_{\rm
\pi NN}}{4M^2_{\pi}}\Bigg]\, \nonumber\\
\hspace{-0.5in}&&\times\, \varepsilon^{\alpha\beta\mu\nu} k_{\alpha}
e^*_{\beta}(k,\lambda) e^*_{\mu}(k_{\rm D},\lambda_{\rm D})
[\bar{u^c}(p_2)(2k_{\rm D\nu} - M_{\rm N}\gamma_{\nu})\gamma^5
u(p_1)]=\nonumber\\
\hspace{-0.5in}&&=e \,\frac{5 g_{\rm V}}{8\pi^2}\,\Bigg[\Bigg(\frac{1}{2}
- Z\Bigg)\,\frac{8}{9}\,\frac{g_{\rm \pi N\Delta}}{g_{\rm \pi
NN}}\,\frac{g_{\rm \gamma N\Delta}}{g_{\rm \pi NN}}\,\frac{g^3_{\rm \pi
NN}}{4M^2_{\pi}}\Bigg] (\vec{k}\times \vec{e}^{\,*}(\vec{k},\lambda))\cdot
\vec{e}^{\,*}(\vec{k}_{\rm D},\lambda_{\rm D})\nonumber\\
\hspace{-0.5in}&&\times [\bar{u^c}(p_2)\gamma^5 u(p_1)].
\end{eqnarray}
In turn, the contribution of the nucleon tensor current [1]
\begin{eqnarray}\label{label4.10}
\delta {\cal L}_{\rm npD}(x) = \frac{g_{\rm T}}{2M_{\rm
N}}\,[\bar{p^c}(x)\sigma^{\mu\nu}n(x) -
\bar{n^c}(x)\sigma^{\mu\nu}p(x)]\,D^{\dagger}_{\mu\nu}(x)
\end{eqnarray}
does not depend on the parameter $Z$ and
reads
\begin{eqnarray}\label{label4.11}
\hspace{-0.5in}&&\delta {\cal M}_{\Delta}({\rm n + p \to D + \gamma}) = e
\,\frac{5 g_{\rm V}}{8\pi^2}\,\Bigg[\frac{1}{5}\,\frac{g_{\rm T}}{g_{\rm V}}\frac{8}{9}\,\frac{g_{\rm \pi N\Delta}}{g_{\rm \pi
NN}}\,\frac{g_{\rm \gamma N\Delta}}{g_{\rm \pi NN}}\,\frac{g^3_{\rm
\pi NN}}{4M^2_{\pi}}\Bigg]\nonumber\\
\hspace{-0.5in}&&\times (\vec{k}\times
\vec{e}^{\,*}(\vec{k},\lambda))\cdot \vec{e}^{\,*}(\vec{k}_{\rm
D},\lambda_{\rm D})\,[\bar{u^c}(p_2)\gamma^5 u(p_1)].
\end{eqnarray}
The coupling constants $g_{\rm T}$ and $g_{\rm V}$ are connected
by the relation [1]
\begin{eqnarray}\label{label4.12}
g_{\rm T} = \sqrt{\frac{3}{8}}\,g_{\rm V} + O(1/\sqrt{N_C}).
\end{eqnarray}
The total amplitude of the neutron--proton radiative capture for
thermal neutrons reads
\begin{eqnarray}\label{label4.13}
\hspace{-0.5in}&&{\cal M}({\rm n + p \to D + \gamma}) = e\, (\mu_{\rm
p} - \mu_{\rm n}) \frac{5 g_{\rm V}}{8\pi^2} C_{\rm NN} (\vec{k}\times
\vec{e}^{\,*}(\vec{k},\lambda))\cdot \vec{e}^{\,*}(\vec{k}_{\rm
D},\lambda_{\rm D})\,[\bar{u^c}(p_2)\gamma^5 u(p_1)]\nonumber\\
\hspace{-0.5in}&&\times\,\Bigg[1 + \frac{g^2_{\rm \pi NN}}{\mu_{\rm p}
- \mu_{\rm
n}}\,\frac{M^2_{\pi}}{8\pi^2}\,\frac{\alpha_{\rho}}{\pi}\Bigg(J_{\rm
\pi a_1 N} + \frac{3}{2 g_{\rm A}}\,J_{\rm \pi VN}\Bigg)
+\frac{1}{\mu_{\rm p} - \mu_{\rm
n}}\,\frac{1}{5}\,\sqrt{\frac{3}{8}}\,\frac{1}{C_{\rm
NN}}\,\frac{8}{9}\,\frac{g_{\rm \pi N\Delta}}{g_{\rm \pi
NN}}\,\frac{g_{\rm \gamma N\Delta}}{g_{\rm \pi NN}}\,\frac{g^3_{\rm
\pi NN}}{4M^2_{\pi}}\nonumber\\ 
\hspace{-0.5in}&&+ \frac{1 - 2Z}{\mu_{\rm p} - \mu_{\rm
n}}\,\frac{1}{C_{\rm NN}}\,\frac{4}{9}\,\frac{g_{\rm \pi
N\Delta}}{g_{\rm \pi NN}}\,\frac{g_{\rm \gamma N\Delta}}{g_{\rm \pi
NN}}\,\frac{g^3_{\rm \pi NN}}{4M^2_{\pi}}\Bigg].
\end{eqnarray}
The total cross section for the neutron--proton radiative capture is
then defined by
\begin{eqnarray}\label{label4.14}
\hspace{-0.5in}&&\sigma({\rm np \to D\gamma})(p)=
\frac{1}{v}\,(\mu_{\rm p}-\mu_{\rm
n})^2\,\frac{25}{64}\,\frac{\alpha}{\pi^2}\,Q_{\rm D}\,C^2_{\rm
NN}\,M_{\rm N}\,\varepsilon^3_{\rm D}\,\nonumber\\
\hspace{-0.5in}&&\times\,\Bigg[1 + \frac{g^2_{\rm \pi NN}}{\mu_{\rm p}
- \mu_{\rm
n}}\,\frac{M^2_{\pi}}{8\pi^2}\,\frac{\alpha_{\rho}}{\pi}\Bigg(J_{\rm
\pi a_1 N} + \frac{3}{2 g_{\rm A}}\,J_{\rm \pi VN}\Bigg)
+\frac{1}{\mu_{\rm p} - \mu_{\rm
n}}\,\frac{1}{5}\,\sqrt{\frac{3}{8}}\,\frac{1}{C_{\rm
NN}}\,\frac{8}{9}\,\frac{g_{\rm \pi N\Delta}}{g_{\rm \pi
NN}}\,\frac{g_{\rm \gamma N\Delta}}{g_{\rm \pi NN}}\,\frac{g^3_{\rm
\pi NN}}{4M^2_{\pi}}\nonumber\\ 
\hspace{-0.5in}&&+ \frac{1 - 2Z}{\mu_{\rm p} - \mu_{\rm
n}}\,\frac{1}{C_{\rm NN}}\,\frac{4}{9}\,\frac{g_{\rm \pi
N\Delta}}{g_{\rm \pi NN}}\,\frac{g_{\rm \gamma N\Delta}}{g_{\rm \pi
NN}}\,\frac{g^3_{\rm \pi NN}}{4M^2_{\pi}}\Bigg]^2.
\end{eqnarray}
The numerical value of the cross section amounts to
\begin{eqnarray}\label{label4.15}
\sigma({\rm np \to D\gamma})(T_{\rm n}) = 325.5\,(1 +
0.246\,(1-2Z))^2\,{\rm m b}.
\end{eqnarray}
Thus, the discrepancy of the theoretical cross section and the
experimental value Eq.(\ref{label1.1}) can by described by the
contribution of the $\Delta(1232)$ resonance. In order to fit the
experimental value of the cross section we should set $Z=0.473$. This
agrees with the experimental bound $|Z|\le 1/2$ [18]. At $Z=1/2$ that
is favoured theoretically [15] we get the cross section $\sigma({\rm
np \to D\gamma})(T_{\rm n}) = 325.5\,{\rm m b}$ agreeing with the experimental
value with accuracy better than 3$\%$.

\section{The photomagnetic disintegration of the deuteron}
\setcounter{equation}{0}

\hspace{0.2in} The amplitude of the photomagnetic disintegration of
the deuteron $\gamma$ + D $\to$ n + p is
related to the amplitude of the neutron--proton radiative capture n +
p $\to$ D + $\gamma$ due to time--reversal invariance and reads
\begin{eqnarray}\label{label5.1}
\hspace{-0.5in}&&{\cal M}({\rm \gamma + D \to n + p}) = e\,(\mu_{\rm
p} - \mu_{\rm n})\,\frac{5 g_{\rm V}}{8\pi^2}\,C_{\rm
NN}\,(\vec{k}\times \vec{e}(\vec{k},\lambda))\cdot
\vec{e}(\vec{k}_{\rm D},\lambda_{\rm D}) \,[\bar{u}(p_2)\gamma^5
u^c(p_1)]\,\nonumber\\
\hspace{-0.5in}&&\times\,\Bigg[1 + \frac{g^2_{\rm \pi NN}}{\mu_{\rm p}
- \mu_{\rm
n}}\,\frac{M^2_{\pi}}{8\pi^2}\,\frac{\alpha_{\rho}}{\pi}\Bigg(J_{\rm
\pi a_1 N} + \frac{3}{2 g_{\rm A}}\,J_{\rm \pi VN}\Bigg) + \frac{1}{\mu_{\rm p} - \mu_{\rm
n}}\,\frac{1}{5}\,\sqrt{\frac{3}{8}}\,\frac{1}{C_{\rm
NN}}\,\frac{8}{9}\,\frac{g_{\rm \pi N\Delta}}{g_{\rm \pi
NN}}\,\frac{g_{\rm \gamma N\Delta}}{g_{\rm \pi NN}}\,\frac{g^3_{\rm
\pi NN}}{4M^2_{\pi}}\nonumber\\ 
\hspace{-0.5in}&&+ \frac{1 - 2Z}{\mu_{\rm p} - \mu_{\rm
n}}\,\frac{1}{C_{\rm NN}}\,\frac{4}{9}\,\frac{g_{\rm \pi
N\Delta}}{g_{\rm \pi NN}}\,\frac{g_{\rm \gamma N\Delta}}{g_{\rm \pi
NN}}\,\frac{g^3_{\rm \pi NN}}{4M^2_{\pi}}\Bigg]^2\,e^{\textstyle i\,\delta_{\rm
np}(p)}\,\frac{\sin\,\delta_{\rm np}(p)}{-a_{\rm np}\,p}\,\frac{1}{1 +
r^2_{\rm D} p^2}.
\end{eqnarray}
The cross section defined by the amplitude Eq.(\ref{label5.1}) is then
given by
\begin{eqnarray}\label{label5.2}
\sigma({\rm \gamma D \to np})(\omega) = \sigma_0
\,\Bigg(\frac{\omega}{\varepsilon_{\rm
D}}\Bigg)\,\frac{1}{\displaystyle \Big(1 - \frac{1}{2} r_{\rm np}
a_{\rm np}p^2\Big)^2 + a^2_{\rm np} p^2}\,\frac{r_{\rm D}p}{(1 +
r^2_{\rm D} p^2)^2},
\end{eqnarray}
where $p=\sqrt{M_{\rm N}(\omega - \varepsilon_{\rm D})}$ is the
relative momentum of the np pair in the ${^1}{\rm S}_0$ state,
$\omega$ is the energy of the photon, and $\sigma_0$ is equal to
\begin{eqnarray}\label{label5.3}
\hspace{-0.3in}&&\sigma_0 = (\mu_{\rm p}-\mu_{\rm n})^2 \frac{25\alpha
Q_{\rm D}}{192\pi^2} C^2_{\rm NN}\,\varepsilon^{3/2}_{\rm
D}M^{5/2}_{\rm N}\,\nonumber\\
\hspace{-0.3in}&&\times\,\Bigg[1 + \frac{g^2_{\rm \pi NN}}{\mu_{\rm p}
- \mu_{\rm
n}}\,\frac{M^2_{\pi}}{8\pi^2}\,\frac{\alpha_{\rho}}{\pi}\Bigg(J_{\rm
\pi a_1 N} + \frac{3}{2 g_{\rm A}}\,J_{\rm \pi VN}\Bigg) + \frac{1}{\mu_{\rm p} - \mu_{\rm
n}}\,\frac{1}{5}\,\sqrt{\frac{3}{8}}\,\frac{1}{C_{\rm
NN}}\,\frac{8}{9}\,\frac{g_{\rm \pi N\Delta}}{g_{\rm \pi
NN}}\,\frac{g_{\rm \gamma N\Delta}}{g_{\rm \pi NN}}\,\frac{g^3_{\rm
\pi NN}}{4M^2_{\pi}}\nonumber\\ 
\hspace{-0.5in}&&+ \frac{1 - 2Z}{\mu_{\rm p} - \mu_{\rm
n}}\,\frac{1}{C_{\rm NN}}\,\frac{4}{9}\,\frac{g_{\rm \pi
N\Delta}}{g_{\rm \pi NN}}\,\frac{g_{\rm \gamma N\Delta}}{g_{\rm \pi
NN}}\,\frac{g^3_{\rm \pi NN}}{4M^2_{\pi}}\Bigg]^2  = 7.10\,{\rm m b}.
\end{eqnarray}
The energy region of the dominance of the photomagnetic
disintegration of the deuteron is restricted by the constraint $r_{\rm
D}p = \sqrt{(\omega -\varepsilon_{\rm D})/\varepsilon_{\rm D}} \ll 1$
or differently $\omega \ge \varepsilon_{\rm D} = 2.225\,{\rm MeV}$.

The numerical values of the cross section for the photomagnetic
disintegration of the deuteron at energies $\omega \le
2\,\varepsilon_{\rm D} = 4.45\,{\rm MeV}$ read
\begin{eqnarray}\label{label5.4}
\sigma({\rm \gamma D \to np})(\omega)\Big|_{\omega = 2.62\,{\rm MeV}}
&=& 0.358\,(0.380)\,{\rm m b},\nonumber\\
\sigma({\rm \gamma D \to np})(\omega)\Big|_{\omega = 2.76\,{\rm MeV}}
&=& 0.302\,(0.327)\,{\rm m b},\nonumber\\
\sigma({\rm \gamma D \to np})(\omega)\Big|_{\omega = 4.45\,{\rm MeV}}
&=& 0.094\,(0.128)\,{\rm m b},
\end{eqnarray}
where in parentheses we have adduced the theoretical values obtained
by Chen and Savage in the EFT [10]. One can see a reasonable agreement
between the results obtained in the NNJL model and the EFT. Thus, the
spatial smearing of the physical deuteron caused by the effective
radius $r_{\rm D}$ and introduced in the NNJL model phenomenologically
in the form of the wave function Eq.(\ref{label2.16}) describes well
the energy dependence of the cross section for the photomagnetic
disintegration of the deuteron at photon energies far from
threshold. Note that in the critical region of the photon energies
$\omega \le 2\,\varepsilon_{\rm D}=4.45\,{\rm MeV}$ the cross section
for the photomagnetic disintegration of the deuteron calculated in the
NNJL model falls steeper with $\omega$ than in the EFT. However, in
this energy region the dominant role is attributed to the
E1--transition which we have not taken into account.

For the correct description of the experimental data on the
photodisintegration of the deuteron [40] (see also Chen and Savage
[10]):
\begin{eqnarray}\label{label5.5}
\sigma({\rm \gamma D \to np})_{\exp}(\omega)\Big|_{\omega = 2.62\,{\rm MeV}}
&=& (1.300\pm 0.029)\,{\rm m b},\nonumber\\
\sigma({\rm \gamma D \to np})_{\exp}(\omega)\Big|_{\omega = 2.76\,{\rm MeV}}
&=& (1.474\pm 0.032)\,{\rm m b},\nonumber\\
\sigma({\rm \gamma D \to np})_{\exp}(\omega)\Big|_{\omega = 4.45\,{\rm MeV}}
&=& (2.430\pm 0.170)\,{\rm m b}
\end{eqnarray}
one must include the contribution of the E1--transition [10].

Since the cross section for the photodisintegration of the deuteron
having been discussed above does not contain the contribution of the
E1--transition, our result can be scarcely compared with the recent
theoretical investigation of the photodisintegration of the deuteron
carried out by Anisovich and Sadovnikova [41] within the dispersion
relation approach based on the dispersion relation technique developed
by Anisovich {\it et al.}  [42]. This investigation represents a
detailed analysis of the saturation of the cross section for the
photodisintegration of the deuteron by different intermediate states
valid mainly for the photon energy region $\omega \ge 50\,{\rm MeV}$,
where the contribution of the E1--transition is important.

\section{The amplitude of the solar proton burning} 
\setcounter{equation}{0}

\hspace{0.2in} The analysis of the reaction p + p $\to$ D + e$^+$ +
$\nu_{\rm e}$ within the NNJL model we should start with the
evaluation of the effective Lagrangian ${\cal L}^{\rm pp \to
De^+\nu_{\rm e}}_{\rm eff}(x)$ of the transition p + p $\to$ D + e$^+$
+ $\nu_{\rm e}$. In the tree--meson approximation\footnote{Below the
analysis of weak nuclear reactions is carried out in the tree--meson
approximation. The inclusion of chiral one--meson loop corrections
goes beyond the scope of this paper and demands a separate
publication.} the effective Lagrangian ${\cal L}^{\rm pp \to
De^+\nu_{\rm e}}_{\rm eff}(x)$ is defined by the one--nucleon loop
diagrams depicted in Fig.3. The detailed evaluation of ${\cal L}^{\rm
pp \to De^+\nu_{\rm e}}_{\rm eff}(x)$ is given in Appendix. The
result reads\footnote{This result is obtained at zero contribution of
the nucleon tensor current (see Appendix and discussion in the
Conclusion).}
\begin{eqnarray}\label{label6.1}
{\cal L}^{\rm pp\to D e^+ \nu_{\rm e}}_{\rm eff}(x) = - i g_{\rm A}C_{\rm NN}
M_{\rm N}\frac{G_{\rm V}}{\sqrt{2}}\frac{3g_{\rm
V}}{4\pi^2}\,D^{\dagger}_{\mu}(x)\,[\bar{p^c}(x)\gamma^5
p(x)]\,[\bar{\psi}_{\nu_{\rm e}}(x)\gamma^{\mu}(1 - \gamma^5) \psi_{\rm
e}(x)].
\end{eqnarray}
where $G_{\rm V} = G_{\rm F}\,\cos \vartheta_C$ with $G_{\rm F} =
1.166\,\times\,10^{-11}\,{\rm MeV}^{-2}$ and $\vartheta_C$ are the
Fermi weak coupling constant and the Cabibbo angle $\cos \vartheta_C =
0.975$

For the derivation of the effective Lagrangian ${\cal L}^{\rm pp \to
De^+\nu_{\rm e}}_{\rm eff}(x)$ we have used  the effective Lagrangian
${\cal L}^{\rm NN \to NN}_{\rm eff}(x)$ responsible for low--energy
transitions N + N $\to$ N + N defined by Eq.(\ref{label1.17}).

The matrix element of the transition p + p $\to$ D + e$^+$ +
$\nu_{\rm e}$ we define by a usual way 
\begin{eqnarray}\label{label6.2}
&&\int d^4x\,\langle D(k_{\rm D})e^+(k_{\rm e^+})\nu_{\rm e}(k_{\nu_{\rm
e}})|{\cal L}^{\rm np\to D e^+\nu_{\rm e}}_{\rm eff}(x)|p(p_1)p(p_2)\rangle =
\nonumber\\ 
&&= (2\pi)^4\delta^{(4)}( k_{\rm D} + k_{\rm e^+} +
k_{\nu_{\rm e}} - p_1 - p_2)\,\frac{{\cal M}({\rm n} + {\rm p}\to {\rm
D} + {\rm e}^+ + \nu_{\rm e})}{\displaystyle \sqrt{2E_1 V\,2E_2 V\,
2E_{\rm D} V\,2 E_{\rm e^+} V\,2 E_{\nu_{\rm e}} V}},
\end{eqnarray}
where $E_i\,(i =1,2,{\rm D}, {\rm e^+}, \nu_{\rm e})$ are the energies
of the protons, the deuteron, the positron and the neutrino, $V$ is the
normalization volume.

The wave functions of the initial $|p(p_1) p(p_2)\rangle$ and final
$\langle D(k_{\rm D})e^+(k_{\rm e^+})\nu_{\rm e}(k_{\nu_{\rm
e}})|$ state we take in the usual form
\begin{eqnarray}\label{label6.3}
|p(p_1) p(p_2)\rangle &=& \frac{1}{\sqrt{2}}\,a^{\dagger}_{\rm
 p}(\vec{p}_1,\sigma_1)\,a^{\dagger}_{\rm
 p}(\vec{p}_2,\sigma_2)|0\rangle,\nonumber\\
\langle D(k_{\rm D})e^+(k_{\rm e^+})\nu_{\rm e}(k_{\nu_{\rm
e}})| &=& \langle 0|a_{\rm D}(\vec{k}_{\rm D},\lambda_{\rm D})\,a_{\rm
 e^+}(\vec{k}_{\rm e^+},\sigma)\,a_{\nu_{\rm e}}(\vec{k}_{\nu_{\rm
e}},\sigma'),
\end{eqnarray}
where $a^{\dagger}_{\rm n}(\vec{p}_1,\sigma_1)$ and $a^{\dagger}_{\rm
p}(\vec{p}_2,\sigma_2)$ are the operators of creation of the
protons. In turn, $a_{\rm D}(\vec{k}_{\rm D},\lambda_{\rm D})$,
$a_{\rm e^+}(\vec{k}_{\rm e^+},\sigma_{\rm e^+})$ and $a_{\nu_{\rm
e}}(\vec{k}_{\nu_{\rm e}},\sigma_{\nu_{\rm e}})$ are the operators of
annihilation of the deuteron, the positron and the neutrino.

The effective Lagrangian Eq.(\ref{label6.1}) defines the effective
vertex of the low--energy nuclear transition p + p $\to$ D + e$^+$ +
$\nu_{\rm e}$
\begin{eqnarray}\label{label6.4}
i{\cal M}({\rm p} + {\rm p} \to {\rm D} + {\rm e}^+ + \nu_{e})&=&
\,g_{\rm A} M_{\rm N}\,C_{\rm NN}\,G_{\rm V}\,\frac{3g_{\rm V}}{4\pi^2}\,
e^*_{\mu}(k_{\rm D})\,[\bar{u}(k_{\nu_{\rm
e}})\gamma^{\mu} (1-\gamma^5) v(k_{\rm e^+})]\nonumber\\
&&\times\,[\bar{u^c}(p_2) \gamma^5
u(p_1)],
\end{eqnarray}
where $e^*_{\mu}(k_{\rm D},\lambda_{\rm D})$ is a 4--vector of a
polarization of the deuteron, $u(k_{\nu_{\rm e}})$, $v(k_{\rm e^+})$,
$u(p_2)$ and $u(p_1)$ are the Dirac bispinors of the neutrino, the
positron and the protons, respectively.

For the evaluation of the matrix element Eq.(\ref{label6.4}) we have
used the wave functions of the protons in the form of the plane
waves. However, a real wave function of the pp pair in the ${^1}{\rm
S}_0$ state is defined by low--energy nuclear forces and Coloumb
repulsion. In order to take into account both low--energy nuclear
forces and Coulomb repulsion for the relative movement of the pp pair
in the ${^1}{\rm S}_0$ state we would sum up an
infinite series of one--proton bubbles. The vertices of these
one--nucleon bubbles are defined by
\begin{eqnarray}\label{label6.5}
V_{\rm pp \to pp}(k',k) = C_{\rm NN}\,\psi^*_{\rm pp}(k'\,)\,
[\bar{u}(p'_2) \gamma^5 u^c(p'_1)]\,[\bar{u^c}(p_2) \gamma^5
u(p_1)]\,\psi_{\rm pp}(k),
\end{eqnarray}
where $\psi_{\rm pp}(k)$ and $\psi^*_{\rm pp}(k'\,)$ are the explicit
Coulomb wave functions of the relative movement of the protons taken
at zero relative distances, and $k$ and $k'$ are relative 3--momenta of
the protons $\vec{k} = (\vec{p}_1 - \vec{p}_2)/2$ and
$\vec{k}^{\,\prime} = (\vec{p}^{\;\prime}_1 - \vec{p}^{\;\prime}_2
)/2$ in the initial and final states. The explicit form of $\psi_{\rm
pp}(k)$ we take following Kong and Ravndal [33] (see also [43])
\begin{eqnarray}\label{label6.6}
\psi_{\rm pp}(k) = e^{\textstyle - \pi/4k r_C}\,\Gamma\Big(1 +
\frac{i}{2k r_C}\Big),
\end{eqnarray}
where $2\,r_C = 2/M_{\rm N}\alpha = 57.64\,{\rm fm}$ and $\alpha = 1/137$
are the Bohr radius of the pp pair in the ${^1}{\rm S}_0$ state and
the fine structure constant. The squared value of the modulo of
$\psi_{\rm pp}(k)$ is given by
\begin{eqnarray}\label{label6.7}
|\psi_{\rm pp}(k)|^2 = C^2_0(k) = \frac{\pi}{k r_C}\,
\frac{1}{\displaystyle e^{\textstyle \pi/k r_C} - 1},
\end{eqnarray}
where $C_0(k)$ is the Gamow penetration factor [3,29,43].  

By taking into account the contribution of the Coulomb wave function
and summing up an infinite series of one--proton bubbles the
expression Eq.(\ref{label6.4}) can be recast into the form
\begin{eqnarray}\label{label6.8}
\hspace{-0.5in}&&i{\cal M}({\rm p} + {\rm p} \to {\rm D} + {\rm e}^+ +
\nu_{e}) = g_{\rm A} M_{\rm N}\,C_{\rm NN}\,G_{\rm V}\,\frac{3g_{\rm V}}{4\pi^2}\, e^*_{\mu}(k_{\rm
D})\,[\bar{u}(k_{\nu_{\rm e}})\gamma^{\mu} (1-\gamma^5) v(k_{\rm
e^+})]\nonumber\\
\hspace{-0.5in}&&\times\,\frac{[\bar{u^c}(p_2) \gamma^5 u(p_1)]\,\psi_{\rm
pp}(k)}{\displaystyle 1 + \frac{C_{\rm NN}}{16\pi^2}\int
\frac{d^4p}{\pi^2i}\,|\psi_{\rm pp}(|\vec{p} + \vec{Q}\,|)|^2 {\rm
tr}\Big\{\gamma^5 \frac{1}{M_{\rm N} - \hat{p} - \hat{P} -
\hat{Q}}\gamma^5 \frac{1}{M_{\rm N} - \hat{p} - \hat{Q}}\Big\}}.
\end{eqnarray}
where $P = p_1 + p_2 = (2\sqrt{k^2 + M^2_{\rm N}}, \vec{0}\,)$ is the
4--momentum of the pp--pair in the center of mass frame; $Q =a\,P +
b\,K = a\,(p_1 + p_2) + b\,(p_1 - p_2)$ is an arbitrary shift of
virtual momentum with arbitrary parameters $a$ and $b$, and in the
center of mass frame $K = p_1 - p_2 = (0,2\,\vec{k}\,)$. The
parameters $a$ and $b$ can be functions of $k$. 

The evaluation of the momentum integral we would carry out at
leading order in the $1/M_{\rm N}$ expansion or differently in the
large $N_C$ expansion [1] due to proportionality $M_{\rm N} \sim
N_C$ valid in QCD with $SU(N_C)$ gauge group at $N_C \to \infty$
[22]. As a result we obtain
\begin{eqnarray}\label{label6.9}
&&\int \frac{d^4p}{\pi^2i}\,|\psi_{\rm
pp}(|\vec{p} + \vec{Q}\,|)|^2 {\rm tr}\Big\{\gamma^5 \frac{1}{M_{\rm N}
- \hat{p} - \hat{P} - \hat{Q}}\gamma^5 \frac{1}{M_{\rm N} - \hat{p} -
\hat{Q}}\Big\} =\nonumber\\
&&= - 8\, a\,(a + 1)\,M^2_{\rm N} + 8\,(b^2 - a\,(a +
1))\,k^2 - i\,8\pi\,M_{\rm N}\,k\,|\psi_{\rm pp}(k)|^2 = \nonumber\\
&&= - 8\, a\,(a + 1)\,M^2_{\rm N} + 8\,(b^2 - a\,(a +
1))\,k^2 - i\,8\pi\,M_{\rm N}\,k\,C^2_0(k).
\end{eqnarray}
Substituting Eq.(\ref{label6.9}) in Eq.(\ref{label6.8}) we get 
\begin{eqnarray}\label{label6.10}
\hspace{-0.7in}&& i{\cal M}({\rm p} + {\rm p} \to {\rm D} + {\rm e}^+
+ \nu_{e}) = g_{\rm A} M_{\rm N}\,C_{\rm NN}\,G_{\rm V}\,\frac{3g_{\rm V}}{4\pi^2}\nonumber\\
\hspace{-0.7in}&&\times \, e^*_{\mu}(k_{\rm D})\,[\bar{u}(k_{\nu_{\rm
e}})\gamma^{\mu} (1-\gamma^5) v(k_{\rm e^+})]\,[\bar{u^c}(p_2)
\gamma^5 u(p_1)]\,e^{\textstyle - \pi/4k
r_C}\,\Gamma\Big(1 + \frac{i}{2k r_C}\Big)\nonumber\\
\hspace{-0.7in}&&\Big[ 1 - a(a+1) \frac{G_{\rm
\pi NN}}{2\pi^2}\,M^2_{\rm N} + \frac{C_{\rm NN}}{2\pi^2}\,(b^2 -
a\,(a + 1))\,k^2 - i\,\frac{C_{\rm NN}M_{\rm
N}}{2\pi}\,k\,C^2_0(k)\Big]^{-1}\!\!\!.
\end{eqnarray}
In order to reconcile the contribution of low--energy elastic pp
scattering with low--energy nuclear phenomenology [43] we should make
a few changes. For this aim we should rewrite Eq.(\ref{label6.10}) in more
convenient form
\begin{eqnarray}\label{label6.11}
\hspace{-0.7in}&& i{\cal M}({\rm p} + {\rm p} \to {\rm D} + {\rm e}^+
+ \nu_{e}) = g_{\rm A} M_{\rm N}\,C_{\rm NN}\,G_{\rm V}\,\frac{3g_{\rm V}}{4\pi^2}\nonumber\\
\hspace{-0.7in}&&\times \, e^*_{\mu}(k_{\rm D})\,[\bar{u}(k_{\nu_{\rm
e}})\gamma^{\mu} (1-\gamma^5) v(k_{\rm e^+})]\,[\bar{u^c}(p_2)
\gamma^5 u(p_1)]\,e^{\textstyle i\sigma_0(k)}\,C_0(k)\nonumber\\
\hspace{-0.7in}&&\Big[ 1 - a(a+1) \frac{G_{\rm
\pi NN}}{2\pi^2}\,M^2_{\rm N} + \frac{C_{\rm NN}}{2\pi^2}\,(b^2 -
a\,(a + 1))\,k^2 - i\,\frac{C_{\rm NN}M_{\rm
N}}{2\pi}\,k\,C^2_0(k)\Big]^{-1}\!\!\!.
\end{eqnarray}
We have denoted
\begin{eqnarray}\label{label6.12}
e^{\textstyle - \pi/4k r_C}\,\Gamma\Big(1 + \frac{i}{2k r_C}\Big)
= e^{\textstyle i\sigma_0(k)}\,C_0(k)\;,\;
\sigma_0(k)&=&{\rm arg}\,\Gamma\Big(1 + \frac{i}{2k r_C}\Big),
\end{eqnarray}
where $\sigma_0(k)$ is a pure Coulomb phase shift.

Now, let us rewrite the denominator of the amplitude
 Eq.(\ref{label6.11}) in the equivalent form
\begin{eqnarray}\label{label6.13}
&& \Big\{\cos\sigma_0(k)\Big[1 - a(a+1) \frac{C_{\rm NN}}{2\pi^2}\,M^2_{\rm N} + \frac{C_{\rm NN}}{2\pi^2}\,(b^2 -
a\,(a + 1))\,k^2\Big]\nonumber\\
&& - \sin\sigma_0(k)\,\frac{C_{\rm NN}M_{\rm
N}}{2\pi}\,k\,C^2_0(k)\Big\}- i\,\Big\{\cos\sigma_0(k)\,\frac{G_{\rm
\pi NN}M_{\rm N}}{2\pi}\,k\,C^2_0(k)\nonumber\\
&&  + \sin\sigma_0(k)\Big[1 - a(a+1) \frac{C_{\rm NN}}{2\pi^2}\,M^2_{\rm N} + \frac{C_{\rm NN}}{2\pi^2}\,(b^2 -
a\,(a + 1))\,k^2\Big]\Big\}=\nonumber\\
&&= \frac{1}{{\cal Z}}\Big[1 -
\frac{1}{2}\,a^{\rm e}_{\rm pp} r^{\rm e}_{\rm pp}k^2 + \frac{a^{\rm
e}_{\rm pp}}{r_C}\,h(2 k r_C) + i\,a^{\rm e}_{\rm pp}\,k\,C^2_0(k)\Big],
\end{eqnarray}
where we have denoted
\begin{eqnarray}\label{label6.14}
\hspace{-0.3in}&& \frac{1}{{\cal Z}}\Big[1 - \frac{1}{2}\,a^{\rm e}_{\rm pp}
r^{\rm e}_{\rm pp}k^2 + \frac{a^{\rm e}_{\rm pp}}{r_C}\,h(2 k r_C)
\Big]= - \sin\sigma_0(k)\,\frac{C_{\rm NN}M_{\rm
N}}{2\pi}\,k\,C^2_0(k)\nonumber\\
\hspace{-0.3in}&& +  \cos\sigma_0(k)\Big[1 - a(a+1) \frac{G_{\rm
\pi NN}}{2\pi^2}\,M^2_{\rm N} + \frac{C_{\rm NN}}{2\pi^2}\,(b^2 -
a\,(a + 1))\,k^2\Big],\nonumber\\
\hspace{-0.3in}&& - \frac{1}{{\cal Z}}\,a^{\rm e}_{\rm
pp}\,k\,C^2_0(k) =\cos\sigma_0(k)\frac{C_{\rm NN}M_{\rm
N}}{2\pi}\,k\,C^2_0(k)\nonumber\\
\hspace{-0.3in}&& + \sin\sigma_0(k)\Big[1 - a(a+1) \frac{C_{\rm NN}}{2\pi^2}\,M^2_{\rm N} + \frac{C_{\rm NN}}{2\pi^2}\,(b^2 -
a\,(a + 1))\,k^2\Big].
\end{eqnarray}
Here ${\cal Z}$ is a constant which we would remove by the
renormalization of the wave functions of the protons, $a^{\rm e}_{\rm
pp} = ( - 7.8196\pm 0.0026)\,{\rm fm}$ and $r^{\rm e}_{\rm pp} =
2.790\pm 0.014\,{\rm fm}$ [44] are the S--wave scattering length and
the effective range of pp scattering in the ${^1}{\rm S}_0$ state
with the Coulomb repulsion, and $h(2 k r_C)$ is defined by [43]
\begin{eqnarray}\label{label6.15}
h(2 k r_C) = - \gamma + {\ell n}(2 k r_C) +
\sum^{\infty}_{n=1}\frac{1}{n(1 + 4n^2k^2r^2_C)}.
\end{eqnarray}
The validity of the relations Eq.(\ref{label6.14}) assumes the
dependence of parameters $a$ and $b$ on the relative momentum $k$.

After the changes Eq.(\ref{label6.13}) and Eq.(\ref{label6.14}) the
amplitude Eq.(\ref{label6.11}) takes the form
\begin{eqnarray}\label{label6.16}
\hspace{-0.2in}&& i{\cal M}({\rm p} + {\rm p} \to {\rm D} + 
{\rm e}^+ + \nu_{e}) =
G_{\rm V}\,g_{\rm A} M_{\rm N}\,C_{\rm NN}\,G_{\rm V}\,\frac{3g_{\rm
V}}{4\pi^2}\,e^*_{\mu}(k_{\rm D})\,[\bar{u}(k_{\nu_{\rm
e}})\gamma^{\mu} (1-\gamma^5) v(k_{\rm e^+})]\nonumber\\ \hspace{-0.2in}&&\times \,\frac{C_0(k)}{\displaystyle1 -
\frac{1}{2}\,a^{\rm e}_{\rm pp} r^{\rm e}_{\rm pp}k^2 + \frac{a^{\rm
e}_{\rm pp}}{r_C}\,h(2 k r_C) + i\,a^{\rm e}_{\rm pp}\,k\,C^2_0(k)
}\,{\cal Z}\,[\bar{u^c}(p_2) \gamma^5 u(p_1)].
\end{eqnarray}
Renormalizing the wave functions of the protons
$\sqrt{{\cal Z}}u(p_2) \to u(p_2)$ and
$\sqrt{{\cal Z}}u(p_1) \to u(p_1)$ we obtain the amplitude of
the solar proton burning 
\begin{eqnarray}\label{label6.17}
\hspace{-0.2in}&& i{\cal M}({\rm p} + {\rm p} \to {\rm D} + {\rm e}^+
+ \nu_{e}) = g_{\rm A} M_{\rm N}\,C_{\rm NN}\,G_{\rm V}\,\frac{3g_{\rm
V}}{4\pi^2}\,e^*_{\mu}(k_{\rm D})\,[\bar{u}(k_{\nu_{\rm
e}})\gamma^{\mu} (1-\gamma^5) v(k_{\rm e^+})]\nonumber\\
\hspace{-0.2in}&&\times \, \frac{\displaystyle C_0(k)}{\displaystyle1
- \frac{1}{2}\,a^{\rm e}_{\rm pp} r^{\rm e}_{\rm pp}k^2 + \frac{a^{\rm
e}_{\rm pp}}{r_C}\,h(2 k r_C) + i\,a^{\rm e}_{\rm pp}\,k\,C^2_0(k)
}\,[\bar{u^c}(p_2) \gamma^5 u(p_1)]\,F_{\rm D}(k^2),
\end{eqnarray}
where $F_{\rm D}(k^2)$ is given by Eq.(\ref{label2.16}) and describes
the spatial smearing of the deuteron coupled to the pp pair in the
${^1}{\rm S}_0$ state.

The real part of the denominator of the amplitude Eq.(\ref{label6.17})
is in complete agreement with a phenomenological relation [43]
\begin{eqnarray}\label{label6.18}
{\rm ctg}\delta^{\rm e}_{\rm pp}(k) = \frac{1}{\displaystyle
C^2_0(k)\,k}\,\Big[ - \frac{1}{a^{\rm e}_{\rm pp}} +  \frac{1}{2}\,r^{\rm
e}_{\rm pp}k^2 -  \frac{1}{r_{\rm C}}\,h(2 k r_{\rm C})\Big],
\end{eqnarray}
describing the phase shift $\delta^{\rm e}_{\rm pp}(k)$ of low--energy
elastic pp scattering in terms of the S--wave scattering length
$a^{\rm e}_{\rm pp}$ and the effective range $r^{\rm e}_{\rm
pp}$. Thus, we argue that the contribution of low--energy elastic pp
scattering to the amplitude of the solar proton burning is described
in agreement with low--energy nuclear phenomenology in terms of the
S--wave scattering length $a^{\rm e}_{\rm pp}$ and the effective range
$r^{\rm e}_{\rm pp}$ taken from the experimental data [44].

\section{The astrophysical factor for the  solar proton burning}
\setcounter{equation}{0}

\hspace{0.2in} The amplitude Eq.(\ref{label6.17}) squared, averaged
over polarizations of the protons and summed over polarizations of
final particles reads
\begin{eqnarray}\label{label7.1}
\hspace{-0.2in}&&\overline{|{\cal M}({\rm p} + {\rm p} \to {\rm D} +
{\rm e}^+ + \nu_{e})|^2} = \,G^2_{\rm V}\, g^2_{\rm A} M^6_{\rm
N}\,C^2_{\rm NN}\,\frac{54 Q_{\rm D}}{\pi^2}\,F^2_{\rm
D}(k^2)\nonumber\\
\hspace{-0.2in}&&\times\,\frac{\displaystyle C^2_0(k)}
{\displaystyle \Big[1 - \frac{1}{2}\,a^{\rm e}_{\rm pp}
r^{\rm e}_{\rm pp}k^2 + \frac{a^{\rm e}_{\rm pp}}{r_C}\,h(2 k r_C)\Big]^2 +
(a^{\rm e}_{\rm pp})^2 k^2 C^4_0(k)}\,\Big( E_{\rm e^+} E_{\nu_{\rm e}} -
\frac{1}{3}\vec{k}_{\rm e^+}\cdot \vec{k}_{\nu_{\rm e}}\Big),
\end{eqnarray}
where $m_{\rm e} = 0.511\,{\rm MeV}$ is the positron mass, and we
have used the relation $g^2_{\rm V} = 2\,\pi^2\,Q_{\rm D}\,M^2_{\rm
N}$.

The cross section for the reaction p + p $\to$ D + e$^+$ + $\nu_{\rm e}$ is defined by
\begin{eqnarray}\label{label7.2}
\hspace{-0.3in}\sigma^{\rm pp\to De^+\nu_{\rm e}}(T_{\rm pp})&=&\frac{1}{v}\,\frac{1}{4E_1E_2}
\int \overline{|{\cal M}({\rm p} + {\rm p} \to {\rm D} +
{\rm e}^+ + \nu_{e})|^2}\nonumber\\
\hspace{-0.3in}&\times&(2\pi)^4\,\delta^{(4)}(k_{\rm D} +
k_{\ell} - p_1 - p_2)\frac{d^3k_{\rm D}}{(2\pi)^3 2E_{\rm
D}}\frac{d^3k_{\rm e^+}}{(2\pi)^3 2E_{\rm e^+}}\frac{d^3k_{\nu_{\rm
e}}}{(2\pi)^3 2 E_{\nu_{\rm e}}},
\end{eqnarray}
where $v$ is a relative velocity of the pp pair and $k_{\ell} = k_{\rm
e^+} + k_{\nu_{\rm e}}$ is a 4--momentum of the leptonic pair.

The integration over the phase volume of the final ${\rm D}{\rm e}^+
\nu_{\rm e}$ state we perform in the non--relativistic limit
\begin{eqnarray}\label{label7.3}
\hspace{-0.5in}&&\int\frac{d^3k_{\rm D}}{(2\pi)^3 2E_{\rm
D}}\frac{d^3k_{\rm e^+}}{(2\pi)^3 2E_{\rm e^+}}\frac{d^3k_{\nu_{\rm
e}}}{(2\pi)^3 2 E_{\nu_{\rm e}}}\,(2\pi)^4\,\delta^{(4)}(k_{\rm D} +
k_{\ell} - p_1 - p_2)\,\Big( E_{\rm e^+} E_{\nu_{\rm e}} -
\frac{1}{3}\vec{k}_{\rm e^+}\cdot \vec{k}_{\nu_{\rm e}}\,\Big)\nonumber\\
\hspace{-0.5in}&&= \frac{1}{32\pi^3 M_{\rm N}}\,\int^{W + T_{\rm
pp}}_{m_{\rm e}}\sqrt{E^2_{\rm e^+}-m^2_{\rm e}}\,E_{\rm e^+}(W + T_{\rm
pp} - E_{\rm e^+})^2\,d E_{\rm e^+} = \frac{(W + T_{\rm pp})^5}{960\pi^3
M_{\rm N}}\,f(\xi),
\end{eqnarray}
where $W = \varepsilon_{\rm D} - (M_{\rm n} - M_{\rm p}) = (2.225
-1.293)\,{\rm MeV} = 0.932\,{\rm MeV}$ and $\xi = m_{\rm e}/(W +
T_{\rm pp})$. The function $f(\xi)$ is defined by the integral
\begin{eqnarray}\label{label7.4}
\hspace{-0.5in}f(\xi)&=&30\,\int^1_{\xi}\sqrt{x^2 -\xi^2}\,x\,(1-x)^2 dx=(1
- \frac{9}{2}\,\xi^2 - 4\,\xi^4)\,\sqrt{1-\xi^2}\nonumber\\
\hspace{-0.5in}&&+ \frac{15}{2}\,\xi^4\,{\ell
n}\Bigg(\frac{1+\sqrt{1-\xi^2}}{\xi}\Bigg)\Bigg|_{T_{\rm pp} = 0} =  0.222
\end{eqnarray}
and normalized to unity at $\xi = 0$.

Thus, the cross section for the solar proton burning is given by
\begin{eqnarray}\label{label7.5}
&&\sigma^{\rm pp\to De^+\nu_{\rm e}}(T_{\rm pp}) = \frac{e^{\displaystyle
- \pi/r_C\sqrt{M_{\rm N}T_{\rm pp}}}}{v^2}\,
\alpha\,\frac{9g^2_{\rm A} G^2_{\rm V} Q_{\rm D}
M^3_{\rm N}}{320\,\pi^4}\,C^2_{\rm NN}\,(W + T_{\rm
pp})^5\,f\Big(\frac{m_{\rm e}}{W + T_{\rm pp}}\Big)\nonumber\\
&&\times\,\frac{\displaystyle F^2_{\rm D}(M_{\rm N}T_{\rm pp})}{
\displaystyle \Big[1 - \frac{1}{2}\,a^{\rm e}_{\rm pp}
r^{\rm e}_{\rm pp}M_{\rm N}T_{\rm pp} + \frac{a^{\rm e}_{\rm
pp}}{r_C}\,h(2 r_C\sqrt{M_{\rm N}T_{\rm pp}})\Big]^2 + (a^{\rm e}_{\rm
pp})^2M_{\rm N}T_{\rm pp} C^4_0(\sqrt{M_{\rm N}T_{\rm pp}})}
\nonumber\\ &&\times\,\frac{1}{\displaystyle 1 - e^{\displaystyle
- \pi/r_C\sqrt{M_{\rm N}T_{\rm pp}}}} = \frac{S_{\rm pp}(T_{\rm pp})}{T_{\rm
pp}}\,e^{\displaystyle - \pi/r_C\sqrt{M_{\rm N}T_{\rm pp}}}.
\end{eqnarray}
The astrophysical factor $S_{\rm pp}(T_{\rm pp})$ reads
\begin{eqnarray}\label{label7.6}
\hspace{-0.5in}&&S_{\rm pp}(T_{\rm pp}) = 
\alpha\,\frac{9g^2_{\rm A}G^2_{\rm V}Q_{\rm
D}M^4_{\rm N}} {1280\pi^4}\,\frac{C^2_{\rm NN}\,(W + T_{\rm
pp})^5}{\displaystyle 1 - e^{\displaystyle
- \pi/r_C\sqrt{M_{\rm N}T_{\rm pp}}}}\,f\Big(\frac{m_{\rm e}}{W + T_{\rm pp}}\Big)\nonumber\\
\hspace{-0.5in}&&\times\, \frac{\displaystyle F^2_{\rm D}(M_{\rm
N}T_{\rm pp})} {\displaystyle
\Big[1 - \frac{1}{2}\,a^{\rm e}_{\rm pp} r^{\rm e}_{\rm pp}M_{\rm
N}T_{\rm pp} + \frac{a^{\rm e}_{\rm pp}}{r_C}\,h(2 r_C\sqrt{M_{\rm
N}T_{\rm pp}})\Big]^2 + (a^{\rm e}_{\rm pp})^2M_{\rm N}T_{\rm pp}
C^4_0(\sqrt{M_{\rm N}T_{\rm pp}})}.
\end{eqnarray}
At zero kinetic energy of the relative movement of the protons $T_{\rm
pp} = 0$ the astrophysical factor $S_{\rm pp}(0)$ is given by
\begin{eqnarray}\label{label7.7}
\hspace{-0.5in}S_{\rm pp}(0) =\alpha\,\frac{9g^2_{\rm A}G^2_{\rm V}Q_{\rm
D}M^4_{\rm N}}{1280\pi^4}\,C^2_{\rm NN}\,W^5\,f\Big(\frac{m_{\rm
e}}{W}\Big) =  4.08\,\times 10^{-25}\,{\rm MeV\,\rm b}.
\end{eqnarray}
The value $S_{\rm pp}(0) = 4.08 \times 10^{-25}\,{\rm MeV\,\rm b}$
agrees well with the recommended value $S_{\rm pp}(0) = 4.00 \times
10^{-25}\,{\rm MeV\,\rm b}$ [29].

Unlike the astrophysical factor obtained by Kamionkowski and Bahcall
[30] the astrophysical factor given by Eq.(\ref{label7.7}) does not
depend explicitly on the S--wave scattering length of low--energy
elastic pp scattering in the ${^1}{\rm S}_0$ state. This is due to the
normalization of the wave function of the pp pair. After the summation
of an infinite series and by using the relation Eq.(\ref{label6.18})
we obtain the wave function of the pp pair in the ${^1}{\rm S}_0$
state in the form
\begin{eqnarray}\label{label7.8}
\psi_{\rm pp}(k)= e^{\textstyle i\,\delta^{\,\rm e}_{\rm
pp}(k)}\,\frac{\sin\,\delta^{\,\rm e}_{\rm pp}(k)}{-a^{\rm e}_{\rm
pp}kC_0(k)},
\end{eqnarray}
that corresponds the normalization of the wave function of the
relative movement of the pp pair used by Schiavilla {\it et al.}
[31]. For the more detailed discussion of this problem we relegate
readers to the paper by Schiavilla {\it et al.} [31]\,\footnote{See the
last paragraph of Sect.\,3 and the first paragraph of Sect.\,5 of
Ref.[31].}.

\section{The reaction $\nu_{\rm e}$ + D $\to$ ${\rm e}^-$ + p + p}
\setcounter{equation}{0}

\hspace{0.2in} The evaluation of the amplitude of the reaction
$\nu_{\rm e}$ + D $\to$ ${\rm e}^-$ + p + p is analogous to that of
the amplitude of the solar proton burning. The result reads
\begin{eqnarray}\label{label8.1}
\hspace{-0.3in}&& i{\cal M}(\nu_{e} + {\rm D}\to {\rm e}^- +
{\rm p} +  {\rm p}) = g_{\rm A} M_{\rm N} \frac{G_{\rm
V}}{\sqrt{2}}\,\frac{3g_{\rm V}}{2\pi^2}\, C_{\rm NN}
\,e^*_{\mu}(k_{\rm D})\,[\bar{u}(k_{\rm e^-})\gamma^{\mu}(1-\gamma^5)
u(k_{\nu_{\rm e}})]\nonumber\\
\hspace{-0.3in} &&\times
\,\frac{C_0(k)}{\displaystyle1 - \frac{1}{2}\,a^{\rm e}_{\rm pp}
r^{\rm e}_{\rm pp}k^2 + \frac{a^{\rm e}_{\rm pp}}{r_C}\,h(2 k r_C) +
i\,a^{\rm e}_{\rm pp}\,k\,C^2_0(k) }\,[\bar{u}(p_2) \gamma^5
u^c(p_1)]\,F_{\rm D}(k^2).
\end{eqnarray}
The amplitude Eq.(\ref{label8.1}) squared, averaged over
polarizations of the deuteron and summed over polarizations of the
final particles reads
\begin{eqnarray}\label{label8.2}
\hspace{-0.5in}&&\overline{|{\cal M}(\nu_{\rm e} + {\rm D} \to {\rm
e}^- + {\rm p} + {\rm p})|^2} = S_{\rm
pp}(0)\,\frac{2^{12}5\pi^2}{\Omega_{\rm D e^+ \nu_{\rm
e}}}\,\frac{r_{\rm C}M^3_{\rm N}}{m^5_{\rm e}}\,F^2_{\rm
D}(k^2)\,F_+(Z, E_{\rm e^-})\nonumber\\
\hspace{-0.5in}&&\times \frac{\displaystyle C^2_0(k)}{\displaystyle
\Big[1 - \frac{1}{2}a^{\rm e}_{\rm pp} r^{\rm e}_{\rm pp} k^2 +
\frac{a^{\rm e}_{\rm pp}}{r_{\rm C}}\,h(2kr_{\rm C})\Big]^2 + (a^{\rm
e}_{\rm pp})^2k^2 C^4_0(k)}\, \Big( E_{{\rm e^-}}E_{{\nu}_{\rm e}} -
\frac{1}{3}\vec{k}_{{\rm e^-}} \cdot \vec{k}_{{\nu}_{\rm e}}\Big).
\end{eqnarray}
where $F_+(Z,E_{\rm e^-})$ is the Fermi function [45] describing the
Coulomb interaction of the electron with the nuclear system having a
charge $Z$. In the case of the reaction $\nu_{\rm e}$ + D $\to$ e$^-$
+ p + p we have $Z = 2$. At $\alpha^2 Z^2 \ll 1$ the Fermi function
$F_+(Z,E_{\rm e^-})$ reads [45]
\begin{eqnarray}\label{label8.3}
_+F(Z,E_{\rm e^-}) = \frac{2\pi \eta_{\rm e^-}}{\displaystyle 1 -
e^{\textstyle -2\pi \eta_{\rm e^-}}},
\end{eqnarray}
where $\eta_{\rm e^-} = Z \alpha/v_{\rm e^-} = Z \alpha E_{\rm
e^-}/\sqrt{E^2_{\rm e^-} -m^2_{\rm e^-} }$ and $v_{\rm e^-}$ is a
velocity of the electron.

For the evaluation of the r.h.s. of Eq.(\ref{label8.2}) we have also
used the expression for the astrophysical factor
\begin{eqnarray}\label{label8.4}
S_{\rm pp}(0) = \frac{9g^2_{\rm A}G^2_{\rm V}Q_{\rm D}M^3_{\rm
N}}{1280\pi^4r_{\rm C}}\,C^2_{\rm NN}\,m^5_{\rm e}\,
\Omega_{\rm D e^+ \nu_{\rm e}},
\end{eqnarray}
where $\Omega_{\rm D e^+ \nu_{\rm e}} = (W/m_{\rm e})^5 f(m_{\rm e}/W)
= 4.481$ at $W = 0.932\,{\rm MeV}$. The function $f(m_{\rm e}/W)$ is
defined by Eq.(\ref{label7.4}).

In the rest frame of the deuteron the cross section for the reaction
$\nu_{\rm e}$ + D $\to$ ${\rm e}^-$ + p + p is defined by
\begin{eqnarray}\label{label8.5}
&&\sigma^{\rm \nu_{\rm e} D\to e^-pp}(E_{\nu_{\rm e}}) = \frac{1}{4M_{\rm
D}E_{\nu_{\rm e}}}\int\,\overline{|{\cal M}(\nu_{\rm e} + {\rm D} \to
{\rm e}^- + {\rm p} + {\rm p})|^2}\nonumber\\
&&\frac{1}{2}\,(2\pi)^4\,\delta^{(4)}(k_{\rm D} + k_{\nu_{\rm e}} -
p_1 - p_2 - k_{\rm e^-})\, \frac{d^3p_1}{(2\pi)^3 2E_1}\frac{d^3
p_2}{(2\pi)^3 2E_2}\frac{d^3k_{{\rm e^-}}}{(2\pi)^3 2E_{{\rm e^-}}},
\end{eqnarray}
where $E_{\nu_{\rm e}}$, $E_1$, $E_2$ and $E_{{\rm e^-}}$ are the
energies of the neutrino, the protons and the electron. The integration over the
phase volume of the (${\rm p p e^-}$) state we perform in the
non--relativistic limit and in the rest frame of the deuteron,
\begin{eqnarray}\label{label8.6}
&&\frac{1}{2}\,\int\frac{d^3p_1}{(2\pi)^3 2E_1}\frac{d^3p_2}{(2\pi)^3
2E_2} \frac{d^3k_{\rm e}}{(2\pi)^3 2E_{\rm
e^-}}(2\pi)^4\,\delta^{(4)}(k_{\rm D} + k_{\nu_{\rm e}} - p_1 - p_2 -
k_{\rm e^-})\,\nonumber\\ &&{\displaystyle \frac{\displaystyle
C^2_0(\sqrt{M_{\rm N}T_{\rm pp}})\,F^2_{\rm D}(M_{\rm N}T_{\rm
pp})\,F_+(Z, E_{\rm e^-})}{\displaystyle \Big[1 - \frac{1}{2}a^{\rm
e}_{\rm pp} r^{\rm e}_{\rm pp}M_{\rm N}T_{\rm pp} + \frac{a^{\rm
e}_{\rm pp}}{r_{\rm C}}\, h(2 r_{\rm C}\sqrt{M_{\rm N}T_{\rm
pp}})\Big]^2 + (a^{\rm e}_{\rm pp})^2 M_{\rm N}T_{\rm pp}
C^4_0(\sqrt{M_{\rm N}T_{\rm pp}})}}\nonumber\\ &&\Big( E_{\rm e^-}
E_{\nu_{\rm e}} - \frac{1}{3} \vec{k}_{\rm e^-} \cdot
\vec{k}_{\nu_{\rm e}}\Big)\, =\frac{E_{\nu_{\rm e}}M^3_{\rm
N}}{128\pi^3}\,\Bigg(\frac{E_{\rm th}}{M_{\rm N}}
\Bigg)^{\!\!7/2}\Bigg(\frac{2 m_{\rm e}}{E_{\rm
th}}\Bigg)^{\!\!3/2}\frac{1}{E^2_{\rm th}}\nonumber\\ &&\int\!\!\!\int
dT_{\rm e^-} dT_{\rm pp}\delta(E_{\nu_{\rm e}}- E_{\rm th} - T_{\rm
e^-} - T_{\rm pp}) \sqrt{T_{\rm e^-}T_{\rm pp}}\Bigg(1 + \frac{T_{\rm
e^-}}{m_{\rm e}}\Bigg)\,{\displaystyle \sqrt{1 + \frac{T_{\rm e^-}}{2
m_{\rm e}}}}\nonumber\\ &&{\displaystyle \frac{\displaystyle
C^2_0(\sqrt{M_{\rm N}T_{\rm pp}}) \,F^2_{\rm D}(M_{\rm N}T_{\rm
pp})\,F_+(Z, E_{\rm e^-})}{\displaystyle \Big[1 - \frac{1}{2}a^{\rm
e}_{\rm pp} r^{\rm e}_{\rm pp}M_{\rm N}T_{\rm pp} + \frac{a^{\rm
e}_{\rm pp}}{r_{\rm C}}\,h(2 r_{\rm C}\sqrt{M_{\rm N}T_{\rm
pp}})\Big]^2 + (a^{\rm e}_{\rm pp})^2 M_{\rm N}T_{\rm pp}
C^4_0(\sqrt{M_{\rm N}T_{\rm pp}})}} \nonumber\\ &&= \frac{E_{\nu_{\rm
e}}M^3_{\rm N}}{128\pi^3} \,\Bigg(\frac{E_{\rm th}}{M_{\rm N}}
\Bigg)^{\!\!7/2}\Bigg(\frac{2 m_{\rm e}}{E_{\rm
th}}\Bigg)^{\!\!3/2}\,(y-1)^2\,\Omega_{\rm p p e^-}(y),
\end{eqnarray}
where $T_{\rm e^-}$ is the kinetic energy of the electron, $E_{\rm
th}$ is the neutrino energy threshold: $E_{\rm th}= \varepsilon_{\rm
D} + m_{\rm e} - (M_{\rm n} - M_{\rm p}) = (2.225 + 0.511 - 1.293) \,
{\rm MeV} = 1.443\,{\rm MeV}$. The function $\Omega_{\rm p p e^-}(y)$,
where $y=E_{\nu_{\rm e}}/E_{\rm th}$ and $E_{\rm th} =
\varepsilon_{\rm D} - (M_{\rm n} - M_{\rm p}) + m_{\rm e} =
1.443\,{\rm MeV}$, is determined by the integral
\begin{eqnarray}\label{label8.7}
\hspace{-0.5in}&&\Omega_{\rm p p e^-}(y) =  \int\limits^{1}_{0} dx
\sqrt{x (1 - x)} \Big(1 + \frac{E_{\rm th}}{m_{\rm
e}}(y-1)(1-x)\Big) \sqrt{1 + \frac{E_{\rm th}}{2 m_{\rm
e}}(y-1)(1-x)}\nonumber\\
\hspace{-0.5in}&&C^2_0(\sqrt{M_{\rm N}E_{\rm th}\,(y - 1)\,x})\,F^2_{\rm
D}(M_{\rm N}E_{\rm th}\,(y - 1)\,x)\,F_+(Z,m_{\rm e} + E_{\rm th}(y -
1)\,(1-x))\nonumber\\
\hspace{-0.5in}&&\Bigg\{\Bigg[1 - \frac{1}{2}a^{\rm e}_{\rm pp} r^{\rm
e}_{\rm pp}M_{\rm N}E_{\rm th}\,(y - 1)\,x + \frac{a^{\rm e}_{\rm
pp}}{r_{\rm C}}\,h(2 r_{\rm C}\sqrt{M_{\rm N}E_{\rm th}\,(y -
1)\,x})\Bigg]^2 \nonumber\\
\hspace{-0.5in}&&\hspace{0.2in} + (a^{\rm e}_{\rm pp})^2\,M_{\rm N}E_{\rm
th}\,(y - 1)\,x C^4_0(\sqrt{M_{\rm N}E_{\rm th}\,(y - 1)\,x})
\Bigg\}^{-1}\!\!\!\!\!,
\end{eqnarray}
where we have changed the variable $T_{\rm pp} = (E_{\nu_{\rm e}} -
E_{\rm th})\,x$.

The cross section for the reaction $\nu_{\rm e}$ + D $\to$ ${\rm e}^-$
+ p + p is defined by
\begin{eqnarray}\label{label8.8}
\hspace{-0.5in}\sigma^{\rm \nu_{\rm e} D\to e^-pp}(E_{\nu_{\rm e}}) &=&
S_{\rm pp}(0)\, \frac{640 r_{\rm C}}{\pi \Omega_{\rm D e^+\nu_{\rm
e}}}\Bigg(\frac{M_{\rm N}}{E_{\rm th}}\Bigg)^{3/2}\Bigg(\frac{E_{\rm
th}}{2m_{\rm e}}\Bigg)^{7/2}(y-1)^2\,\Omega_{\rm p p
e^-}(y)=\nonumber\\ 
&=&3.69\times 10^5\,S_{\rm
pp}(0)\,(y-1)^2\,\Omega_{\rm p p e^-}(y),
\end{eqnarray}
where $S_{\rm pp}(0)$ is measured in ${\rm MeV}\,{\rm cm}^2$. For
$S_{\rm pp}(0) = 4.08\times 10^{-49}\,{\rm MeV}\,{\rm cm}^2$
Eq.(\ref{label7.7}) the cross section $\sigma^{\rm \nu_{\rm e} D\to e^-pp}(E_{\nu_{\rm e}})$ reads
\begin{eqnarray}\label{label8.9}
\sigma^{\rm \nu_{\rm e} D\to e^-pp}(E_{\nu_{\rm e}}) =
1.50\,(y-1)^2\,\Omega_{\rm p p e^-}(y)\,10^{-43}\,{\rm cm}^2.
\end{eqnarray}
Recently the calculation of the cross section for the reaction
$\nu_{\rm e}$ + D $\to$ e$^-$ + p + p has been carried in Ref.\,[46]
within the PMA and tabulated for the neutrino energies ranging over
the region from threshold up to 170$\,{\rm MeV}$. Since our result is
valid for much lower neutrino energies, we give the numerical values
of the cross section only for the energies from the region $4\,{\rm MeV}
\le E_{\nu_{\rm e}} \le 10\,{\rm MeV}$:
\begin{eqnarray}\label{label8.10}
\sigma^{\rm \nu_{\rm e} D\to e^-pp}(E_{\nu_{\rm e}}= 4\,{\rm MeV}) &=&
2.46\,(1.577)\times 10^{-43}\,{\rm cm}^2,\nonumber\\
\sigma^{\rm \nu_{\rm e} D\to e^-pp}(E_{\nu_{\rm e}}= 6\,{\rm MeV}) &=&
10.04\,(6.239)\times 10^{-43}\,{\rm cm}^2,\nonumber\\
\sigma^{\rm \nu_{\rm e} D\to e^-pp}(E_{\nu_{\rm e}}= 8\,{\rm MeV}) &=&
2.37\,(1.463)\times 10^{-42}\,{\rm cm}^2,\nonumber\\
\sigma^{\rm \nu_{\rm e} D\to e^-pp}(E_{\nu_{\rm e}}= 10\,{\rm MeV}) &=&
4.39\,(2.708)\times 10^{-43}\,{\rm cm}^2.
\end{eqnarray}
The data in parentheses are taken from Table 1 of Ref.\,[46]. Thus, on
the average the cross section $\sigma^{\rm \nu_{\rm e} D\to
e^-pp}(E_{\nu_{\rm e}})$ calculated in the NNJL model by a factor of
1.6 is larger compared with the PMA ones.

Since the amplitude of the reaction $\nu_{\rm e}$ + D $\to$ e$^-$ + p
+ p is completely defined by the amplitude of the solar proton burning
p + p $\to$ D + e$^+$ + $\nu_{\rm e}$ that is described in the NNJL
model in agreement with other theoretical approaches, our prediction
for the cross section for the neutrino disintegration of the deuteron
Eq.(\ref{label8.10}) is a challenge to the experiments planned at SNO
[35].

In order to compare the cross section for the neutrino disintegration
of the deuteron $\nu_{\rm e}$ + D $\to$ e$^-$ + p + p with
experimental data planned to be obtained by SNO Collaboration we
should average it over the ${^8}{\rm B}$ solar neutrino energy
spectrum produced by the $\beta$ decay ${^8}{\rm B} \to {^8}{\rm
Be^*}$ + e$^+$ + $\nu_{\rm e}$ in the solar core. Using the ${^8}{\rm
B}$ solar neutrino energy spectrum [47] and integrating over the
region $5\,{\rm MeV} \le E_{\nu_{\rm e}} \le 15\,{\rm MeV}$ [48],
where the lower bound is related to the experimental threshold of
experiments at SNO and the upper one is defined by the kinematics of
the decay ${^8}{\rm B} \to {^8}{\rm Be^*}$ + e$^+$ + $\nu_{\rm e}$, we
get
\begin{eqnarray}\label{label8.11}
\langle \sigma^{\rm \nu_{\rm e} D\to e^-pp}(E_{\nu_{\rm
e}})\rangle_{\Phi({^8}{\rm B})} = 2.62\times 10^{-42}\,{\rm cm}^2.
\end{eqnarray}
For the comparison of the theoretical cross section with the
experimental one measured at SNO one should take into account a
possible decrease of the experimental value in the case of the
existence of neutrino flavour oscillations [27,48].

\section{The astrophysical factor for the pep process}
\setcounter{equation}{0}

\hspace{0.2in} In the NNJL model the amplitude of the reaction p +
e$^-$ + p $\to$ D + $\nu_{\rm e}$ or the pep--process is related to
the amplitude of the solar proton burning p + p $\to$ D + e$^+$ +
$\nu_{\rm e}$ Eq.(\ref{label6.17}) as well as to the amplitude of the
neutrino disintegration of the deuteron $\nu_{\rm e}$ + D $\to$ e$^-$
+ p + p. Indeed, the effective Lagrangians ${\cal L}^{\rm pe^-p \to
D\nu_{\rm e}}_{\rm eff}(x)$, ${\cal L}^{\rm pp \to De^+\nu_{\rm
e}}_{\rm eff}(x)$ and ${\cal L}^{\rm \nu_{\rm e}D \to e^-pp}_{\rm
eff}(x)$ are defined by the anomaly of the one--nucleon loop triangle
$AAV$--diagrams with two axial--vector $(A)$ and one vector $(V)$
vertices (see Appendix). The amplitude of the pep--process reads
\begin{eqnarray}\label{label9.1}
&& i{\cal M}({\rm p} + {\rm e}^- + {\rm p} \to {\rm D} + \nu_{e}) =
g_{\rm A} M_{\rm N}\,C_{\rm NN}\,G_{\rm V}\,\frac{3g_{\rm
V}}{4\pi^2}\,e^*_{\mu}(k_{\rm D})\,[\bar{u}(k_{\nu_{\rm
e}})\gamma^{\mu} (1-\gamma^5) u(k_{\rm e^-})]\,\nonumber\\ 
&&\times \,\frac{\displaystyle C_0(k)}{\displaystyle1 -
\frac{1}{2}\,a^{\rm e}_{\rm pp} r^{\rm e}_{\rm pp}k^2 + \frac{a^{\rm
e}_{\rm pp}}{r_C}\,h(2 k r_C) + i\,a^{\rm e}_{\rm pp}\,k\,C^2_0(k)
}\,[\bar{u^c}(p_2) \gamma^5 u(p_1)]\,F_{\rm D}(k^2).
\end{eqnarray}
The amplitude Eq.(\ref{label9.1}) squared, averaged and
summed over polarizations of the interacting particles is defined by
\begin{eqnarray}\label{label9.2}
&&\overline{|{\cal M}({\rm p} + {\rm e}^- + {\rm p} \to {\rm D} +
\nu_{e})|^2} = g^2_{\rm A} M^6_{\rm N}\,C^2_{\rm
NN}\,G^2_{\rm V}\,\frac{27 Q_{\rm D}}{\pi^2}\,F^2_{\rm D}(k^2)\,F_+(Z, E_{\rm
e^-})\nonumber\\ &&\times\,\frac{\displaystyle C^2_0(k)}
{\displaystyle \Big[1 - \frac{1}{2}\,a^{\rm e}_{\rm pp} r^{\rm e}_{\rm
pp}k^2 + \frac{a^{\rm e}_{\rm pp}}{r_C}\,h(2 k r_C)\Big]^2 + (a^{\rm
e}_{\rm pp})^2 k^2 C^4_0(k)}\,\Big( E_{\rm e^+} E_{\nu_{\rm e}} -
\frac{1}{3}\vec{k}_{\rm e^+}\cdot \vec{k}_{\nu_{\rm e}}\Big),
\end{eqnarray}
where $F_+(Z, E_{\rm e^-})$ is the Fermi function given by
Eq.(\ref{label8.3}).

At low energies the cross section $\sigma^{\rm pe^-p \to D\nu_{\rm
e}}(T_{\rm pp})$ for the pep--process can be determined as follows
[49]
\begin{eqnarray}\label{label9.3}
&&\sigma^{\rm pe^-p \to D\nu_{\rm e}}(T_{\rm pp}) =
\frac{1}{v}\frac{1}{4M^2_{\rm N}}\int \frac{d^3k_{\rm e^-}}{(2\pi)^3 2
E_{\rm e^-}}\,g\, n(\vec{k}_{\rm e^-})\int \overline{|{\cal M}({\rm p}
+ {\rm e}^- + {\rm p} \to {\rm D} + \nu_{\rm e})|^2}\nonumber\\
&&(2\pi)^4 \delta^{(4)}(k_{\rm D} + k_{\nu_{\rm e}} - p_1 - p_2 -
k_{\rm e^-}) \frac{d^3k_{\rm D}}{(2\pi)^3 2M_{\rm
D}}\frac{d^3k_{\nu_{\rm e}}}{(2\pi)^3 2E_{\nu_{\rm e}}},
\end{eqnarray}
where $g = 2$ is the number of the electron spin states and $v$ is a
relative velocity of the pp pair.  The electron distribution function
$n(\vec{k}_{\rm e^-})$ can be taken in the form [45]
\begin{eqnarray}\label{label9.4}
n(\vec{k}_{\rm e^-}) = e^{\displaystyle \bar{\nu} - T_{\rm e^-}/kT_c},
\end{eqnarray}
where $k = 8.617\times 10^{-11}\,{\rm MeV\,K^{-1}}$, $T_c$ is a
temperature of the core of the Sun.  The distribution function
$n(\vec{k}_{\rm e^-})$ is normalized by the condition
\begin{eqnarray}\label{label9.5}
g\int \frac{d^3k_{\rm e^-}}{(2\pi)^3}\,n(\vec{k}_{\rm e^-}) = n_{\rm e^-},
\end{eqnarray}
where $n_{\rm e^-}$ is the electron number density. From the
normalization condition Eq.(\ref{label9.5}) we derive
\begin{eqnarray}\label{label9.6}
e^{\displaystyle \bar{\nu}} = \frac{\displaystyle  4\,\pi^3\, n_{\rm
e^-}}{\displaystyle (2\pi\,m_{\rm e}\,kT_c)^{3/2}}.
\end{eqnarray}
The astrophysical factor $S_{\rm pep}(0)$ is then defined
by
\begin{eqnarray}\label{label9.7}
S_{\rm pep}(0) = S_{\rm pp}(0)\,\frac{15}{2\pi}\,\frac{1}{\Omega_{\rm D
e^+ \nu_{\rm e}}}\,\frac{1}{m^3_{\rm e}}\,\Bigg(\frac{E_{\rm
th}}{m_{\rm e}}\Bigg)^2\,e^{\displaystyle \bar{\nu}}\,\int d^3k_{\rm
e^-} \,e^{\displaystyle - T_{\rm e^-}/kT_c}\,F_+(Z, E_{\rm e^-}).
\end{eqnarray}
For the ratio $S_{\rm pep}(0)/S_{\rm pp}(0)$ we obtain
\begin{eqnarray}\label{label9.8}
\frac{S_{\rm pep}(0)}{S_{\rm pp}(0)} = \frac{2^{3/2}\pi^{5/2}}{f_{\rm
pp}(0)}\,\Bigg(\frac{\alpha Z n_{\rm e^-}}{m^3_{\rm
e}}\Bigg)\,\Bigg(\frac{E_{\rm th}}{m_{\rm
e}}\Bigg)^2\,\sqrt{\frac{m_{\rm e}}{k T_c}}\,I\Bigg(Z\sqrt{\frac{2
m_{\rm e}}{k T_c}}\Bigg).
\end{eqnarray}
We have set $f_{\rm pp}(0) = \Omega_{\rm D e^+ \nu_{\rm e}}/30 =
0.149$ [45] and the function $I(x)$ introduced by Bahcall and May [45]
reads
\begin{eqnarray}\label{label9.9}
I(x) = \int\limits^{\infty}_0 {\displaystyle \frac{\displaystyle
du\,e^{\displaystyle -u}}{\displaystyle 1 - e^{\displaystyle
-\pi\alpha\,x/\sqrt{u}}}}.
\end{eqnarray}
The relation between the astrophysical factors $S_{\rm pep}(0)$ and
$S_{\rm pp}(0)$ given by Eq.(\ref{label9.8}) is in complete agreement
with that obtained by Bahcall and May [45].

By virtue of the direct relation between the amplitudes of the
pep--process and the reaction for the neutrino disintegration of the
deuteron $\nu_{\rm e}$ + D $\to$ e$^-$ + p + p that we have in the
NNJL model the agreement with the result obtained by Bahcall and May
[45] is on favour of our predictions for the cross section for the
reaction $\nu_{\rm e}$ + D $\to$ e$^-$ + p + p.

\section{The reaction $\bar{\nu}_{\rm e}$ + D $\to$ e$^+$ + n + n}
\setcounter{equation}{0}

\hspace{0.2in} The effective Lagrangian ${\cal L}^{\rm \bar{\nu}_{\rm
e} D \to e^+ nn }_{\rm eff}(x)$ of the low--energy nuclear transition
$\bar{\nu}_{\rm e}$ + D $\to$ e$^+$ + n + n we evaluate by
analogy with ${\cal L}^{\rm pp \to D e^+\bar{\nu}_{\rm e}}_{\rm
eff}(x)$ (see Eq.({\rm A}.27) of Appendix) through the one--nucleon
loop exchanges and at leading order in the large $N_C$ expansion. The
effective Lagrangian ${\cal L}^{\rm \bar{\nu}_{\rm e}D \to e^+ nn
}_{\rm eff}(x)$ is defined by the anomaly of the one--nucleon
loop triangle $AAV$--diagram as well as the effective Lagrangian
${\cal L}^{\rm pp \to D e^+\nu_{\rm e}}_{\rm eff}(x)$.  The result
reads
\begin{eqnarray}\label{label10.1}
{\cal L}^{\rm \bar{\nu}_{\rm e}D \to e^+ nn }_{\rm eff}(x) = 
i g_{\rm A}M_{\rm N}
C_{\rm NN}\frac{G_{\rm V}}{\sqrt{2}}\frac{3g_{\rm
V}}{4\pi^2}\,D_{\mu}(x)\,[\bar{n}(x)\gamma^5
n^c(x)]\,[\bar{\psi}_{\nu_{\rm e}}(x)\gamma^{\mu}(1 - \gamma^5) \psi_{\rm
e}(x)].
\end{eqnarray}
The amplitude of the reaction $\bar{\nu}_{\rm e}$ + D $\to$ e$^+$ + n
+ n we obtain in the form
\begin{eqnarray}\label{label10.2}
\hspace{-0.5in}i{\cal M}(\bar{\nu}_{\rm e} + {\rm D} \to {\rm e}^+ +
{\rm n} + {\rm n}) &=& -\, g_{\rm A} M_{\rm N}\,C_{\rm
NN}\,\frac{G_{\rm V}}{\sqrt{2}}\,\frac{3g_{\rm
V}}{2\pi^2}\,\frac{1}{\displaystyle 1 - \frac{1}{2}r_{\rm nn}a_{\rm
nn}k^2 + i\,a_{\rm nn}\,k}\,F_{\rm D}(k^2) \nonumber\\ &&\times
e_{\mu}(k_{\rm D})\,[\bar{v}(k_{\bar{\nu}_{\rm
e}})\gamma^{\mu}(1-\gamma^5) v(k_{{\rm e}^+})]\,[\bar{u}(p_2) \gamma^5
u^c(p_1)],
\end{eqnarray}
where the form factor $F_{\rm D}(k^2)$ describes a spatial smearing of
the deuteron Eq.(\ref{label2.16}).  The factor $1/(1 -
\frac{1}{2}r_{\rm nn}a_{\rm nn}k^2 + i\,a_{\rm nn}\,k)$ gives the
contribution of low--energy nuclear forces to the relative movement of
the nn pair in the ${^1}{\rm S}_0$ state. The result is obtained by
the summation of an infinite series of the one--neutron bubbles
evaluated at leading order in the large $N_C$ expansion. Since we
work in the isotopical limit, we set $a_{\rm nn} = a_{\rm np} = -
23.75\,{\rm fm}$ and $r_{\rm nn} = r_{\rm np} = 2.75\,{\rm fm}$. The
recent experimental values of the S--wave scattering length and the
effective range of low--energy elastic nn scattering are equal to
$a_{\rm nn}=( - 18.8\pm 0.3)\,{\rm fm}$ and $r_{\rm nn} = (2.75\pm
0.11)\,{\rm fm}$ [50,51].

The amplitude Eq.\,(\ref{label10.2}),
squared, averaged over polarizations of the deuteron and summed over
polarizations of the final particles, reads
\begin{eqnarray}\label{label10.3}
\hspace{-0.2in}\overline{|{\cal M}(\bar{\nu}_{\rm e} + {\rm D} \to
{\rm e}^+ + {\rm n} + {\rm n})|^2} =\frac{144}{\pi^2}\,\frac{Q_{\rm
D}g^2_{\rm A}G^2_{\rm V}C^2_{\rm NN}M^6_{\rm N}\,F^2_{\rm
D}(k^2)}{\displaystyle \Big(1 - \frac{1}{2}r_{\rm nn} a_{\rm nn}
k^2\Big)^2 + a^2_{\rm nn}k^2} \Big( E_{{\rm e}^+} E_{\bar{\nu}_{\rm
e}} - \frac{1}{3}\vec{k}_{{\rm e}^+}\cdot \vec{k}_{\bar{\nu}_{\rm
e}}\Big).
\end{eqnarray}
The cross section for the reaction $\bar{\nu}_{\rm e}$ + D $\to$ e$^+$
+ n + n is defined by
\begin{eqnarray}\label{label10.4}
&&\sigma^{\rm \bar{\nu}_{\rm e} D \to e^+nn}(E_{\bar{\nu}_{\rm e}}) =
\frac{1}{4E_{\rm D}E_{\bar{\nu}_{\rm e}}}\int\,\overline{|{\cal
M}(\bar{\nu}_{\rm e} +
{\rm D} \to  {\rm e}^+ + {\rm n} + {\rm n})|^2}\nonumber\\
&&\frac{1}{2}\,(2\pi)^4\,\delta^{(4)}(Q + k_{{\bar{\nu}_{\rm e}}} - p_1 -
p_2 - k_{{\rm e}^+})\,
\frac{d^3p_1}{(2\pi)^3 2E_1}\frac{d^3 p_2}{(2\pi)^3 2E_2}\frac{d^3k_{{\rm
e}^+}}{(2\pi)^3
2E_{{\rm e}^+}},
\end{eqnarray}
where $E_{\rm D}$, $E_{\bar{\nu}_{\rm e}}$, $E_1$, $E_2$ and $E_{{\rm
e}^+}$ are the energies of the deuteron, the anti--neutrino, the
neutrons and the positron. The integration over the phase volume of
the (${\rm n n e^+}$) state we perform in the non--relativistic limit
and in the rest frame of the deuteron
\begin{eqnarray}\label{label10.5}
&&\frac{1}{2}\,\int\frac{d^3p_1}{(2\pi)^3 2E_1}\frac{d^3p_2}{(2\pi)^3
2E_2} \frac{d^3k_{{\rm e}^+}}{(2\pi)^3 2E_{{\rm
e}^+}}(2\pi)^4\,\delta^{(4)}(Q + k_{{\bar{\nu}_{\rm e}}} - p_1 - p_2 -
k_{{\rm e}^+})\,\nonumber\\ &&\times\, \frac{\displaystyle \Big(
E_{{\rm e}^+} E_{\bar{\nu}_{\rm e}} - \frac{1}{3} \vec{k}_{{\rm
e}^+}\cdot \vec{k}_{\bar{\nu}_{\rm e}}\Big)\,F^2_{\rm D}(M_{\rm N}\,T_{\rm
nn})}{\displaystyle \Big(1 -
\frac{1}{2} r_{\rm nn} a_{\rm nn} M_{\rm N}T_{\rm nn}\Big)^2 +
a^2_{\rm nn}M_{\rm N}\,T_{\rm nn}} = \frac{E_{\bar{\nu}_{\rm e}}M^3_{\rm
N}}{1024\pi^2}\,\Bigg(\frac{E_{\rm th}}{M_{\rm N}}
\Bigg)^{\!\!7/2}\Bigg(\frac{2 m_{\rm e}}{E_{\rm
th}}\Bigg)^{\!\!3/2}\frac{8}{\pi E^2_{\rm th}} \nonumber\\ &&\times
\int\!\!\!\int dT_{\rm e^+} dT_{\rm nn}\, \frac{\displaystyle
\sqrt{T_{\rm e^+}T_{\rm nn}}\,F^2_{\rm D}(M_{\rm N}\,T_{\rm
nn})}{\displaystyle \Big(1 -
\frac{1}{2} r_{\rm nn} a_{\rm nn} M_{\rm N}T_{\rm nn}\Big)^2 +
a^2_{\rm nn}M_{\rm N}\,T_{\rm nn}}\,\nonumber\\
&&\times\,\Bigg(1 + \frac{T_{\rm e^+}}{m_{\rm e}}
\Bigg)\,{\displaystyle \sqrt{1 + \frac{T_{\rm e^+}}{2 m_{\rm e}} }}\,
\delta\Big(E_{\bar{\nu}_{\rm e}}- E_{\rm th} - T_{\rm e^+} - T_{\rm
nn}\Big) = \nonumber\\ 
&&= \frac{E_{\bar{\nu}_{\rm e}}M^3_{\rm
N}}{1024\pi^2}\,\Bigg(\frac{E_{\rm th}}{M_{\rm N}}
\Bigg)^{\!\!7/2}\Bigg(\frac{2 m_{\rm e}}{E_{\rm
th}}\Bigg)^{\!\!3/2}\,(y -1)^2\,\Omega_{\rm nn\bar{\nu}_{\rm e}}(y),
\end{eqnarray}
where $T_{\rm nn}$ is the kinetic energy of the nn pair in the
${^1}{\rm S}_0$ state, $T_{\rm e^+}$ and $m_{\rm e} = 0.511\,{\rm
MeV}$ are the kinetic energy and the mass of the positron, $y =
E_{\bar{\nu}_{\rm e}}/E_{\rm th}$ and $E_{\rm th}$ is the
anti--neutrino energy threshold of the reaction $\bar{\nu}_{\rm e}$ +
D $\to$ e$^+$ + n + n: $E_{\rm th}= \varepsilon_{\rm D} + m_{\rm e} +
(M_{\rm n} - M_{\rm p}) = (2.225 + 0.511 + 1.293) \, {\rm MeV} =
4.029\,{\rm MeV}$. The function $\Omega_{\rm nn\bar{\nu}_{\rm e}}(y)$
is defined by
\begin{eqnarray}\label{label10.6}
\Omega_{\rm nn\bar{\nu}_{\rm e}}(y) &=&
\frac{8}{\pi}\,\int\limits^{1}_{0} dx\, \frac{\sqrt{x\,(1 -
x)}\,F^2_{\rm D}(M_{\rm N}E_{\rm th}\,(y - 1)\,x)}{\displaystyle
\Big(1 - \frac{1}{2}r_{\rm nn} a_{\rm nn}M_{\rm N}E_{\rm th}\,(y -
1)\,x \Big)^2 + a^2_{\rm nn}M_{\rm N}E_{\rm th}\,(y -
1)\,x}\nonumber\\ &&\times\,\Big(1 + \frac{E_{\rm th}}{m_{\rm
e}}(y-1)(1-x)\Big) \,\sqrt{1 + \frac{E_{\rm th}} {2 m_{\rm
e}}(y-1)(1-x)},
\end{eqnarray}
where we have changed the variable $T_{\rm nn} = (E_{\bar{\nu}_{\rm
e}} - E_{\rm th})\,x$.  The function $f(y)$ is normalized to unity at
$y=1$, i.e.  at threshold $E_{\bar{\nu}_{\rm e}} = E_{\rm th}$. Thus,
the cross section for the reaction $\bar{\nu}_{\rm e}$ + D $\to$ e$^+$
+ n + n reads
\begin{eqnarray}\label{label10.7}
\sigma^{\rm \bar{\nu}_{\rm e} D \to e^+nn}(E_{\bar{\nu}_{\rm e}}) =
\sigma_0\,(y - 1)^2\,\Omega_{\rm nn\bar{\nu}_{\rm e}}(y),
\end{eqnarray}
where $\sigma_0$ amounts to
\begin{eqnarray}\label{label10.8}
\hspace{-0.5in}\sigma_0 = Q_{\rm D}\,C^2_{\rm NN}\,\frac{9g^2_{\rm A}
G^2_{\rm V} M^8_{\rm N}}{512\pi^4}\,\Bigg(\frac{E_{\rm th}}{M_{\rm
N}}\Bigg)^{\!\!7/2}\Bigg(\frac{2 m_{\rm e}}{E_{\rm th}}\Bigg)^{\!\!3/2}
= 4.58\,\times \,10^{-43}\,{\rm cm}^2.
\end{eqnarray}
The experimental data on the anti--neutrino disintegration of the
deuteron are given in terms of the cross section averaged over the
anti--neutrino energy spectrum [36]:
\begin{eqnarray}\label{label10.9}
 \langle\sigma^{\rm \bar{\nu}_{\rm e} D \to e^+nn}(E_{\bar{\nu}_{\rm
e}})\rangle_{\exp} = (9.83\pm 2.04)\times 10^{-45}\,{\rm cm}^2.
\end{eqnarray}
In order to average the theoretical cross section Eq.(\ref{label10.7})
over the anti--neutrino spectrum we should use the spectrum given by
Table YII of Ref.\,[36].  This yields
\begin{eqnarray}\label{label10.10}
\langle\sigma^{\rm \bar{\nu}_{\rm e} D \to e^+nn}(E_{\bar{\nu}_{\rm e}})\rangle = 
11.56\times 10^{-45}\,{\rm cm}^2.
\end{eqnarray}
The theoretical value Eq.(\ref{label10.10})\footnote{This result is
obtained at zero contribution of the nucleon tensor current (see
Appendix  and the discussion in the Conclusion).} agrees well with
the experimental one Eq.(\ref{label10.9}).

\section{The reactions $\nu_{\rm e}(\bar{\nu}_{\rm e})$ + D
$\to$ $\nu_{\rm e}(\bar{\nu}_{\rm e})$ + n + p} 
\setcounter{equation}{0}

\hspace{0.2in} The amplitude of the neutrino disintegration of the
deuteron caused by neutral weak current $\nu_{\rm e}$ + D $\to$
$\nu_{\rm e}$ + n + p evaluated through one--nucleon loop exchanges
and at leading order in the large $N_C$ expansion reads (see Eq.({\rm
A}.30) of Appendix):
\begin{eqnarray}\label{label11.1}
&&i{\cal M}(\nu_{\rm e} + {\rm D} \to \nu_{\rm e} + {\rm n} + {\rm p})
= -\,g_{\rm A}\, M_{\rm N}\,C_{\rm NN}\,\frac{G_{\rm
F}}{\sqrt{2}}\,\frac{3g_{\rm V}}{4\pi^2}\, \frac{1}{\displaystyle 1 - \frac{1}{2}r_{\rm np}a_{\rm np}k^2 +
i\,a_{\rm np}\,k}\,F_{\rm
D}(k^2)\nonumber\\ &&\times\,e_{\mu}(k_{\rm
D})\,[\bar{u}(k^{\prime}_{\nu_{\rm e}})\gamma^{\mu}(1-\gamma^5)
u(k_{\nu_{\rm e}})] [\bar{u}(p_2) \gamma^5 u^c(p_1)].
\end{eqnarray}
The amplitude Eq.(\ref{label11.1}) squared, averaged over
polarizations of the deuteron, summed over polarizations of the
nucleons reads
\begin{eqnarray}\label{label11.2}
\overline{|{\cal M}(\nu_{\rm e} + {\rm D} \to \nu_{\rm e} + {\rm n}
+{\rm p})|^2} = \frac{36 }{\pi^2}\, \frac{Q_{\rm D} g^2_{\rm
A}G^2_{\rm F}C^2_{\rm NN} M^6_{\rm N}F^2_{\rm D}(k^2)}{\displaystyle
\Big(1 - \frac{1}{2} r_{\rm np} a_{\rm np} k^2 \Big)^2 + a^2_{\rm
np}k^2 }\Big( E^{\prime}_{\nu_{\rm e}} E_{\nu_{\rm e}} - \frac{1}{3}
\vec{k}^{\prime}_{\nu_{\rm e}} \cdot \vec{k}_{\nu_{\rm e}}\Big).
\end{eqnarray}
In the rest frame of the deuteron the cross section for the reaction
$\nu_{\rm e}$ + D $\to$ $\nu_{\rm e}$ + n + p is defined by
\begin{eqnarray}\label{label11.3}
&&\sigma^{\rm \nu_{\rm e} D \to \nu_{\rm e}np}(E_{\bar{\nu}_{\rm e}}) =
\frac{1}{4M_{\rm D}E_{\nu_{\rm e}}}\int\,\overline{|{\cal
M}(\nu_{\rm e} +
{\rm D} \to \nu_{\rm e} + {\rm n} + {\rm p})|^2}\nonumber\\
&&(2\pi)^4\,\delta^{(4)}(k_{\rm D} + k_{\nu_{\rm e}} - p_1 -
p_2 - k^{\prime}_{\nu_{\rm e}})\,
\frac{d^3p_1}{(2\pi)^3 2E_1}\frac{d^3 p_2}{(2\pi)^3
2E_2}\frac{d^3k^{\prime}_{\nu_{\rm e}}}{(2\pi)^3
2E^{\prime}_{\nu_{\rm e}}}.
\end{eqnarray}
The integration over the phase volume of the (${\rm n p \nu_{\rm
e}}$) state we perform in the non--relativistic limit and in the rest
frame of the deuteron,
\begin{eqnarray}\label{label11.4}
&&\int\frac{d^3p_1}{(2\pi)^3 2E_1}\frac{d^3p_2}{(2\pi)^3 2E_2}
\frac{d^3k^{\prime}_{\nu_{\rm e}}}{(2\pi)^3 2E^{\prime}_{\nu_{\rm
e}}}(2\pi)^4\,\delta^{(4)}(k_{\rm D} + k_{\nu_{\rm e}} - p_1 - p_2 -
k^{\prime}_{\nu_{\rm e}})\,\nonumber\\
&&\Big( E_{\nu_{\rm e}} E^{\prime}_{\nu_{\rm e}} - \frac{1}{3}
\vec{k}_{\nu_{\rm e}}\cdot
\vec{k}^{\prime}_{\nu_{\rm e}}\Big)\,\frac{F^2_{\rm D}(M_{\rm N}T_{\rm
np})}{\displaystyle \Big(1 - \frac{1}{2}r_{\rm np}a_{\rm np} M_{\rm
N}T_{\rm np} \Big)^2 + a^2_{\rm np}M_{\rm N}T_{\rm np}} = \nonumber\\
&&=\frac{E_{\nu_{\rm e}}M^3_{\rm N}}{210\pi^3}\,\Bigg(\frac{E_{\rm
th}}{M_{\rm N}}\Bigg)^{\!\!7/2}\,(y - 1)^{7/2}\,\Omega_{\rm np\nu_{\rm e}}(y).
\end{eqnarray}
The function $\Omega_{\rm np\nu_{\rm e}}(y)$, where
$y=E_{\bar{\nu}_{\rm e}}/E_{\rm th}$ and $E_{\rm th}= \varepsilon_{\rm
D}=2.225\,{\rm MeV}$ is threshold of the reaction, is defined by the
integral
\begin{eqnarray}\label{label11.5}
\Omega_{\rm np\nu_{\rm e}}(y) = \frac{105}{16}\,\int\limits^{1}_{0} dx
\frac{\sqrt{x}\,(1 - x)^2}{\displaystyle \Big(1 -
\frac{1}{2}\frac{r_{\rm np}a_{\rm np} }{r^2_{\rm D}}(y-1) x \Big)^2 +
\frac{a^2_{\rm np}}{r^2_{\rm D}}(y-1)\, x}\frac{1}{(1 + (y - 1)\,
x)^2},
\end{eqnarray}
where we have changed the variable $T_{\rm np} = (E_{\bar{\nu}_{\rm
e}} - E_{\rm th})\,x$ and used the relation $M_{\rm N}E_{\rm th} =
1/r^2_{\rm D}$ at $E_{\rm th}=\varepsilon_{\rm D}$. The function
$\Omega_{\rm np\nu_{\rm e}}(y)$ is normalized to unity at $y=1$, i.e. 
at threshold $E_{\bar{\nu}_{\rm e}} = E_{\rm th}$.

The cross section for the reaction $\nu_{\rm e}$ + D $\to$ $\nu_{\rm
e}$ + n + p reads
\begin{eqnarray}\label{label11.6}
\sigma^{\rm \nu_{\rm e} D \to \nu_{\rm e}np}(E_{\nu_{\rm e}}) = \sigma_0\,(y -
1)^{7/2}\,\Omega_{\rm np\nu_{\rm e}}(y),
\end{eqnarray}
where $\sigma_0$ amounts to
\begin{eqnarray}\label{label11.7}
\sigma_0 = Q_{\rm D}\,C^2_{\rm NN}\,\frac{3 g^2_{\rm A}\,G^2_{\rm
F}  M^8_{\rm N}}{140\pi^5}\,
\Bigg(\frac{E_{\rm th}}{M_{\rm N}}\Bigg)^{\!\!7/2}= 1.84 \times
10^{-43}\,{\rm cm}^2.
\end{eqnarray}
In our approach the cross section for the disintegration of the
deuteron by neutrinos $\nu_{\rm e}$ + D $\to$ $\nu_{\rm e}$ + n + p
coincides with the cross section for the disintegration of the
deuteron by anti--neutrinos $\bar{\nu}_{\rm e}$ + D $\to$
$\bar{\nu}_{\rm e}$ + n + p, $\sigma^{\rm \nu_{\rm e} D \to \nu_{\rm
e}np}(E_{\nu_{\rm e}}) = \sigma^{\rm \bar{\nu}_{\rm e}D \to
\bar{\nu}_{\rm e}np}(E_{\bar{\nu}_{\rm e}})$. Therefore, we have an
opportunity to compare our result with the experimental data on the
disintegration of the deuteron by anti--neutrinos [36]. The
experimental value of the cross section for the anti--neutrino
disintegration of the deuteron $\bar{\nu}_{\rm e}$ + D $\to$
$\bar{\nu}_{\rm e}$ + n + p averaged over the anti--neutrino spectrum
reads [36]:
\begin{eqnarray}\label{label11.8}
\langle\sigma^{\rm \bar{\nu}_{\rm e}D \to \bar{\nu}_{\rm
e}np}(E_{\bar{\nu}_{\rm e}})\rangle_{\exp}= (6.08\pm 0.77)\times
10^{-45}\,{\rm cm}^2.
\end{eqnarray}
By using the anti--neutrino spectrum given by Table YII of Ref.\,[36]
for the calculation of the average value of the theoretical cross
section Eq.(\ref{label11.6}) we obtain
\begin{eqnarray}\label{label11.9}
\langle\sigma^{\rm \bar{\nu}_{\rm e}D \to \bar{\nu}_{\rm
e}np}(E_{\bar{\nu}_{\rm e}})\rangle = 
6.28\times 10^{-45}\,{\rm cm}^2.
\end{eqnarray}
The theoretical value Eq.(\ref{label11.9})\footnote{This result is
obtained at zero contribution of the nucleon tensor current (see
Appendix and the discussion in the Conclusion).} agrees well with
the experimental one Eq.(\ref{label11.8}).

The cross section for the neutrino disintegration of the deuteron
$\nu_{\rm e}$ + D $\to$ $\nu_{\rm e}$ + n + p averaged over the
${^8}{\rm B}$ solar neutrino spectrum [47] for energy region $5\,{\rm
MeV} \le E_{\nu_{\rm e}}\le 15\,{\rm MeV}$ is given by
\begin{eqnarray}\label{label11.10}
\langle \sigma^{\rm \nu_{\rm e} D\to \nu_{\rm e}np}(E_{\nu_{\rm
e}})\rangle_{\Phi({^8}{\rm B})} = 1.85\times 10^{-43}\,{\rm cm}^2.
\end{eqnarray}
This result can be directly compared with the experimental data
obtained by SNO, since the averaged value for the cross section for
the reaction $\nu_{\rm e}$ + D $\to$ $\nu_{\rm e}$ + n + p caused by
the neutral weak current should not depend on whether neutrino
flavours oscillate or not [27,48]. Of course, if none {\it sterile}
neutrinos exist [27,48].

\section{Conclusion}
\setcounter{equation}{0}

\hspace{0.2in} We have considered the description of a dynamics of
low--energy nuclear forces within the Nambu--Jona--Lasinio model of
light nuclei (the NNJL model) for low--energy electromagnetic and weak
nuclear reactions with the deuteron. We have shown that the NNJL model
enables to describe in a reasonable agreement with both experimental
data and other theoretical approaches all variety of low--energy
electromagnetic and weak nuclear reactions with the deuteron in the
final or initial state coupled to a nucleon--nucleon (NN) pair in the
${^1}{\rm S}_0$ state. In the bulk the reaction rates for the
neutron--proton radiative capture n + p $\to$ D + $\gamma$ for thermal
neutrons, the photomagnetic disintegration of the deuteron $\gamma$ +
D $\to$ n + p, the solar proton burning p + p $\to$ D + e$^+$ +
$\nu_{\rm e}$, the pep--process p + e$^-$ + p $\to$ D + $\nu_{\rm e}$,
the disintegration of the deuteron by neutrinos and anti--neutrinos
caused by charged $\nu_{\rm e}$ + D $\to$ e$^-$ + p + p and
$\bar{\nu}_{\rm e}$ + D $\to$ e$^+$ + n + n and neutral $\nu_{\rm
e}(\bar{\nu}_{\rm e})$ + D $\to$ $\nu_{\rm e}(\bar{\nu}_{\rm e})$ + n
+ p weak currents are calculated in agreement with other theorical
approaches and experimental data.

 When matching our results with those obtained in the PMA we would
like to emphasize a much more simple description of the NN interaction
responsible for the transition N + N $\to$ N + N and a substantial
simplification of the evaluation of matrix elements of low--energy
nuclear transitions near thresholds of the reactions where the NN pair
in the ${^1}{\rm S}_0$ state is produced or absorbed with a relative
momentum $p$ comparable with zero. Such a simplification is rather
clear in the NNJL model where with a good accuracy the deuteron can be
considered within a quantum field theoretic approach as a point--like
particle. Indeed, the spatial region of the localization of the NN
pair is of order of $O(1/p)$.  Near thresholds the effective radius of
the deuteron $r_{\rm D}= 4.319\,{\rm MeV}$ is much smaller than $1/p$,
$r_{\rm D} \ll 1/p$. This yields that the NN pair does not feel the
spatial smearing of the deuteron and couples to the deuteron as to a
point--like particle. A correct description of strong interactions of
the point--like deuteron coupled to the NN pair in the ${^1}{\rm S}_0$
state is guaranteed then by the one--nucleon loop exchanges with
dominant contributions of the nucleon--loop anomalies. This implies
that the effective Lagrangians ${\cal L}^{\rm np \to D\gamma}_{\rm
eff}(x)$, ${\cal L}^{\rm pp \to De^+\nu_{\rm e}}_{\rm eff}(x)$, ${\cal
L}^{\rm \nu_{\rm e}D \to e^-pp}_{\rm eff}(x)$, ${\cal L}^{\rm pe^- p
\to D\nu_{\rm e}}_{\rm eff}(x)$, ${\cal L}^{\rm \bar{\nu}_{\rm e}D \to
e^+nn}_{\rm eff}(x)$ and ${\cal L}^{\rm \bar{\nu}_{\rm e}D
\to\bar{\nu}_{\rm e} np}_{\rm eff}(x)$ are well defined.  Thus, the
procedure of the derivation of effective Lagrangians of low--energy
nuclear transitions in the NNJL model treating the deuteron as a
point--like particle coupled to nucleons and other particles through
one--nucleon loop exchanges seems to be good established. We argue
that the application of this procedure should get correct results for
the derivation of effective Lagrangians of any other low--energy
nuclear transitions, effective vertices, of nuclear reactions.

Some problems occur for the evaluation of the amplitudes of nuclear
reactions demanding the continuation of matrix elements of low--energy
nuclear transitions defined by the effective
Lagrangians to the energy region far from thresholds.

The continuation of matrix elements of low--energy nuclear transitions
demands in the NNJL model: 1) the spatial smearing of the NN pair in
the ${^1}{\rm S}_0$ state and 2) the spatial smearing of the deuteron
caused by the finite value of the effective radius $r_{\rm D}$.  The
spatial smearing of the NN pair in the ${^1}{\rm S}_0$ state can be
carried out by a summation of an infinite series of one--nucleon
bubbles describing rescattering or differently a relative movement of
the NN pair either in an initial or a final state of a nuclear
reaction. The result of the NN rescattering can be expressed in terms
of S--wave scattering lengths and effective ranges in complete
agreement with nuclear phenomenology of low--energy elastic NN
scattering in the ${^1}{\rm S}_0$ state. However, for the description
of the spatial smearing of the deuteron the abilities of the NNJL
model are rescticted and most what one can do at present level of the
development of the model is to introduce the spatial smearing of the
deuteron phenomenologically in the form of the Fourier transform of
the approximate ${^3}{\rm S}_1$ wave function of the deuteron
normalized to unity at $p=0$: $F_{\rm D}(p^2) = 1/(1 + r^2_{\rm D}
p^2)$. We have chosen a simplest form of the spatial smearing of the
deuteron. Of course, $F_{\rm D}(p^2)$ can be taken in the more
complicated form of the Fourier transform of the explicit wave
function of the deuteron. Of course, such a dependence is not absolute
and the spatial smearing of the deuteron can be taken into account in
the form of phenomenological form factors as it has been done by Mintz
[52], for example.

We would like to emphasize that the cross section for the M1--capture
n + p $\to$ D + $\gamma$ for thermal neutrons has been calculated by
accounting for the contributions of chiral one--meson loop corrections
and the $\Delta(1232)$ resonance. The total cross section for the
M1--capture has been found dependent on the parameter $Z$ defining the
${\rm \pi N \Delta}$ coupling constant off--mass shell of the
$\Delta(1232)$ resonance: $\sigma({\rm np \to D\gamma})(T_{\rm n})=
325.5\,(1 + 0.246\,(1-2Z))^2\,{\rm m b}$ Eq.(\ref{label4.15}). In
order to fit the experimental value of the cross section we should set
$Z=0.473$. This agrees with the experimental bound $|Z|\le 1/2$
[18]. At $Z=1/2$ the contribution of the $\Delta(1232)$ resonance to
the amplitude of the M1--capture is defined only by the nucleon tensor
current. Setting $Z=1/2$ that is favoured theoretically [15] we
calculate the cross section for the M1--capture $\sigma({\rm np \to
D\gamma})(T_{\rm n}) = 325.5\,{\rm m b}$ agreeing with the
experimental data with accuracy better than 3$\%$.

When matching our result for the cross section for the M1--capture
with the recent one obtained in the EFT approach by Chen, Rupak and
Savage [9]: $\sigma({\rm np \to D\gamma})(T_{\rm n}) = ( 287.1 + 6.51
{^{\not\pi}}L_1)\,{ \rm m b}$ (see Eq.(3.49) of Ref.\,[9]), we
accentuate the dependence of the cross section on the parameter
${^{\not\pi}}L_1$ undefined in the approach. This parameter has to be
fixed from the experimental data [9]. In the NNJL model the parameter
${^{\not\pi}}L_1$ can be expressed in terms of the $Z$ parameter as
follows: ${^{\not\pi}}L_1 = 5.90 + 24.60\,(1-2Z) +
3.03\,(1-2Z)^2$. Thus, in the NNJL model the parameter
${^{\not\pi}}L_1$ acquires a distinct meaning of the contribution of
the $\Delta(1232)$ resonance.

The obtained result for the M1--capture n + p $\to$ D + $\gamma$ we
have applied to the analysis of the photomagnetic disintegration of
the deuteron $\gamma$ + D $\to$ n + p. Due to time--reversal
invariance the cross section for the photomagnetic disintegration of
the deuteron can be directly expressed through the cross section for
the M1--capture. We have compared the numerical values of the cross
section $\sigma({\rm \gamma D \to n p})(\omega)$ calculated in the
NNJL model with the results obtained by Chen and Savage [10] and found
a good agreement.  Nevertheless, it should be emphasized
that in the critical region of the photon energies $\omega \le
2\,\varepsilon_{\rm D} = 4.45\,{\rm MeV}$ restricting the energy
region of the dominance of the photomagnetic disintegration of the
deuteron the cross section calculated in the NNJL model falls steeper
than the cross section obtained in the EFT. However, in this region
the dominant role should be attributed to the E1--transition [10],
which we have not considered. The comparison of our results on the
photomagnetic disintegration of the deuteron with the experimental
data demands the inclusion of the E1--transition as well. This is
planned in our forthcoming publications.

The effective coupling constants of low--energy weak transitions p + p
$\to$ D + e$^+$ + $\nu_{\rm e}$, p + e$^-$ + p $\to$ D + $\nu_{\rm
e}$, $\nu_{\rm e}$ + D $\to$ e$^-$ + p + p, $\bar{\nu}_{\rm e}$ + D
$\to$ e$^+$ + n + n and $\nu_{\rm e}(\bar{\nu}_{\rm e})$ + D $\to$
$\nu_{\rm e}(\bar{\nu}_{\rm e})$ + n + p have been found dependent on
an arbitrary parameter $\bar{\xi}$ in the form $g_{\rm V}\to g_{\rm
V}\,(1 + \bar{\xi})$ caused by the contribution of the nucleon tensor
current [1].

Since at $\bar{\xi} = 0$ we get the value of the astrophysical factor
$S_{\rm pp}(0) = 4.08\times 10^{-25}\,{\rm MeV\, b}$ in complete
agreement with the recommended one $S_{\rm pp}(0) = 4.00\times
10^{-25}\,{\rm MeV\, b}$ related to the Standard Solar Model [27,28],
any non--zero value of $\bar{\xi}$ should lead to an Alternative Solar
Model (ASM). The problem of the formulation of the ASM goes beyond the
scope of this paper. Therefore, in order to dwell within the Standard
Solar Model [27,28] we have to set $\bar{\xi} = 0$.

Setting $\bar{\xi} = 0$ we have shown that all low--energy weak
nuclear reactions of astrophysical interest (i) the solar proton
burning p + p $\to$ D + e$^+$ + $\nu_{\rm e}$, (ii) the pep--process p
+ e$^-$ + p $\to$ D + $\nu_{\rm e}$ and (iii) the disintegration of
the deuteron by neutrinos and anti--neutrinos caused by charged
$\nu_{\rm e}$ + D $\to$ e$^-$ + p + p and $\bar{\nu}_{\rm e}$ + D
$\to$ e$^+$ + n + n and neutral $\nu_{\rm e}(\bar{\nu}_{\rm e})$ + D
$\to$ $\nu_{\rm e}(\bar{\nu}_{\rm e})$ + n + p weak currents are
described in the bulk in agreement with other theorical approaches and
experimental data. The effective Lagrangians of low--energy weak
nuclear transitions p + p $\to$ D + e$^+$ + $\nu_{\rm e}$, p + e$^-$ +
p $\to$ D + $\nu_{\rm e}$, $\nu_{\rm e}$ + D $\to$ e$^-$ + p + p,
$\bar{\nu}_{\rm e}$ + D $\to$ e$^+$ + n + n and $\nu_{\rm
e}(\bar{\nu}_{\rm e})$ + D $\to$ $\nu_{\rm e}(\bar{\nu}_{\rm e})$ + n
+ p are determined by the anomaly of the one--nucleon loop triangle
$AAV$--diagrams. This confirms the statement argued in the NNJL model
concerning the dominant role of the one--nucleon loop anomalies for
the description of low--energy nuclear forces within a quantum field
theoretic approach.

We have shown that the contributions of low--energy elastic pp
scattering in the ${^1}{\rm S}_0$ state with the Coulomb repulsion to
the amplitudes of the reactions p + p $\to$ D + e$^+$ + $\nu_{\rm e}$,
$\nu_{\rm e}$ + D $\to$ e$^-$ + p + p and p + e$^-$ + p $\to$ D +
$\nu_{\rm e}$ are described in the NNJL model in full agreement with
low--energy nuclear phenomenology in terms of the S--wave scattering
length and the effective range. The amplitude of low--energy elastic
pp scattering has been obtained by summing up an infinite series of
one--proton loop diagrams and evaluating the result of the summation
at leading order in the large $N_C$ expansion. The same method has
been applied to the evaluation of the contribution of low--energy
nuclear forces to the relative movements of the nn and np pairs,
respectively, for the reactions $\bar{\nu}_{\rm e}$ + D $\to$ e$^+$ +
n + n and $\nu_{\rm e}(\bar{\nu}_{\rm e})$ + D $\to$ $\nu_{\rm
e}(\bar{\nu}_{\rm e})$ + n + p. This has given the amplitudes of
low--energy elastic nn and np scattering described in terms of S--wave
scattering lengths and effective ranges in agreement with low--energy
nuclear phenomenology [26] as well.

The astrophysical factor $S_{\rm pep}(0)$ for the pep--process, p +
e$^-$ + p $\to$ D + $\nu_{\rm e}$, evaluated relative to $S_{\rm
pp}(0)$ is found in full agreement with the result obtained by Bahcall
and May [45].

The cross sections for the anti--neutrino disintegration of the
deuteron caused by charged $\bar{\nu}_{\rm e}$ + D $\to$ e$^+$ + n + n
and neutral $\bar{\nu}_{\rm e}$ + D $\to$ $\bar{\nu}_{\rm e}$ + n + p
weak currents and averaged over the anti--neutrino spectrum
$\langle\sigma^{\rm \bar{\nu}_{\rm e} D \to e^+nn}(E_{\bar{\nu}_{\rm
e}})\rangle = 11.56\times 10^{-45}\,{\rm cm}^2$ and $\langle\sigma^{\rm
\bar{\nu}_{\rm e} D \to \bar{\nu}_{\rm e} n p }(E_{\bar{\nu}_{\rm
e}})\rangle = 6.28\times 10^{-45}\,{\rm cm}^2$ agree well with recent
experimental data 
\begin{eqnarray*}
\langle\sigma^{\bar{\nu}_{\rm e}D \to e^+ n n
}(E_{\bar{\nu}_{\rm e}})\rangle_{\exp} &=& (9.83\pm 2.04) \times
10^{-45}\,{\rm cm}^2,\nonumber\\
\langle\sigma^{\bar{\nu}_{\rm e}D \to
\bar{\nu}_{\rm e} n p }(E_{\bar{\nu}_{\rm e}})\rangle_{\exp}&=&
(6.08\pm 0.77) \times 10^{-45}\,{\rm cm}^2
\end{eqnarray*}
obtained by the Reines's experimental group [36].

The cross sections for the reactions $\bar{\nu}_{\rm e}$ + D $\to$
e$^+$ + n + n and $\bar{\nu}_{\rm e}$ + D $\to$ $\bar{\nu}_{\rm e}$ +
n + p have been recently calculated by Butler and Chen [53] in the EFT
approach. The obtained results have been written in the following
general form $\sigma = (a + b\,L_{1,\rm A})\times 10^{-42}\,{\rm
cm}^2$ (see Table I of Ref.\,[53]), where $a$ and $b$ are the
parameters which have been calculated in the approach, whereas
$L_{1,\rm A}$ is a free one. In the NNJL model the appearance of the
free parameter is related to the contribution of the nucleon tensor
current [1] that effectively leads to the change of the coupling
constant $g_{\rm V} \to g_{\rm V}\,(1 + \bar{\xi})$, where $\bar{\xi}$
is an arbitrary parameter. The best agreement with the recommended
value of the astrophysical factor for the solar proton burning [29]
and the contemporary experimental data [36] on the cross sections for
the anti--neutrino disintegration of the deuteron $\bar{\nu}_{\rm e}$
+ D $\to$ e$^+$ + n + n and $\bar{\nu}_{\rm e}$ + D $\to$
$\bar{\nu}_{\rm e}$ + n + p caused by charged and neutral weak
current, respectively, we obtain at $\bar{\xi} = 0$ (see Appendix).

We would like to emphasize that the contribution of the $\Delta(1232)$
resonance to the amplitudes of the low--energy weak nuclear reactions
p + p $\to$ D + e$^+$ + $\nu_{\rm e}$, p + e$^-$ + p $\to$ $\nu_{\rm
e}$ + D, $\nu_{\rm e}$ + D $\to$ e$^-$ + p + p, $\bar{\nu}_{\rm e}$ +
D $\to$ e$^+$ + n + n and $\bar{\nu}_{\rm e}$ + D $\to$
$\bar{\nu}_{\rm e}$ + n + p can be neglected. In fact, the
contribution of the $\Delta(1232)$ resonance to the amplitudes of
these reactions is of order of the momentum of the leptonic pair. This
is due to the gauge invariance of the effective interactions $\Delta N
W^+$ and $\Delta N Z$ which should be proportional to $W^+_{\mu\nu} =
\partial_{\mu} W^+_{\nu} - \partial_{\nu} W^+_{\mu}$ and
$Z_{\mu\nu}=\partial_{\mu} Z_{\nu} - \partial_{\nu} Z_{\mu}$,
respectively.  Since the amplitudes of the low--energy weak nuclear
reactions under consideration are defined by the Gamow--Teller
transitions, the terms proportional to the momentum of the leptonic
pair give negligible small contributions. Thus, the cross sections for
low--energy weak nuclear reactions enumerated above do not depend
practically on the uncertainties of the parameter $Z$.

This distinguishes low--energy weak nuclear reactions with the
deuteron from the neutron--proton radiative capture analysed in the
NNJL model. Unlike the low--energy weak nuclear reactions the
contribution of the $\Delta(1232)$ resonance to the amplitude of the
neutron--proton radiative capture is essential for the explanation of
the experimental data.

The cross section for the reaction $\nu_{\rm e}$ + D $\to$ e$^-$ + p +
p has been evaluated with respect to $S_{\rm pp}(0)$. We have found an
enhancement of the cross section by a factor of 1.6 on the average
relative to the results obtained in the PMA.  It would be important to
verify this result for the reaction $\nu_{\rm e}$ + D $\to$ e$^-$ + p
+ p in solar neutrino experiments planned by SNO. Indeed, first, this
should provide an experimental study of $S_{\rm pp}(0)$ and, second,
the cross sections for the anti--neutrino disintegration of the
deuteron caused by charged $\bar{\nu}_{\rm e}$ + D $\to$ e$^+$ + n + n
and neutral $\bar{\nu}_{\rm e}$ + D $\to$ $\bar{\nu}_{\rm e}$ + n + p
weak currents have been found in good agreement with recent
experimental data [36].

For the comparison of the cross sections for the reactions $\nu_{\rm
e}$ + D $\to$ e$^-$ + p + p and $\nu_{\rm e}$ + D $\to$ $\nu_{\rm e}$
+ n + p caused by charged and neutral weak currents, respectively, we
have averaged the cross sections over the ${^8}{\rm B}$ solar neutrino
energy spectrum at energy region $5\,{\rm MeV} \le E_{\nu_{\rm e}} \le
15\,{\rm MeV}$, where the lower bound is related to the experimental
threshold of the experiments at SNO and the upper one is defined by
the kinematics of the $\beta$ decay ${^8}{\rm B} \to {^8}{\rm Be^*}$ +
e$^+$ + $\nu_{\rm e}$ being the source of the ${^8}{\rm B}$ neutrinos
in the solar core. We have obtained
\begin{eqnarray*}
\langle \sigma^{\rm \nu_{\rm e} D\to e^-pp}(E_{\nu_{\rm
e}})\rangle_{\Phi({^8}{\rm B})}&=& 2.62\times 10^{-42}\,{\rm cm}^2,\nonumber\\
\langle \sigma^{\rm \nu_{\rm e} D\to \nu_{\rm e}np}(E_{\nu_{\rm
e}})\rangle_{\Phi({^8}{\rm B})}&=& 1.85\times 10^{-43}\,{\rm cm}^2.
\end{eqnarray*}
The experimental value for the cross section for the reaction
$\nu_{\rm e}$ + D $\to$ e$^-$ + p + p caused by the charged weak
current can, in principle, differ from the theoretical one due to a
possible contribution of the neutrino flavour oscillations [27,48]. In
turn, the averaged value of the cross section for the reaction
$\nu_{\rm e}$ + D $\to$ $\nu_{\rm e}$ + n + p caused by the neutral
weak current should be directly compared with the experimental data,
since it should not depend on whether neutrino flavours oscillate or
not. Of course, the former is valid only if there is no so--called
{\it sterile} neutrino [27,48] having no interactions with Standard
Model particles [39].

Concluding the paper we would like to emphasize that the NNJL model
conveys the idea of a dominant role of one--fermion loop (one--nucleon
loop) anomalies from elementary particle physics to the nuclear
one. This is a new approach to the description of low--energy nuclear
forces in physics of light nuclei. In spite of almost 30 years have
passed after the discovery of one--fermion loop anomalies and
application of these anomalies to the evaluation of effective chiral
Lagrangians of low--energy interactions of hadrons, in nuclear physics
fermion--loop anomalies have not been applied to the analysis of
low--energy nuclear interactions and properties of light nuclei. The
contributions of one--nucleon loop anomalies are strongly related to
high--energy $N\bar{N}$ fluctuations of virtual nucleon fields [1,54].
An important role of $N\bar{N}$ fluctuations for the correct
description of low--energy properties of finite nuclei has been
understood in Ref.\,[55]. Moreover, $N\bar{N}$ fluctuations have been
described in terms of one--nucleon loop diagrams within quantum field
theoretic approaches, but the contributions of one--nucleon loop
anomalies have not been considered.  The NNJL model allows to fill
this blank. Within the framework of the NNJL model we aim to
understand, in principle, the role of nucleon--loop anomalies for the
description of a dynamics of low-energy nuclear forces at the quantum
field theoretic level.

\section*{Acknowledgement}
 
\hspace{0.2in} We are grateful to Prof. M. Kamionkowski for helpful
remarks and encouragement for further applications of the expounded in
the paper technique to the evaluation of the astrophysical factor for
pp fusion and the cross section for the neutrino disintegration of the
deuteron $\nu_{\rm e}$ + D $\to$ e$^-$ + p + p by accounting for the
Coulomb repulsion between the protons.  We thank Prof. J. N. Bahcall
for discussions concerning the ${^8}{\rm B}$ neutrino energy spectrum
and experiments at SNO.  We thank Dr. V. A. Sadovnikova for many
helpful and interesting discussions.

We thank Dr. J. Beacom for calling our attention to the experimental
data [36]. Discussions of the experimental data [36] with
Prof. H. Sobel and Dr. L. Price are greatly appreciated.

\newpage

\section*{Appendix. The effective Lagrangian of the transition 
p + p $\to$ D + e$^+$ + $\nu_{\rm e}$}

\hspace{0.2in} The reaction p + p $\to$ D + e$^+$ + $\nu_{\rm e}$ runs
through the intermediate W--boson exchange, p + p $\to$ D + W$^+$
$\to$ D + e$^+$ + $\nu_{\rm e}$. In the NNJL model we determine this
transition in terms of the following effective interactions [1,39]
$$
{\cal L}_{\rm npD}(x) = -ig_{\rm V}[\bar{p^c}(x)\gamma^{\mu}n(x) -
\bar{n^c}(x)\gamma^{\mu}p(x)]\,D^{\dagger}_{\mu}(x),
$$
$$
{\cal L}^{\rm pp \to pp}_{\rm eff}(x) =
\frac{1}{2}\,C_{\rm NN}\, \{[\bar{p}(x)\,\gamma^{\mu}\gamma^5 p^c
(x)]\,[\bar{p^c}(x)\,\gamma_{\mu}\gamma^5 p(x)] +
[\bar{p}(x)\,\gamma^5 p^c (x)]\,[\bar{p^c}(x)\,\gamma^5 p(x)]\},
$$
$$
{\cal L}_{\rm npW}(x) = - \frac{g_{\rm
W}}{2\sqrt{2}}\,\cos\vartheta_C\,[\bar{n}(x)\gamma^{\nu}(1 - g_{\rm
A}\gamma^5) p(x)]\,W^-_{\nu}(x).\eqno({\rm A}.1)
$$
The transition W$^+$ $\to$ e$^+$ + $\nu_{\rm e}$ is defined by the
Lagrangian [39]
$$
{\cal L}_{\rm \nu_{\rm e} e^+ W}(x) = - \frac{g_{\rm
W}}{2\sqrt{2}}\,[\bar{\psi}_{\nu_{\rm e}}(x)\gamma^{\nu}(1 - \gamma^5)
\psi_{\rm e}(x)]\,W^+_{\nu}(x). \eqno({\rm A}.2)
$$
The electroweak coupling constant $g_{\rm W}$ is connected with the
Fermi weak constant $G_{\rm F}$ and the mass of the W boson $M_{\rm
W}$ through the relation [39]
$$
\frac{g^2_{\rm W}}{8M^2_{\rm W}} = \frac{G_{\rm F}}{\sqrt{2}}.\eqno({\rm A}.3)
$$
For the evaluation of the effective Lagrangian ${\cal
L}^{\rm pp \to D e^+\nu_{\rm e}}_{\rm eff}(x)$ it is convenient to use
the interaction
$$
{\cal L}_{\rm npW}(x) = [\bar{n}(x)\gamma^{\nu}(1 - g_{\rm A}\gamma^5)
p(x)]\,W^-_{\nu}(x),\eqno({\rm A}.4)
$$
and for the description of the subsequent weak transition  W$^+$ $\to$
e$^+$ + $\nu_{\rm e}$ to replace the operator of the W--boson field by the
operator of the leptonic weak current
$$
W^-_{\nu}(x) \to -\frac{G_{\rm V}}{\sqrt{2}}\,[\bar{\psi}_{\nu_{\rm
e}}(x)\gamma_{\nu}(1 - \gamma^5) \psi_{\rm e}(x)],\eqno({\rm A}.5)
$$
where $G_{\rm V} = G_{\rm F}\,\cos \vartheta_C$.

The S matrix describing the transitions like p + p $\to$ D + W$^+$ is
defined by
$$
{\rm S} = {\rm T}e^{\textstyle i\int d^4x\,[{\cal L}_{\rm npD}(x)  +
{\cal L}_{\rm npW}(x) + {\cal L}^{\rm pp \to pp}_{\rm eff}(x) +
\ldots]},\eqno({\rm A}.6)
$$
where T is the time--ordering operator and the ellipses denote the
contribution of interactions irrelevant for the problem.

For the evaluation of the effective Lagrangian ${\cal L}^{\rm pp \to
DW^+}_{\rm eff}(x)$ we have to consider the third order term of the S
matrix which reads
$$
{\rm S}^{(3)} = \frac{i^3}{3!}\int d^4x_1 d^4x_2 d^4x_3\,{\rm T}([{\cal
L}_{\rm npD}(x_1)  + {\cal L}_{\rm npW}(x_1) + {\cal L}^{\rm pp\to pp}_{\rm
eff}(x_1) + \ldots]
$$
$$
\times\,[{\cal L}_{\rm npD}(x_2)  + {\cal L}_{\rm npW}(x_2) + {\cal L}^{\rm
pp\to pp}_{\rm eff}(x_2) + \ldots]
$$
$$
\times\,[{\cal L}_{\rm npD}(x_3)  + {\cal L}_{\rm npW}(x_3) + {\cal L}^{\rm
pp\to pp}_{\rm eff}(x_3) + \ldots]) =
$$
$$
= -i \int d^4x_1 d^4x_2 d^4x_3\,{\rm T}({\cal L}^{\rm pp\to pp}_{\rm
eff}(x_1){\cal L}_{\rm npD}(x_2){\cal L}_{\rm npW}(x_3)) + \ldots
\eqno({\rm A}.7)
$$
The ellipses denote the terms which do not contribute to the
transition p + p $\to$ D + W$^+$ and the interaction ${\cal L}_{\rm
npW}(x)$.  The S matrix element ${\rm
S}^{(3)}_{\rm pp\to DW^+}$ describing the transition p + p $\to$ D +
W$^+$ we determine as follows
$$
{\rm S}^{(3)}_{\rm pp\to DW^+} = -i \int d^4x_1 d^4x_2 d^4x_3\,{\rm
T}({\cal L}^{\rm pp\to pp}_{\rm eff}(x_1){\cal L}_{\rm npD}(x_2){\cal
L}_{\rm npW}(x_3)).\eqno({\rm A}.8)
$$
For the derivation of the effective Lagrangian ${\cal L}^{\rm pp\to
DW^+}_{\rm eff}(x)$ from the S matrix element Eq.({\rm A}.8) we should
make all necessary contractions of the operators of the proton and the
neutron fields. These contractions we denote by the brackets as
$$
\langle{\rm S}^{(3)}_{\rm pp\to DW^+}\rangle = -i \int d^4x_1 d^4x_2 d^4x_3\,\langle{\rm
T}({\cal L}^{\rm pp\to pp}_{\rm eff}(x_1){\cal L}_{\rm npD}(x_2){\cal
L}_{\rm npW}(x_3))\rangle.\eqno({\rm A}.9)
$$
Now the effective Lagrangian ${\cal L}^{\rm pp\to DW^+}_{\rm eff}(x)$
related to the S matrix element $\langle{\rm S}^{(3)}_{\rm pp\to DW^+}\rangle$ can
be defined by
$$
\langle{\rm S}^{(3)}_{\rm pp\to DW^+}\rangle = i\int d^4x\,
{\cal L}^{\rm pp\to DW^+}_{\rm eff}(x) =
$$
$$
= -i \int d^4x_1 d^4x_2 d^4x_3\,\langle{\rm T}({\cal L}^{\rm pp\to pp}_{\rm
eff}(x_1){\cal L}_{\rm npD}(x_2){\cal L}_{\rm npW}(x_3))\rangle. \eqno({\rm A}.10)
$$
In terms of the operators of the interacting fields the effective
Lagrangian ${\cal L}^{\rm pp\to DW^+}_{\rm eff}(x)$ reads
$$
\int d^4x\,{\cal L}^{\rm pp\to DW^+}_{\rm eff}(x) =  - \int d^4x_1 d^4x_2
d^4x_3\,\langle{\rm T}({\cal L}^{\rm pp\to pp}_{\rm eff}(x_1){\cal L}_{\rm
npD}(x_2){\cal L}_{\rm npW}(x_3))\rangle
$$
$$
= -\,\frac{1}{2}\,C_{\rm NN}\,\times\,(-ig_{\rm V})\,\times \,(-g_{\rm
A})\int d^4x_1 d^4x_2 d^4x_3\,{\rm T}([\bar{p^c}(x_1)\,\gamma_{\alpha}\gamma^5 p(x_1)]\,D^{\dagger}_{\mu}(x_2)\,W^-_{\nu}(x_3))
$$
$$
\times \langle0|{\rm T}([\bar{p}(x_1)\,\gamma^{\alpha}\gamma^5 p^c
(x_1)][\bar{p^c}(x_2)\gamma^{\mu}n(x_2) -
\bar{n^c}(x_2)\gamma^{\mu}p(x_2)]\,[\bar{n}(x_3)\gamma^{\nu}\gamma^5
p(x_3)])|0\rangle
$$
$$
 -\,\frac{1}{2}\,C_{\rm NN}\,\times\,(-ig_{\rm V})\,\times
\,(-g_{\rm A})\int d^4x_1 d^4x_2 d^4x_3\,{\rm T}([\bar{p^c}(x_1)
\gamma^5 p(x_1)]\,D^{\dagger}_{\mu}(x_2)\,W^-_{\nu}(x_3))
$$
$$
\times \langle0|{\rm T}([\bar{p}(x_1)
\gamma^5 p^c (x_1)][\bar{p^c}(x_2)\gamma^{\mu}n(x_2) -
\bar{n^c}(x_2)\gamma^{\mu}p(x_2)]\,[\bar{n}(x_3)\gamma^{\nu}\gamma^5 p(x_3)])|0\rangle.\eqno({\rm A}.11)
$$
Since p + p $\to$ D + W$^+$ is the Gamow--Teller transition, we have
taken into account the W--boson coupled to the axial--vector nucleon
current.

Due to the relation $\bar{n^c}(x_2)\gamma^{\mu}p(x_2) = -
\bar{p^c}(x_2)\gamma^{\mu}n(x_2)$ the r.h.s. of Eq.({\rm A}.11) can be
reduced as follows
$$
\int d^4x\,{\cal L}^{\rm pp\to DW^+}_{\rm eff}(x) =  - \int d^4x_1 d^4x_2
d^4x_3\,\langle{\rm T}({\cal L}^{\rm pp\to pp}_{\rm eff}(x_1){\cal L}_{\rm
npD}(x_2){\cal L}_{\rm npW}(x_3))\rangle
$$
$$
=  C_{\rm NN}\,\times\,(-ig_{\rm V})\,\times \,g_{\rm A}\int d^4x_1
d^4x_2 d^4x_3\,{\rm T}([\bar{p^c}(x_1)\,\gamma_{\alpha}\gamma^5 p(x_1)]\,D^{\dagger}_{\mu}(x_2)\,W^-_{\nu}(x_3))
$$
$$
\times\,\langle0|{\rm T}([\bar{p}(x_1)\,\gamma^{\alpha}\gamma^5 p^c
(x_1)][\bar{p^c}(x_2)\gamma^{\mu}n(x_2)][\bar{n}(x_3)\gamma
^{\nu}\gamma^5 p(x_3)])|0\rangle
$$
$$
 + C_{\rm NN}\,\times\,(-ig_{\rm V})\,\times \,g_{\rm A}\int d^4x_1
d^4x_2 d^4x_3\,{\rm T}([\bar{p^c}(x_1)
\gamma^5 p(x_1)]\,D^{\dagger}_{\mu}(x_2)\,W^-_{\nu}(x_3))
$$
$$
\times\,\langle0|{\rm T}([\bar{p}(x_1)
\gamma^5 p^c (x_1)][\bar{p^c}(x_2)\gamma^{\mu}n(x_2)][\bar{n}(x_3)\gamma
^{\nu}\gamma^5 p(x_3)])|0\rangle.\eqno({\rm A}.12)
$$
Making the necessary contractions we arrive at the expression
$$
\int d^4x\,{\cal L}^{\rm pp\to DW^+}_{\rm eff}(x) =  - \int d^4x_1 d^4x_2
d^4x_3\,\langle{\rm T}({\cal L}^{\rm pp\to pp}_{\rm eff}(x_1){\cal L}_{\rm
npD}(x_2){\cal L}_{\rm npW}(x_3))\rangle =
$$
$$
= 2\,\times\,C_{\rm NN}\,\times\,(-ig_{\rm V})\,\times \,g_{\rm A}\int
d^4x_1 d^4x_2 d^4x_3\,{\rm T}([\bar{p^c}(x_1)\,\gamma_{\alpha}\gamma^5 p(x_1)]\,D^{\dagger}_{\mu}(x_2)\,W^-_{\nu}(x_3))
$$
$$
\times\,(-1)\,{\rm tr}\{\gamma^{\alpha}\gamma^5 (-i) S^c_F(x_1 - x_2) \gamma^{\mu} (-i)
S_F(x_2 - x_3) \gamma^{\nu}\gamma^5 (-i) S_F(x_3 - x_1)\}
$$
$$
+ 2\,\times\,C_{\rm NN}\,\times\,(-ig_{\rm V})\,\times \,g_{\rm A}\int
d^4x_1 d^4x_2 d^4x_3\,{\rm T}([\bar{p^c}(x_1)
\gamma^5 p(x_1)]\,D^{\dagger}_{\mu}(x_2)\,W^-_{\nu}(x_3))
$$
$$
\times\,(-1)\,{\rm tr}\{\gamma^5 (-i) S^c_F(x_1 - x_2) \gamma^{\mu}
(-i) S_F(x_2 - x_3) \gamma^{\nu}\gamma^5 (-i) S_F(x_3 -
x_1)\},\eqno({\rm A}.13)
$$
where the combinatorial factor 2 takes into account the fact that the
protons are identical particles in the nucleon loop. 

In the momentum representation of the Green functions we get
$$
\int d^4x\,{\cal L}^{\rm pp\to DW^+}_{\rm eff}(x) =
$$
$$
= - i\,g_{\rm A}C_{\rm NN}\frac{g_{\rm V}}{8\pi^2}\int d^4x_1 \int
\frac{d^4x_2 d^4k_2}{(2\pi)^4}\frac{d^4x_3
d^4k_3}{(2\pi)^4}\,e^{\displaystyle -i k_2\cdot (x_2-x_1)}e^{\displaystyle
-i k_3\cdot (x_3-x_1)}
$$
$$
\times\,{\rm
T}([\bar{p^c}(x_1)\,\gamma_{\alpha}\gamma^5 p(x_1)]\,D^{\dagger}_{\mu}(x_2)\,W^-_{\nu}(x_3))\,J^{\alpha\mu\nu}(k_2, k_3; Q)
$$
$$
- i\,g_{\rm A}C_{\rm NN}\frac{g_{\rm V}}{8\pi^2}\int d^4x_1 \int
\frac{d^4x_2 d^4k_2}{(2\pi)^4}\frac{d^4x_3
d^4k_3}{(2\pi)^4}\,e^{\displaystyle -i k_2\cdot (x_2-x_1)}e^{\displaystyle
-i k_3\cdot (x_3-x_1)}
$$
$$
\times\,{\rm T}([\bar{p^c}(x_1)\,\gamma^5 p(x_1)]\,
D^{\dagger}_{\mu}(x_2)\,W^-_{\nu}(x_3))\,\,J^{\mu\nu}(k_2, k_3; Q).\eqno({\rm A}.14)
$$
The structure functions ${\cal J}^{\alpha\mu\nu}(k_2, k_3;
Q)$ and ${\cal J}^{\mu\nu}(k_2, k_3; Q)$ are defined by the
momentum integrals
$$
{\cal J}^{\alpha\mu\nu}(k_2, k_3; Q) =
$$
$$
=\int\frac{d^4k}{\pi^2i} \,{\rm tr}
\Bigg\{\gamma^{\alpha}\gamma^5\frac{1}{M_{\rm N} - \hat{k} - \hat{Q} +
\hat{k}_2}\gamma^{\mu}\frac{1}{M_{\rm N} - \hat{k} -
\hat{Q}}\gamma^{\nu}\gamma^5 \frac{1}{M_{\rm N} - \hat{k} - \hat{Q} -
\hat{k}_3}\Bigg\},
$$
$$
{\cal J}^{\mu\nu}(k_2, k_3; Q) =
$$
$$
=\int\frac{d^4k}{\pi^2i} \,{\rm tr}\Bigg\{\gamma^5\frac{1}{M_{\rm N} -
\hat{k} - \hat{Q} + \hat{k}_2}\gamma^{\mu}\frac{1}{M_{\rm N} - \hat{k}
- \hat{Q}}\gamma^{\nu}\gamma^5 \frac{1}{M_{\rm N} - \hat{k} - \hat{Q}
- \hat{k}_3}\Bigg\}.\eqno({\rm A}.15)
$$
We have introduced a 4--vector $Q = a\,k_2 + b\,k_3$ caused by
an arbitrary shift of a virtual momentum  with arbitrary parameters $a$ and
$b$.

In order to obtain the effective Lagrangian ${\cal L}^{\rm pp \to D
e^+\nu_{\rm e}}_{\rm eff}(x)$ of the transition p + p $\to$ D + e$^+$
+ $\nu_{\rm e}$ we have to replace the operator of the W--boson field
by the operator of the leptonic weak current Eq.({\rm A}.5):
$$
\int d^4x\,{\cal L}^{\rm pp\to D e^+ \nu_{\rm e}}_{\rm eff}(x) = $$
$$
= i\,g_{\rm A}C_{\rm NN}\frac{G_{\rm V}}{\sqrt{2}} \frac{g_{\rm
V}}{8\pi^2}\int d^4x_1 \int \frac{d^4x_2 d^4k_2}{(2\pi)^4}\frac{d^4x_3
d^4k_3}{(2\pi)^4}\,e^{\displaystyle -i k_2\cdot (x_2-x_1)}e^{\displaystyle
-i k_3\cdot (x_3-x_1)}
$$
$$
\times\,{\rm T}([\bar{p^c}(x_1)\,\gamma_{\alpha}\gamma^5
p(x_1)]\,D^{\dagger}_{\mu}(x_2)\,[\bar{\psi}_{\nu_{\rm
e}}(x_3)\gamma_{\nu}(1 - \gamma^5) \psi_{\rm
e}(x_3)])\,J^{\alpha\mu\nu}(k_2, k_3; Q)
$$
$$
+ i\,g_{\rm A}C_{\rm NN}\frac{G_{\rm V}}{\sqrt{2}}\frac{g_{\rm
V}}{8\pi^2}\int d^4x_1 \int \frac{d^4x_2 d^4k_2}{(2\pi)^4}\frac{d^4x_3
d^4k_3}{(2\pi)^4}\,e^{\displaystyle -i k_2\cdot (x_2-x_1)}e^{\displaystyle
-i k_3\cdot (x_3-x_1)}
$$
$$
\times\,{\rm T}([\bar{p^c}(x_1)\,\gamma^5 p(x_1)]\,
D^{\dagger}_{\mu}(x_2)\,[\bar{\psi}_{\nu_{\rm e}}(x_3)\gamma_{\nu}(1 -
\gamma^5) \psi_{\rm e}(x_3)])\,J^{\mu\nu}(k_2, k_3; Q).\eqno({\rm
A}.16)
$$
Thus, the problem of the evaluation of the effective Lagrangian of the
transition p + p $\to$ D + e$^+$ + $\nu_{\rm e}$ reduces itself to the
problem of the evaluation of the structure functions Eq.({\rm
A}.15). The momentum $k_3$ is related to the 4--momentum of the
leptonic pair. Due to the Gamow--Teller type of the transition p + p
$\to$ D + W$^+$ the contribution proportional to the 4--momentum of
the leptonic pair turns out to be much smaller with respect to the
contribution proportional to the 4--momentum of the deuteron
$k_2$. Therefore, without loss of generality we can set $k_3 = 0$ in
the integrand.  This gives
$$
{\cal J}^{\alpha\mu\nu}(k_2, k_3; Q) =
$$
$$
=\int\frac{d^4k}{\pi^2i} \,{\rm tr}
\Bigg\{\gamma^{\alpha}\gamma^5\frac{1}{M_{\rm N} - \hat{k} - \hat{Q} +
\hat{k}_2}\gamma^{\mu}\frac{1}{M_{\rm N} - \hat{k} -
\hat{Q}}\gamma^{\nu}\gamma^5 \frac{1}{M_{\rm N} - \hat{k} - \hat{Q}}\Bigg\},
$$
$$
{\cal J}^{\mu\nu}(k_2, k_3; Q) =
$$
$$
=\int\frac{d^4k}{\pi^2i} \,{\rm tr}\Bigg\{\gamma^5\frac{1}{M_{\rm N} -
\hat{k} - \hat{Q} + \hat{k}_2}\gamma^{\mu}\frac{1}{M_{\rm N} -
\hat{k} - \hat{Q}}\gamma^{\nu}\gamma^5 \frac{1}{M_{\rm N} - \hat{k} -
\hat{Q}}\Bigg\},\eqno({\rm A}.17)
$$
The evaluation of the momentum integrals at leading order in the
$1/M_{\rm N}$ expansion corresponding the leading order in the large
$N_C$ expansion due to the proportionality $M_{\rm N} \sim N_C$ in QCD
with $SU(N_C)$ gauge group at $N_C \to \infty$ [22] (see also
Ref.\,[1]) yields
$$
{\cal J}^{\alpha\mu\nu}(k_2, k_3; Q) = 3\,(k^{\alpha}_2 g^{\nu\mu} -
k^{\nu}_2 g^{\mu\alpha}) + \frac{1}{9}\,(1+2a)\,(k^{\alpha}_2
g^{\nu\mu} + k^{\nu}_2 g^{\mu\alpha}),
$$
$$
{\cal J}^{\mu\nu}(k_2, k_3; Q) = g^{\mu\nu} 4\,M_{\rm
N}\,J_2(M_{\rm N}) \sim O(1/N^2_C), \eqno({\rm A}.18)
$$
where the terms proportional to $k^{\mu}_2$ have been dropped, because
 they produce the contributions to the effective Lagrangian multiplied
 by $\partial^{\mu}D_{\mu}(x)$ vanishing by virtue of the constraint
 $\partial^{\mu}D_{\mu}(x) = 0$. Then, $J_2(M_{\rm N})$ is a
 logarithmically divergent integral defined in the NNJL model in terms
 of the cut--off $\Lambda_D = 46.172\,{\rm MeV}$ such as $\Lambda_{\rm
 D} \ll M_{\rm N}$ [1]:
$$
J_2(M_{\rm N}) = \int\frac{d^4k}{\pi^2i} \frac{1}{(M^2_{\rm N} -
 k^2)^2} = 2\int\limits^{\Lambda_{\rm D}}_0
 \frac{d|\vec{k}\,|\vec{k}^{\,2}}{(M^2_{\rm N} + \vec{k}^{\,2})^{3/2}}
 = \frac{2}{3}\,\Bigg(\frac{\Lambda_{\rm D}}{M_{\rm
 N}}\Bigg)^{\!3}\sim O(1/N^3_C).\eqno({\rm A}.19)
$$
The cut--off $\Lambda_{\rm D}$ restricts 3--momenta of the virtual
nucleon fluctuations forming the physical deuteron [1]. Due to the
uncertainty relation $\Delta r\,\Lambda_{\rm D}\sim 1$ the spatial
region of virtual nucleon fluctuations forming the physical deuteron
is defined by $\Delta r \sim 4.274\,{\rm fm}$. This agrees with the
effective radius of the deuteron $r_{\rm D} = 1/\sqrt{\varepsilon_{\rm
D}M_{\rm N}} = 4.319\,{\rm fm}$.

Keeping the terms of the same order in the large $N_C$ expansion we
get the structure functions
$$
{\cal J}^{\alpha\mu\nu}(k_2, k_3; Q) = 3\,(k^{\alpha}_2 g^{\nu\mu} -
k^{\nu}_2 g^{\mu\alpha}) + \frac{1}{9}\,(1+2a)\,(k^{\alpha}_2
g^{\nu\mu} + k^{\nu}_2 g^{\mu\alpha}),
$$
$$
{\cal J}^{\mu\nu}(k_2, k_3; Q) = 0, \eqno({\rm A}.20)
$$
The structure functions Eq.({\rm A}.20) define the effective Lagrangian
 ${\cal L}^{\rm pp\to D e^+
\nu_{\rm e}}_{\rm eff}(x)$:
$$
{\cal L}^{\rm pp\to D e^+ \nu_{\rm e}}_{\rm eff}(x) = g_{\rm A} C_{\rm
NN}\frac{G_{\rm V}}{\sqrt{2}}\frac{3g_{\rm
V}}{8\pi^2}\,[\bar{p^c}(x)\gamma^{\mu}\gamma^5
p(x)]\,[\bar{\psi}_{\nu_{\rm e}}(x)\gamma^{\nu}(1 - \gamma^5) \psi_{\rm
e}(x)]
$$
$$
\times\,\Big[(\partial_{\mu}D^{\dagger}_{\nu}(x) -
\partial_{\nu}D^{\dagger}_{\mu}(x)) +
\frac{1}{27}\,(1+2a)\,(\partial_{\mu}D^{\dagger}_{\nu}(x) +
\partial_{\nu}D^{\dagger}_{\mu}(x))\Big].\eqno({\rm A}.21)
$$
In order to remove uncertainty caused by the shift of virtual momenta
of the one--nucleon loop diagrams we demand the invariance of the
effective Lagrangian ${\cal L}^{\rm pp\to D e^+ \nu_{\rm e}}_{\rm
eff}(x)$ under gauge transformations of the deuteron field
$$
D^{\dagger}_{\mu}(x) \to D^{\dagger}_{\mu}(x) +
\partial_{\mu}\Omega(x),\eqno({\rm A}.22)
$$
where $\Omega(x)$ is a gauge function. The requirement of the
invariance of the effective Lagrangian of the low--energy transition p
+ p $\to$ D + e$^+$ + $\nu_{\rm e}$ under gauge transformation
Eq.({\rm A}.22) imposes the constraint $a = -1/2$. This reduces the
effective Lagrangian ${\cal L}^{\rm pp\to D e^+ \nu_{\rm e}}_{\rm
eff}(x)$ to the form
$$
{\cal L}^{\rm pp\to D e^+ \nu_{\rm e}}_{\rm eff}(x) = g_{\rm A} C_{\rm 
NN}\frac{G_{\rm V}}{\sqrt{2}}\frac{3g_{\rm
V}}{8\pi^2}\,D^{\dagger}_{\mu\nu}(x)\,[\bar{p^c}(x)\gamma^{\mu}\gamma^5
p(x)]\,[\bar{\psi}_{\nu_{\rm e}}(x)\gamma^{\nu}(1 - \gamma^5) \psi_{\rm
e}(x)].\eqno({\rm A}.23)
$$
This effective Lagrangian is defined by the structure function ${\cal
J}^{\alpha\mu\nu}(k_2, k_2; Q)$:
$$
{\cal J}^{\alpha\mu\nu}(k_2, k_3; Q) = 3\,(k^{\alpha}_2 g^{\nu\mu} -
k^{\nu}_2 g^{\mu\alpha}).\eqno({\rm A}.24)
$$
The structure function ${\cal J}^{\alpha\mu\nu}(k_2, k_2; Q)$ does not
depend on the mass of virtual nucleons and according to Gertsein and
Jackiw [37] can be valued as the anomaly of the one--nucleon triangle
$AAV$--diagram. The requirement of gauge invariance applied to remove
ambiguities of the structure function ${\cal J}^{\alpha\mu\nu}(k_2,
k_3; Q)$ and to fix the contribution of the anomaly of the
one--nucleon loop $AAV$--diagrams is in complete agreement with the
derivation of the Adler--Bell--Jackiw axial--vector anomaly performed
in terms of one--nucleon loop $AVV$--diagrams (see {\it Jackiw} [54]).

The effective Lagrangian ${\cal L}^{\rm \bar{\nu}_{\rm e}D \to e^+ nn
}_{\rm eff}(x)$ describing the low--energy transition $\bar{\nu}_{\rm
e}$ + D $\to$ e$^+$ + n + n can be obtained by the way analogous to
${\cal L}^{\rm pp\to D e^+ \nu_{\rm e}}_{\rm eff}(x)$ and reads
$$
{\cal L}^{\rm \bar{\nu}_{\rm e}D \to e^+ nn }_{\rm eff}(x) = -\,g_{\rm A}\,C_{\rm 
NN}\,\frac{G_{\rm V}}{\sqrt{2}}\frac{3g_{\rm
V}}{8\pi^2}\,D_{\mu\nu}(x)\,[\bar{n}(x)\gamma^{\mu}\gamma^5
n^c(x)]\,[\bar{\psi}_{\nu_{\rm e}}(x)\gamma^{\nu}(1 - \gamma^5) \psi_{\rm
e}(x)].\eqno({\rm A}.25)
$$
In the low--energy limit when
$D^{\dagger}_{\mu\nu}(x)\,[\bar{p^c}(x)\gamma^{\mu}\gamma^5 p(x)] \to
-\,2\,i\,M_{\rm N}D^{\dagger}_{\nu}(x)[\bar{p^c}(x)\gamma^5 p(x)]$ the
effective Lagrangian Eq.({\rm A}.23) reduces itself to the form
$$
{\cal L}^{\rm pp\to D e^+ \nu_{\rm e}}_{\rm eff}(x) =-\,i\, g_{\rm A}\,
M_{\rm N}\,C_{\rm NN}\,\frac{G_{\rm V}}{\sqrt{2}}\frac{3g_{\rm
V}}{4\pi^2}\,D^{\dagger}_{\nu}(x)\,[\bar{p^c}(x)\gamma^5
p(x)]\,[\bar{\psi}_{\nu_{\rm e}}(x)\gamma^{\nu}(1 - \gamma^5)
\psi_{\rm e}(x)].\eqno({\rm A}.26)
$$
The low--energy reduction $D_{\mu\nu}[\bar{n}(x)\gamma^{\mu}\gamma^5
n^c(x)] \to -\,2\,i\,M_{\rm N}D_{\nu}(x)\bar{n}(x)\gamma^5 n^c(x)$
applied to the effective Lagrangian ${\cal L}^{\rm \bar{\nu}_{\rm e}D
\to e^+ nn }_{\rm eff}(x)$ gives
$$
{\cal L}^{\rm \bar{\nu}_{\rm e}D \to e^+ nn }_{\rm eff}(x) = i\,g_{\rm
A}\,M_{\rm N}\,C_{\rm NN}\,\frac{G_{\rm V}}{\sqrt{2}}\frac{3g_{\rm
V}}{4\pi^2}\,D_{\nu}(x)\,[\bar{n}(x)\gamma^5
n^c(x)]\,[\bar{\psi}_{\nu_{\rm e}}(x)\gamma^{\nu}(1 - \gamma^5)
\psi_{\rm e}(x)].\eqno({\rm A}.27)
$$
The effective Lagrangians Eq.({\rm A}.23) and Eq.({\rm A}.25) as well
as Eq.({\rm A}.26) and Eq.({\rm A}.27) testify distinctly that the
transitions p + p $\to$ D + e$^+$ + $\nu_{\rm e}$ and $\bar{\nu}_{\rm
e}$ + D $\to$ e$^+$ n + n are governed by the same dynamics of
low--energy nuclear forces in agreement with charge independence of
weak interaction strength.

For the evaluation of the effective Lagrangian ${\cal L}^{\rm
{\nu}_{\rm e}D \to {\nu}_{\rm e} n p }_{\rm eff}(x)$ of the transition
$\nu_{\rm e}$ + D $\to$ $\nu_{\rm e}$ + n + p in the NNJL model one
should use the following Lagrangians
$$
{\cal L}^{\dagger}_{\rm npD}(x) = -ig_{\rm V}[\bar{p}(x)\gamma^{\mu}n^c(x) -
\bar{n}(x)\gamma^{\mu}p^c(x)]\,D_{\mu}(x),
$$
$$
{\cal L}^{\rm np \to np}_{\rm eff}(x) =
C_{\rm NN}\, \{[\bar{p}(x)\,\gamma^{\mu}\gamma^5 n^c
(x)]\,[\bar{n^c}(x)\,\gamma_{\mu}\gamma^5 p(x)] +
[\bar{p}(x)\,\gamma^5 n^c (x)]\,[\bar{n^c}(x)\,\gamma^5 p(x)]\},
$$
$$
{\cal L}_{\rm NNZ}(x) = g_{\rm A}\,[\bar{p}(x)\gamma^{\nu}\gamma^5
p(x) - \bar{n}(x)\gamma^{\nu}\gamma^5 n(x)]\,Z_{\nu}(x),
$$
$$
Z_{\nu}(x) \to \frac{G_{\rm F}}{2\sqrt{2}}\, [\bar{\psi}_{\nu_{\rm
e}}(x)\gamma^{\nu}(1 - \gamma^5) \psi_{\rm e}(x)].\eqno({\rm A}.28)
$$
The effective Lagrangian ${\cal L}^{\rm {\nu}_{\rm e}D \to {\nu}_{\rm
e} n p }_{\rm eff}(x)$ of the low--energy transition $\nu_{\rm e}$ + D
$\to$ $\nu_{\rm e}$ + n + p is then defined by
$$
{\cal L}^{\rm {\nu}_{\rm e}D \to {\nu}_{\rm e} n p }_{\rm eff}(x) =
-\,g_{\rm A}\,C_{\rm NN}\,\frac{G_{\rm F}}{\sqrt{2}}\frac{3g_{\rm
V}}{8\pi^2}\,D_{\mu\nu}(x)\,[\bar{p}(x)\gamma^{\mu}\gamma^5
n^c(x)]\,[\bar{\psi}_{\nu_{\rm e}}(x)\gamma^{\nu}(1 - \gamma^5)
\psi_{\nu_{\rm e}}(x)].\eqno({\rm A}.29)
$$
In the low--energy limit when
$D_{\mu\nu}(x)[\bar{p}(x)\gamma^{\mu}\gamma^5 n^c(x)] \to -2iM_{\rm
N}D_{\nu}(x)\,[\bar{p}(x)\gamma^5 n^c(x)]$ the effective Lagrangian
reduces itself to the from
$$
{\cal L}^{\rm {\nu}_{\rm e}D \to {\nu}_{\rm e} n p }_{\rm eff}(x) =
i\,g_{\rm A}\, M_{\rm N}\, C_{\rm NN}\,\frac{G_{\rm
F}}{\sqrt{2}}\frac{3g_{\rm V}}{4\pi^2}\,D_{\nu}(x)\,[\bar{p}(x)\gamma^5
n^c(x)]\,[\bar{\psi}_{\nu_{\rm e}}(x)\gamma^{\nu}(1 - \gamma^5)
\psi_{\nu_{\rm e}}(x)].\eqno({\rm A}.30)
$$
For the evaluation of the matrix element of the transition $\nu_{\rm
e}$ + D $\to$ $\nu_{\rm e}$ + n + p one would use the wave function of
the np pair in the standard form $|n(p_1)p(p_2)\rangle =
a^{\dagger}_{\rm n}(p_1,\sigma_1)\,a^{\dagger}_{\rm
p}(p_2,\sigma_2)|0\rangle$. The contribution of low--energy nuclear
forces to the relative movement of the np pair in the ${^1}{\rm S}_0$
state should be described by the infinite series of one--nucleon
bubbles evaluated at leading order in the large $N_C$ expansion. The
result should be expressed in terms of the S--wave scattering length
$a_{\rm np}$ and the effective range $r_{\rm np}$ of low--energy
elastic np scattering in the ${^1}{\rm S}_0$ state.

Now let us obtain the contribution of the nucleon tensor current
Eq.(\ref{label4.10}). In terms of the structure functions the
effective Lagrangian $\delta {\cal L}^{\rm pp \to D e^+\nu_{\rm e}}_{\rm eff}(x)$
$$
\int d^4x\,\delta {\cal L}^{\rm pp\to D e^+ \nu_{\rm e}}_{\rm eff}(x) = 
$$
$$
=-\, g_{\rm A}\frac{C_{\rm NN}}{8\pi^2}\frac{G_{\rm V}}{\sqrt{2}} \frac{g_{\rm
T}}{2M_{\rm N}}\int d^4x_1 \int \frac{d^4x_2 d^4k_2}{(2\pi)^4}\frac{d^4x_3
d^4k_3}{(2\pi)^4}\,e^{\displaystyle -i k_2\cdot (x_2-x_1)}e^{\displaystyle
-i k_3\cdot (x_3-x_1)}
$$
$$
\times\,{\rm T}([\bar{p^c}(x_1)\,\gamma_{\alpha}\gamma^5
p(x_1)]\,D^{\dagger}_{\mu\nu}(x_2)\,[\bar{\psi}_{\nu_{\rm
e}}(x_3)\gamma_{\lambda}(1 - \gamma^5) \psi_{\rm
e}(x_3)])\,J^{\alpha\mu\nu\lambda}(k_2, k_3; Q)
$$
$$
-\, g_{\rm A}\frac{C_{\rm NN}}{8\pi^2}\frac{G_{\rm V}}{\sqrt{2}} \frac{g_{\rm
T}}{2M_{\rm N}}\int d^4x_1 \int \frac{d^4x_2 d^4k_2}{(2\pi)^4}\frac{d^4x_3
d^4k_3}{(2\pi)^4}\,e^{\displaystyle -i k_2\cdot (x_2-x_1)}e^{\displaystyle
-i k_3\cdot (x_3-x_1)}
$$
$$
\times\,{\rm T}([\bar{p^c}(x_1)\,\gamma^5 p(x_1)]\,
D^{\dagger}_{\mu\nu}(x_2)\,[\bar{\psi}_{\nu_{\rm e}}(x_3)\gamma_{\lambda}(1 -
\gamma^5) \psi_{\rm e}(x_3)])\,J^{\mu\nu\lambda}(k_2, k_3; Q).\eqno({\rm
A}.31)
$$
The structure functions are given by
$$
{\cal J}^{\alpha\mu\nu\lambda}(k_2, k_3; Q) =
$$
$$
=\int\frac{d^4k}{\pi^2i} \,{\rm tr}
\Bigg\{\gamma^{\alpha}\gamma^5\frac{1}{M_{\rm N} - \hat{k} - \hat{Q} +
\hat{k}_2}\sigma^{\mu\nu}\frac{1}{M_{\rm N} - \hat{k} -
\hat{Q}}\gamma^{\lambda}\gamma^5 \frac{1}{M_{\rm N} - \hat{k} - \hat{Q} -
\hat{k}_3}\Bigg\},
$$
$$
{\cal J}^{\mu\nu\lambda}(k_2, k_3; Q) =
$$
$$
=\int\frac{d^4k}{\pi^2i} \,{\rm tr}\Bigg\{\gamma^5\frac{1}{M_{\rm N} -
\hat{k} - \hat{Q} + \hat{k}_2}\sigma^{\mu\nu}\frac{1}{M_{\rm N} - \hat{k}
- \hat{Q}}\gamma^{\lambda}\gamma^5 \frac{1}{M_{\rm N} - \hat{k} - \hat{Q}
- \hat{k}_3}\Bigg\},\eqno({\rm A}.32)
$$
where a 4--vector $Q = a\,k_2 + b\,k_3$ is an arbitrary shift of a
virtual momentum with arbitrary parameters $a$ and $b$.

The evaluation of the structure functions Eq.({\rm A}.32) at leading
order in the large $N_C$ expansion gives the following effective
Lagrangian $\delta {\cal L}^{\rm pp\to D e^+ \nu_{\rm e}}_{\rm eff}(x)$
caused by the contribution of the nucleon tensor current 
$$
\delta {\cal L}^{\rm pp\to D e^+ \nu_{\rm e}}_{\rm eff}(x)=g_{\rm A}\,C_{\rm NN}\,\frac{G_{\rm V}}{\sqrt{2}} \frac{g_{\rm
T}}{2\pi^2}\,\Big\{D^{\dagger}_{\mu\nu}(x)\,[\bar{p^c}(x)
\gamma^{\mu}\gamma^5 p(x)] - \frac{i a}{2M_{\rm
N}}\,\partial^{\mu}D^{\dagger}_{\mu\nu}(x)[\bar{p^c}(x)\gamma^5
p(x)]\Big\}
$$
$$
\times [\bar{\psi}_{\nu_{\rm e}}(x)\gamma^{\nu}(1 - \gamma^5)
\psi_{\rm e}(x)].\eqno({\rm A}.33)
$$
It is seen that the coupling constants of the effective Lagrangian
depend on the arbitrary parameter $a$ caused by a shift of a virtual
momentum.

The analogous expression one can get for the effective Lagrangian of
the transition $\bar{\nu}_{\rm e}$ + D $\to$ e$^+$ + n + n caused by
the contribution of the nucleon tensor current as well
$$
\delta {\cal L}^{\rm \bar{\nu}_{\rm e}D \to e^+ nn }_{\rm eff}(x)=g_{\rm A}\,C_{\rm NN}\,\frac{G_{\rm V}}{\sqrt{2}} \frac{g_{\rm
T}}{2\pi^2}\,\Big\{D_{\mu\nu}(x)\,[\bar{n}(x)
\gamma^{\mu}\gamma^5 n^c(x)] - \frac{i a}{2M_{\rm
N}}\,\partial^{\mu}D_{\mu\nu}(x)[\bar{n}(x)\gamma^5
n^c(x)]\Big\}
$$
$$
\times [\bar{\psi}_{\nu_{\rm e}}(x)\gamma^{\nu}(1 - \gamma^5)
\psi_{\rm e}(x)].\eqno({\rm A}.34)
$$
In the low--energy limit when 
$$
D^{\dagger}_{\mu\nu}(x)\,[\bar{p^c}(x)
\gamma^{\mu}\gamma^5 p(x)] \to -\,2\,i\,M_{\rm N} D^{\dagger}_{\nu}(x)\,[\bar{p^c}(x)
\gamma^5 p(x)],
$$
$$
D_{\mu\nu}(x)\,[\bar{n}(x)
\gamma^{\mu}\gamma^5 n^c(x)] \to -\,2\,i\,M_{\rm N} D_{\nu}(x)\,[\bar{n}(x)
\gamma^5 n(x)]\eqno({\rm A}.35)
$$
the effective Lagrangians Eq.({\rm A}.33) and Eq.({\rm A}.34) can be
recast into the form
$$
\delta {\cal L}^{\rm pp\to D e^+ \nu_{\rm e}}_{\rm eff}(x)= -\,i\,g_{\rm
A}\,M_{\rm N}\,C_{\rm NN}\,\frac{G_{\rm V}}{\sqrt{2}} \frac{3g_{\rm
T}}{4\pi^2}\,\xi\,D^{\dagger}_{\nu}(x)[\bar{p^c}(x)\gamma^5
p(x)]\,[\bar{\psi}_{\nu_{\rm e}}(x)\gamma^{\nu}(1 - \gamma^5)
\psi_{\rm e}(x)],
$$
$$
\delta {\cal L}^{\rm \bar{\nu}_{\rm e}D \to e^+ nn }_{\rm eff}(x)=-\,i\, g_{\rm A}\,M_{\rm N}\,C_{\rm NN}\,\frac{G_{\rm V}}{\sqrt{2}} \frac{3g_{\rm
T}}{4\pi^2}\,\xi\,D_{\nu}(x)[\bar{n}(x)\gamma^5
n^c(x)]\,[\bar{\psi}_{\nu_{\rm e}}(x)\gamma^{\nu}(1 - \gamma^5)
\psi_{\rm e}(x)],\eqno({\rm A}.36)
$$
where $\xi$ is an arbitrary parameter related to the parameter $a$ as
follows
$$
\xi =\frac{1}{3}\,(4\,-\,a).\eqno({\rm A}.37)
$$
The total effective Lagrangians of the transitions p + p $\to$ D +
e$^+$ + $\nu_{\rm e}$, $\bar{\nu}_{\rm e}$ + D $\to$ e$^+$ + n + n and
$\bar{\nu}_{\rm e}$ + D $\to$ $\bar{\nu}_{\rm e}$ + n + p are defined
by
$$
{\cal L}^{\rm pp\to D e^+ \nu_{\rm e}}_{\rm eff, tc}(x)= -\,i\,(1 +
\bar{\xi})\,g_{\rm A}\,M_{\rm N}\,C_{\rm NN}\,\frac{G_{\rm
V}}{\sqrt{2}} \frac{3g_{\rm
V}}{4\pi^2}\,D^{\dagger}_{\nu}(x)[\bar{p^c}(x)\gamma^5
p(x)]\,[\bar{\psi}_{\nu_{\rm e}}(x)\gamma^{\nu}(1 - \gamma^5)
\psi_{\rm e}(x)],
$$
$$
{\cal L}^{\rm \bar{\nu}_{\rm e}D \to e^+ nn }_{\rm eff}(x)=-\,i\,(1 +
\bar{\xi})\, g_{\rm A}\,M_{\rm N}\,C_{\rm NN}\,\frac{G_{\rm V}}{\sqrt{2}}
\frac{3g_{\rm V}}{4\pi^2}\,D_{\nu}(x)[\bar{n}(x)\gamma^5
n^c(x)]\,[\bar{\psi}_{\nu_{\rm e}}(x)\gamma^{\nu}(1 - \gamma^5)
\psi_{\rm e}(x)],
$$
$$
{\cal L}^{\rm {\nu}_{\rm e}D \to {\nu}_{\rm e} n p }_{\rm eff}(x) =
i\,(1 + \bar{\xi})\,g_{\rm A}\, M_{\rm N}\, C_{\rm NN}\,\frac{G_{\rm
F}}{\sqrt{2}}\frac{3g_{\rm V}}{4\pi^2}\,D_{\nu}(x)\,[\bar{p}(x)\gamma^5
n^c(x)]\,[\bar{\psi}_{\nu_{\rm e}}(x)\gamma^{\nu}(1 - \gamma^5)
\psi_{\nu_{\rm e}}(x)],\eqno({\rm A}.38)
$$
where $\bar{\xi}$ is obtained by using the relation $g_{\rm T} =
\sqrt{3/8}\,g_{\rm V}$ and is defined by
$$
\bar{\xi} = \sqrt{\frac{3}{8}}\,\xi.\eqno({\rm A}.39)
$$
Under the assumption of isotopical invariance of low--energy nulcear
forces the best agreement with the recommended value for the
astrophysical factor for the solar proton burning [29] and the
contemporary experimental data [36] on the cross sections for the
anti--neutrino disintegration of the deuteron $\bar{\nu}_{\rm e}$ + D
$\to$ e$^+$ + n + n and $\bar{\nu}_{\rm e}$ + D $\to$ $\bar{\nu}_{\rm
e}$ + n + p caused by charged and neutral weak current, respectively,
we obtain at $\bar{\xi} = 0$.

\newpage

\section*{Figure caption}

\begin{itemize}
\item 

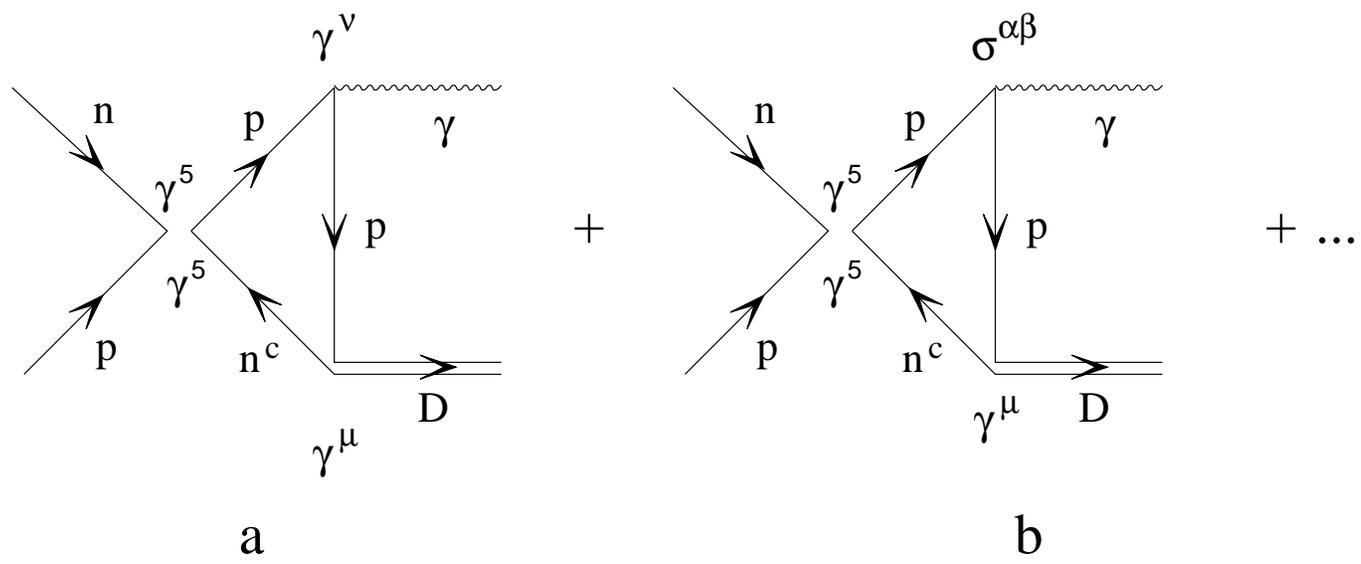
Fig.1. One--nucleon loop diagrams of the contribution of the effective
coupling $[\bar{p}(x)\gamma^5n^c(x)][\bar{n^c}(x)\gamma^5 p(x)]$ to
the effective Lagrangian of the M1 transition n +  
p $\to$ D + $\gamma$.

\item 

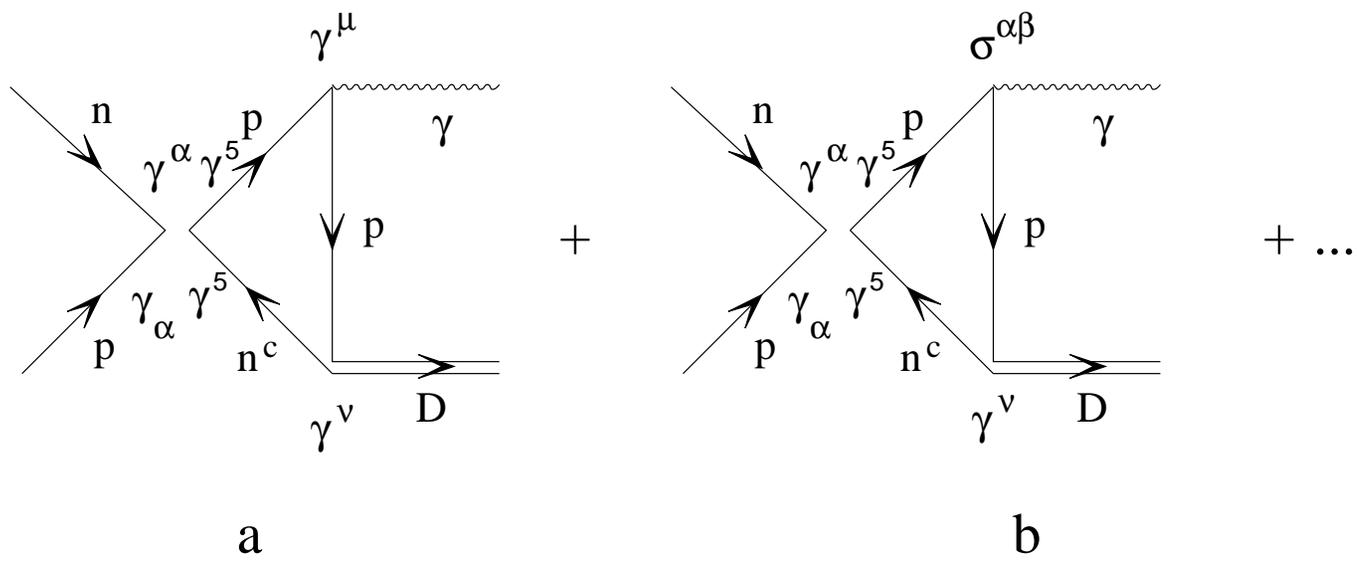
Fig.2. One--nucleon loop diagrams of the contribution of the effective
coupling
$[\bar{p}(x)\gamma^{\alpha}\gamma^5n^c(x)]
[\bar{n^c}(x)\gamma_{\alpha}\gamma^5
p(x)]$ to the effective Lagrangian of the M1 transition 
n + p $\to$ D + $\gamma$.

\item 

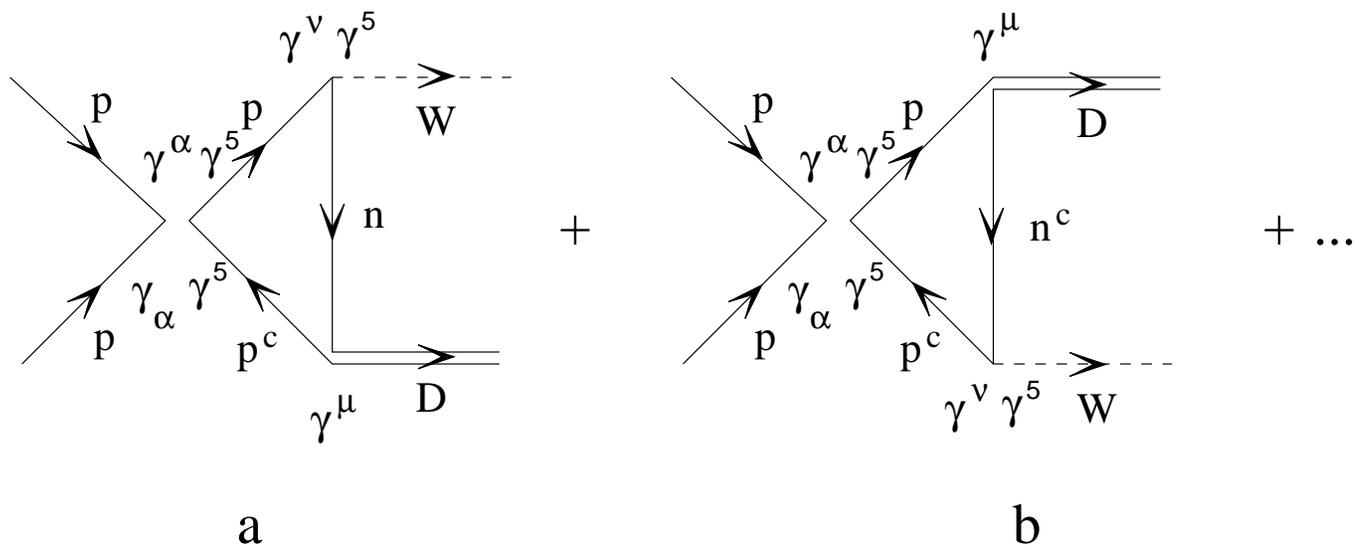
Fig.3. One--nucleon loop diagrams describing the effective Lagrangian
${\cal L}^{\rm pp \to De^+\nu_{\rm e}}_{\rm eff}(x)$ of the
low--energy transition p + p $\to$ D + e$^+$ + $\nu_{\rm e}$.

\end{itemize}

\end{document}